\documentclass[11pt,nofootinbib,superscriptaddress,preprint]{revtex4-1}
\pdfoutput=1
\usepackage[T1]{fontenc}
\usepackage[figuresright]{rotating}
\usepackage[colorlinks=true,citecolor={blue},linktocpage=true]{hyperref}
\usepackage[super]{nth}
\usepackage{times,graphicx,tabularx,dcolumn,bm,makecell,makebox,multirow,amsmath,amstext,amssymb,xfrac,amsfonts,longtable,booktabs,url,subfigure,dsfont,amsbsy,amsthm,esint,cleveref,mathrsfs,color,extarrows,datetime,comment,tikz-cd,float,braket}
\usepackage{lipsum}

\begin{document}
\title{Diagrammatics, Pentagon Equations, and Hexagon Equations of Topological Orders with Loop- and Membrane-like Excitations}
\author{Yizhou Huang}
\affiliation{School of Physics, Sun Yat-sen University, Guangzhou, 510275, China}
\author{Zhi-Feng Zhang}
\affiliation{Max Planck Institute for the Physics of Complex Ststems, Nöthnitzer Straße 38, Dresden 01187, Germany}
\author{Peng Ye}
\thanks{Corresponding author}\email{yepeng5@mail.sysu.edu.cn}
\affiliation{School of Physics, Sun Yat-sen University, Guangzhou, 510275, China}
\date{\today}
\begin{abstract} 
In spacetime dimensions of 4 (i.e., 3+1) and higher, topological orders exhibit spatially extended excitations like loops and membranes, which support diverse topological data characterizing braiding, fusion, and shrinking processes, despite the absence of anyons. Our understanding of these topological data remains less mature compared to 3D, where anyons have been extensively studied and can be fully described through diagrammatic representations. Inspired by recent advancements in field theory descriptions of higher-dimensional topological orders, this paper systematically constructs diagrammatic representations for 4D and 5D topological orders, generalizable to higher dimensions. We introduce elementary diagrams for fusion and shrinking processes, treating them as vectors in fusion and shrinking spaces, respectively, and build complex diagrams by combining these elementary diagrams. Within these vector spaces, we design unitary operations represented by \(F\)-, \(\Delta\)-, and \(\Delta^2\)-symbols to transform between different bases. We discover \textit{pentagon equations} and \textit{(hierarchical) shrinking-fusion hexagon equations} that impose constraints on the legitimate forms of these unitary operations. We conjecture that all anomaly-free higher-dimensional topological orders must satisfy these conditions and any violations indicate a quantum anomaly. This work opens promising avenues for future research, including the exploration of diagrammatic representations involving braiding and the study of non-invertible symmetries and symmetry topological field theories in higher spacetime dimensions.

	\end{abstract}
\maketitle
\newpage
\tableofcontents{}

\section{Introduction}\label{s1} 

The concept of topological order was introduced in condensed matter physics to describe exotic phases of matter characterized by long-range entanglement~\cite{zeng2018quantum}. Going beyond traditional symmetry-breaking theory, topological order has captured significant interest across various fields, including condensed matter physics, high-energy physics, mathematical physics, and quantum information~\cite{wenZootopoRMP}. Literature has explored diverse properties of topological orders, encompassing the existence of topological excitations, fusion rules, braiding statistics, and chiral central charge, collectively constituting the ``topological data'' crucial for their characterization and classification. One method of acquiring these data is through the path-integral formalism of topological quantum field theory (TQFT)~\cite{Witten1989,Turaev2016}. A notable example of TQFT is the $3$D\footnote{Unless otherwise specified, $3$D, $4$D, and $5$D always represent dimensions of  spacetime. That is, $3$D=(2+1)D, $4$D=(3+1)D, and $5$D=(4+1)D.} Chern-Simons theory~\cite{PhysRevB.42.8133,Nayak:2008aa,wen2004quantum}, which describes the low-energy, long-wavelength physics of $3$D topological orders (known as anyon models) like fractional quantum Hall states. As a side note,  the Chern-Simons theory is also applied to describe 3D symmetry-protected topological phases (SPT) and symmetry enriched topological phases (SET) by implementing global symmetry transformations, see, e.g., Refs.~\cite{PhysRevB.86.125119,YW12,PhysRevB.93.115136,PhysRevB.87.195103}. Anyon properties, such as fusion rules and braiding statistics, can be systematically computed within the field-theoretical framework. Moreover, fusion and braiding can be depicted diagrammatically, with the corresponding diagrammatic algebra closely linked to TQFT and topological quantum computation~\cite{KITAEV20062,PhysRevB.71.045110,PhysRevB.82.155138,bonderson2007non,Ardonne_2010,BONDERSON20082709,PhysRevB.90.195130,simon2023topological,kong2014braided,PhysRevB.100.115147}. Within anyon diagrams, $F$-symbols and $R$-symbols can be defined to facilitate transformations between different diagrams, subject to consistency conditions like the pentagon and hexagon equations in anomaly-free topological orders\footnote{For more details about anyon diagrams, we recommend the relevant section in the research article~\cite{PhysRevB.100.115147} as well as the   book~\cite{simon2023topological}. Appendix~\ref{ap1} is also supplemented for an introduction to   pentagon equations and braiding-fusion hexagon equations in anyon models.}.

Topological orders in 3D have been extensively explored through the lenses of TQFT and category theory. How about higher-dimensional topological orders? In 4D, topological excitations include both point-like particles and one-dimensional loop excitations, hereafter referred to as ``particles'' and ``loops'' for brevity. While mutual braidings among particles are trivial in 4D and higher dimensions~\cite{Leinaas1977,wu84,PhysRevLett.62.1071,PhysRevLett.62.1221,PhysRevLett.51.2250,goldhaber89}, the presence of topological excitations with spatially extended shapes significantly expands the potential for braiding and fusion properties. Moreover, the existence of spatially extended shapes allows for the consideration of shrinking rules \cite{Zhang2023fusion,Huang2023}, which dictate how loops can be shrunk into particles across one or more channels. Notably, nontrivial shrinking rules can only manifest in topological orders in 4D and higher, where topological excitations with spatially extended shapes exist. In 3D, a loop excitation is topologically indistinguishable from anyons since the braiding process is unable to detect the geometric hole of the loop excitation.

Analogous to the Chern-Simons field theory description of 3D topological orders, when particles and loops in 4D are respectively represented by charges and fluxes of an Abelian finite gauge group, the $BF$ theory and its twisted variants~\cite{horowitz_quantum_1990,ye16a,Moy_Fradkin2023,YW13a,2016arXiv161209298P,PhysRevB.99.235137,YeGu2015,ypdw,PhysRevLett.121.061601,zhang_compatible_2021,bti2,bti6} have been applied to describe 4D topological orders  as well as \textit{gauged} SPTs, which also attract a lot of investigations from non-invertible symmetry and symmetry topological field theory (SymTFT)~\cite{schafernameki2023ictp,Heidenreich2021,PhysRevLett.128.111601,PhysRevD.105.125016,Roumpedakis2023,2022JHEP08053K,Kaidi2023,2023JHEP10053K,Choi2023,antinucci2024anomalies,10.21468/SciPostPhys.14.4.067,argurio2024symmetry,cao2024symmetry,brennan2024symtft}. The $BF$ term ($BdA$  in which $B$ and $A$ are respectively $2$- and $1$-form gauge fields) equipped with various twisted terms (e.g., $AAdA$, $AAAA$, $AAB$, $dAdA$, and $BB$) enables systematic computation of topological data, such as particle-loop braiding~\cite{hansson_superconductors_2004,PRESKILL199050,PhysRevLett.62.1071,PhysRevLett.62.1221,ALFORD1992251}, multi-loop braiding~\cite{wang_levin1,PhysRevLett.114.031601,2016arXiv161209298P,string4,jian_qi_14,string5,PhysRevX.6.021015,Tiwari:2016aa,corbodism3,string6,3loop_ryu}, particle-loop-loop braiding (i.e., Borromean rings braiding)~\cite{PhysRevLett.121.061601}, emergent fermionic statistics~\cite{Kapustin:2014gua,bti2,PhysRevB.99.235137,Zhang2023Continuum}, and topological response~\cite{RevModPhys.83.1057,lapa17,YW13a,ye16a,Ye:2017aa,bti6,RevModPhys.88.035001,PhysRevB.99.205120}. Refs.~\cite{Ning2018prb,ye16_set,2016arXiv161008645Y} field-theoretically demonstrate how global symmetry is fractionalized on loop excitations. There, the concept of ``mixed three-loop braiding'' processes is introduced, leading to a classification scheme of SETs in higher dimensions.    Importantly,    braiding data discovered above are not always allowed to coexist. It has been shown that there exists important  compatibility among these braiding processes~\cite{zhang_compatible_2021}, ruling out gauge-non-invariant combinations of braiding processes. Furthermore,   loop excitations in Borromean rings topological order~\cite{PhysRevLett.121.061601} can exhibit non-Abelian fusion and shrinking rules~\cite{Zhang2023fusion}, despite the gauge charges carried by loops being Abelian. At the same time,  all shrinking rules are consistent with fusion rules  in the sense that   fusion coefficients and shrinking coefficients together should satisfy consistency conditions~\cite{Zhang2023fusion}. On the other hand, the emergence of fermions is also a fundamental issue. By studying correlation functions of Wilson operators equipped with  the framing regularization, Ref.~\cite{Zhang2023Continuum} investigates how fermionic statistics of particle excitations can emerge in the low-energy gauge theory, successfully incorporating the data of self-statistics of particle excitations.

When considering topological orders in $5$D, the topological data become even more exotic. Topological excitations now include particles, loops, and two-dimensional membranes, highlighting unexplored features of topological orders. From a field-theoretical perspective, we can still employ $BF$ field theory to investigate braiding statistics, fusion rules, and shrinking rules. The $BF$ term can take the form of either $CdA$ or $\tilde{B}dB$, where $C$ represents $3$-form gauge fields, $\tilde{B}$ and $B$ represent different $2$-form gauge fields. These terms can be further enriched with various twisted terms, such as $AAAAA$, $AAAdA$, $AdAdA$, $AAC$, $AAAB$, $BBA$, $AdAB$, $AAdB$, and $BC$~\cite{zhang_topological_2022}.  Recently, several exotic braiding processes have been computed via these $5$D topological terms~\cite{zhang_topological_2022}, resulting in mathematically nontrivial links formed by closed spacetime trajectories of particles, loops, and membranes. Technically, braiding statistics can be systematically computed by evaluating correlation functions of Wilson operators, whose expectation values are related to intersections of sub-manifolds embedded in the $5$D spacetime manifold. Additionally, exotic fusion and shrinking rules in $5$D have been explored~\cite{Huang2023}. Some membranes exhibit hierarchical shrinking rules, where a membrane shrinks into particles and loops, followed by the subsequent shrinking of loops into particles. Such nontrivial \textit{hierarchical shrinking} structures can only exist in $5$D and higher.

We have mentioned that the topological data of anyon models in $3$D can be schematically represented by diagrams, with their corresponding algebraic structure related to TQFT and category theory. Using diagrammatic representations is beneficial for understanding the universal properties of topological orders and their underlying mathematical structures, which plays a critical role in unitary operations for realizing topological quantum computation~\cite{simon2023topological}. Important consistency conditions, such as the celebrated \textit{pentagon and hexagon equations} can be efficiently handled via diagrams. However, there is still a lack of research on diagrammatic representations of higher-dimensional topological orders, which hinders  a systematic exploration of topologically ordered phases in higher dimensions. Especially, the presence of loop excitations and membrane excitations is expected to significantly enrich the potential constraints beyond 3D anyon theory. Inspired by the aforementioned progress in field theory in $4$D and $5$D, in this work, we aim to construct diagrammatic representations capable of describing the exotic properties of higher-dimensional topological orders in this paper.

In our diagrammatic representations, excitations are categorized into different sets, and we use different types of straight lines (e.g., single-line, double-line) to represent them. Given that fusion rules still satisfy the associativity condition, we can define fusion diagrams and $F$-symbols, similar to the case in $3$D. Changing the order of fusion processes in a diagram is implemented by $F$-symbols, which can be understood as basis transformations. To describe (hierarchical) shrinking processes and the consistent relation between (hierarchical) shrinking and fusion, we define (hierarchical) shrinking diagrams and $\Delta$-symbols ($\Delta^2$-symbols). Changing the order of fusion and (hierarchical) shrinking processes in a diagram is implemented by such $\Delta$-symbols ($\Delta^2$-symbols), which can also be viewed as basis transformations. Having defined $F$-symbols and $\Delta$-symbols ($\Delta^2$-symbols), we further discover a series of   equations that must be implemented for these symbols. First, $F$-symbols still satisfy the ``\textit{pentagon equation}''; second, $F$-symbols and $\Delta$-symbols ($\Delta^2$-symbols) jointly satisfy the so-called ``\textit{(hierarchical) shrinking-fusion hexagon equations}''. We conjecture that all anomaly-free topological orders in higher-dimensional spacetime should satisfy these equations. A quantum anomaly appears if one of these equations is not satisfied. 
It should be noted that the diagrams we construct in this paper involve fusion and (hierarchical) shrinking processes, thereby reflecting the algebraic structure of consistent fusion and (hierarchical) shrinking rules. 
As for braiding processes, since they may  involve more than two excitations and thus form various kinds of exotic links, the diagrammatic representations may be considerably more complicated, and we leave them for future exploration.

This paper is organized as follows. 
In section~\ref{s221}, we first review some basic concepts and properties of fusion and (hierarchical) shrinking rules, with a particular focus on their path-integral formalism established in our previous studies. We conclude that fusion rules respect (hierarchical) shrinking rules, which lays the foundation for our construction of diagrammatic representations of fusion and (hierarchical) shrinking rules. In section~\ref{s22}, we study fusion and shrinking in $4$D and $5$D as mappings between different sets of excitations. Since excitations from different sets play distinct roles in (hierarchical) shrinking processes, we need to treat them differently in our diagrams.
In section~\ref{s3}, we focus on 4D topological orders. Based on what we have established in section~\ref{s22} and~\ref{s3}, we define basic fusion and shrinking diagrams and combine them to generate more complex diagrams. By introducing $F$- and $\Delta$-symbols, we can transform diagrams and ultimately obtain the pentagon equation and (hierarchical) shrinking-fusion hexagon equation, which are key equations in our diagrammatic representations.
In section~\ref{s4}, we extend our construction to $5$D topological orders, introducing hierarchical shrinking diagrams and unitary operations denoted as $\Delta^2$-symbols. Following a similar strategy, we derive a new consistency relation called the ``hierarchical shrinking-fusion hexagon equation''.
A brief summary is provided in section~\ref{s5}, in which several promising future directions are discussed. 
Two appendices are supplemented. In appendix~\ref{ap1}, we offer a short review of diagrammatic representations of $3$D topological orders (anyon models).  Appendix~\ref{ap2} discusses the completeness of pentagon equations and shrinking-fusion hexagon equations.

\section{Path-integral formalism of topological data}\label{s221}
In this section, we will review several  field-theoretical results established before~\cite{Zhang2023fusion,Huang2023}, where path-integral formalism is applied. We first present two typical twisted terms in $BF$ theory, both of which can lead to non-Abelian fusion rules and nontrivial (hierarchical) shrinking rules. Then we give several examples of constructing gauge invariant Wilson operators for topological excitations. Meanwhile, we will explain the meaning of several useful notations that were introduced in our previous papers~\cite{Zhang2023fusion,Huang2023}, which will be frequently used in establishing the notion of ``set of excitations'' in section~\ref{s22}. Finally, we elaborate on how to calculate fusion and (hierarchical) shrinking rules in path-integral formalism. By exhausting all possible fusion and (hierarchical) shrinking processes, we conclude that in 4D $BF$ theory, fusion rules respect shrinking rules, i.e. satisfy eq.~(\ref{eq_consistency_relation_4D}). While in 5D, there exist nontrivial hierarchical shrinking rules. Fusion rules no longer respect shrinking rules in 5D generally. Instead, fusion rules respect hierarchical shrinking rules, i.e., satisfy eq.~(\ref{eq_consistency_relation_5D}). The two relations~(\ref{eq_consistency_relation_4D}) and~(\ref{eq_consistency_relation_5D}) are the key results in our previous works~\cite{Zhang2023fusion,Huang2023}. Our construction of diagrammatic representations is essentially based on these two relations.

\subsection{$BF$ theory with twisted terms}\label{ss221}
We consider $BF$ theory as the underlying TQFT for higher-dimensional topological orders. The topological action consists of $BF$ terms and twisted terms. The $BF$ term $BdA$\footnote{The wedge product ``$\wedge$'' is omitted for the notational convenience.} in $4$D is written as a wedge product of $B$ and $dA$, where $B$ and $A$ are respectively $2$- and $1$-form gauge fields. Twisted terms in $4$D consist of $AAdA$, $AAAA$, $AAB$, $dAdA$ and $BB$ terms. While in $5$D, $BF$ terms include $CdA$ and $\tilde{B}dB$ terms, where $C$ is $3$-form, $B$ and $\tilde{B}$ are two different $2$-form, $A$ is $1$-form. Twisted terms now include $AAAAA$, $AAAdA$, $AdAdA$, $AAC$, $AAAB$, $BBA$, $AdAB$, $AAdB$ and $BC$ terms. A $BF$ action can contain several different twisted terms as long as the action is gauge invariant~\cite{zhang_compatible_2021,Zhang2023Continuum}.

To better illustrate fusion and shrinking rules from TQFT, below we introduce two typical $BF$ actions first~\cite{Zhang2023fusion,Huang2023}. For simplicity, we only consider one twisted term and take the discrete Abelian gauge group as $G=\prod_{i=1}^n{\mathbb{Z} _{N_i}}=\left( \mathbb{Z} _2 \right) ^3$. In $4$D, we choose the twisted term to be the $AAB$ term and write the action as:
\begin{align}
S=\int{\frac{N_1}{2\pi}B^1dA^1+}\frac{N_2}{2\pi}B^2dA^2+\frac{N_3}{2\pi}B^3dA^3+qA^1A^2B^3,\label{eq_aab}
\end{align}
where $A^i$ and $B^i$ are $1$- and $2$-form gauge fields respectively. The coefficient $q=\frac{pN_1N_2N_3}{\left( 2\pi \right) ^2N_{123}}$, where $p\in\mathbb{Z}_{N_{123}}$ and $N_{123}$ is the greatest common divisor of $N_1$, $N_2$, and $N_3$. $A^1$, $A^2$ and $B^3$ satisfy the flat-connection conditions: $dA^1=0$, $dA^2=0$, and $dB^3=0$. The gauge transformations are:
\begin{align}
	A^1\!\rightarrow &A^1+d\chi ^1,\quad A^2\!\rightarrow A^2+d\chi ^2,\quad B^3\!\rightarrow B^3+dV^3\,,\nonumber
	\\
	B^1\!\rightarrow &B^1+dV^1-\frac{2\pi q}{N_1}\left( \chi ^2B^3-A^2V^3+\chi ^2dV^3 \right) \,,\nonumber
	\\
	B^2\!\rightarrow &B^2+dV^2+\frac{2\pi q}{N_2}\left( \chi ^1B^3-A^1V^3+\chi ^1dV^3 \right) \,,\nonumber
	\\
	A^3\!\rightarrow &A^3+d\chi ^3-\frac{2\pi q}{N_3}\left( \chi ^1A^2-\chi ^2A^1+\frac{1}{2}\chi ^1d\chi ^2-\frac{1}{2}\chi ^2d\chi ^1 \right) \,,
\end{align}
where $\chi^i$ and $V^i$ are $0$-form and $1$-form gauge parameters respectively. They satisfy the following compactness conditions: $\int d\chi^i\in2\pi\mathbb{Z}$ and $\int dV^i\in2\pi\mathbb{Z}$. Another example is the action that contains the $BBA$ twisted term in $5$D, we still consider $G=\prod_{i=1}^n{\mathbb{Z} _{N_i}}=\left( \mathbb{Z} _2 \right) ^3$ and write the action as:
\begin{align}
	S=\int{\frac{N_1}{2\pi}\tilde{B}^1dB^1+}\frac{N_2}{2\pi}\tilde{B}^2dB^2+\frac{N_3}{2\pi}C^3dA^3+qB^1B^2A^3,\label{eq_bba}
\end{align}
where $A^3$, $B^i$, $\tilde{B}^i$, and $C^3$ are $1$-, $2$-, $2$-, and $3$-form gauge fields respectively. The coefficient $q=\frac{pN_1N_2N_3}{\left( 2\pi \right) ^2N_{123}}$, where $p\in \mathbb{Z} _{N_{123}}$ and $N_{123}$ is the greatest common divisor of $N_{1}$, $N_{2}$ and $N_{3}$. The flat-connection conditions are $dB^1=0$, $dB^2=0$, and $dA^3=0$. The gauge transformations are given by:
\begin{align}
	B^1\!\rightarrow &B^1+dV^1,\quad B^2\!\rightarrow B^2+dV^2,\quad A^3\!\rightarrow A^3+d\chi ^3,\nonumber
	\\
	\tilde{B}^1\!\rightarrow &\tilde{B}^1+d\tilde{V}^1+\frac{2\pi q}{N_1}\left( V^2A^3+B^2\chi ^3+V^2d\chi ^3 \right) \,,\nonumber
	\\
	\tilde{B}^2\!\rightarrow &\tilde{B}^2+d\tilde{V}^2+\frac{2\pi q}{N_2}\left( V^1A^3+B^1\chi ^3+V^1d\chi ^3 \right) \,,\nonumber
	\\
	C^3\!\rightarrow &C^3+dT^3-\frac{2\pi q}{N_3}\left( V^1B^2+B^1V^2+\frac{1}{2}V^1dV^2+\frac{1}{2}V^2dV^1 \right) \,,
\end{align}
where $\chi^3$, $V^i$, $\tilde{V}^i$ and $T^3$ are $0$-, $1$-, $1$- and $2$-form gauge parameters respectively. The compactness conditions are: $\int d\chi^3\in2\pi\mathbb{Z}$, $\int dV^i\in2\pi\mathbb{Z}$, $\int d\tilde{V}^i\in2\pi\mathbb{Z}$, and $\int dT^3\in2\pi\mathbb{Z}$. The two examples above both have non-Abelian fusion and shrinking rules. Besides, the $BBA$ twisted term also admits hierarchical shrinking rules. 

\subsection{Wilson operators for topological excitations}\label{ss222}
In TQFT, topological excitations carry gauge charges minimally coupled to gauge fields. Thus, by respectively using gauge fields of $1$-, $2$-, and $3$-form, we can construct Wilson loop operators for particle excitations, Wilson surface operators for loop excitations, and  Wilson volume operators for membrane excitations. We use the Wilson operator $\mathcal{O}_\mathsf{a}$ to represent the topological excitation $\mathsf{a}$ in the path integral formalism. Following the notations in refs.~\cite{Zhang2023fusion,Huang2023}, we present some examples of these Wilson operators.

A particle can only carry gauge charges minimally coupled to 1-form gauge fields $A$ and we generally use $\mathsf{P}_{n_{1}n_{2}\cdots n_{k}}$ to represent a particle simultaneously carrying $n_{1}$, $n_{2}$, $\cdots$ , $n_{k}$ units of $\mathbb{Z} _{N_1}$, $\mathbb{Z} _{N_2}$, $\cdots$, $\mathbb{Z} _{N_k}$ gauge charge. The Wilson operator for such excitation should be constructed by using $k$ gauge fields ${A^1, A^2, \cdots, A^k}$. Consider $BBA$ twisted terms as an example, a particle carrying unit gauge charge minimally coupled to $1$-form gauge field $A^3$ is denoted as $\mathsf{P}_{001}$, which can be represented by the following gauge invariant Wilson operator $\mathcal{O}_{\mathsf{P}_{001}}$:
\begin{align}
	\mathcal{O}_{\mathsf{P}_{001}}=\exp \left( i\int_{\gamma}{A^3} \right) \,,
\end{align}
where $\gamma=S^1$ is the closed world-line of the particle. Since  the action~(\ref{eq_bba}) contains only one $1$-form gauge field, i.e., $A^3$, only $n_3$ can be nonzero. Since $G=\left(\mathbb{Z} _{2}\right)^{3}$, $n_3$ only takes $0$ or $1$ mod $2$ here.

A loop in 5D can carry gauge charges minimally coupled to 2-form gauge fields and be decorated by particles, i.e., the loop can be decorated by gauge charges minimally coupled to 1-form gauge fields. We generally use $\mathsf{L}_{n_{1}n_{2}\cdots n_{k},\tilde{n}_{1}\tilde{n}_{2}\cdots\tilde{n}_{k}}^{m_1m_2\cdots m_l}$ to represent a loop. Here $n_{1}$, $n_{2}$, $\cdots$ , $n_{k}$ in the subscript denote $n_{1}$, $n_{2}$, $\cdots$ , $n_{k}$ units of $\mathbb{Z} _{N_1}$, $\mathbb{Z} _{N_2}$, $\cdots$, $\mathbb{Z} _{N_k}$ gauge charge minimally coupled to $B^1$, $B^2$, $\cdots$, $B^k$ fields respectively. $\tilde{n}_{1}$, $\tilde{n}_{2}$, $\cdots$ , $\tilde{n}_{k}$ in the subscript denote $\tilde{n}_{1}$, $\tilde{n}_{2}$, $\cdots$ , $\tilde{n}_{k}$ units of $\mathbb{Z} _{N_1}$, $\mathbb{Z} _{N_2}$, $\cdots$, $\mathbb{Z} _{N_k}$ gauge charge minimally coupled to $\tilde{B}^1$, $\tilde{B}^2$, $\cdots$, $\tilde{B}^k$ fields respectively. $m_{1}$, $m_{2}$, $\cdots$ , $m_{l}$ in the superscript denote $m_{1}$, $m_{2}$, $\cdots$ , $m_{l}$ units of $\mathbb{Z} _{N_1}$, $\mathbb{Z} _{N_2}$, $\cdots$, $\mathbb{Z} _{N_l}$ gauge charge minimally coupled to $A^1$, $A^2$, $\cdots$, $A^l$ fields respectively. In $BBA$ twisted term, a loop simultaneously carrying gauge charges minimally coupled to $B^1$ and $B^2$ and decorated by gauge charge minimally coupled to $A^3$ is denoted as $\mathsf{L}_{110,000}^{001}$. The gauge invariant Wilson operator is given by
\begin{align}
	\mathcal{O}_{\mathsf{L}_{110,000}^{001}}=\exp \left( i\int_{\sigma}{B^1}+i\int_{\sigma}{B^2}+i\int_{\gamma}{A^3} \right)\,,
\end{align}
where $\sigma=S^1\times S^1=T^2$ is the closed world sheet of the loop excitation. $\gamma$ is an arbitrary closed line on $\sigma$, i.e., $\gamma\in \sigma$. For the notational simplicity, we omit 000 in the superscript and subscript, thus $\mathsf{L}_{110,000}^{001}$ can be rewritten as $\mathsf{L}_{110,}^{001}$. Note that the comma in the subscript should not be omitted. The numbers on the left side of the comma correspond to the $B$ fields while the numbers on the right side of the comma correspond to the $\tilde{B}$ fields.

As for membranes, we consider two different shapes: the membrane denoted by $\mathsf{M^S}$ is in the shape of $S^2$ and the membrane denoted by $\mathsf{M^T}$ is in the shape of $T^2$. Both of them can carry gauge charges minimally coupled to 3-form fields and they can be decorated by loops and particles. The loop and particle decorations on the membrane carry gauge charges minimally coupled to 2-form and 1-form fields respectively. We generally use $\mathsf{M^S}_{n_1n_2\cdots n_k}^{p_1p_2\cdots p_k;m_1m_2\cdots m_l,\tilde{m}_1\tilde{m}_2\cdots\tilde{m}_l}$ or $\mathsf{M^T}_{n_1n_2\cdots n_k}^{p_1p_2\cdots p_k;m_1m_2\cdots m_l,\tilde{m}_1\tilde{m}_2\cdots\tilde{m}_l}$ to represent them. Here $n_{1}$, $n_{2}$, $\cdots$ , $n_{k}$ in the subscript denote $n_{1}$, $n_{2}$, $\cdots$ , $n_{k}$ units of $\mathbb{Z} _{N_1}$, $\mathbb{Z} _{N_2}$, $\cdots$, $\mathbb{Z} _{N_k}$ gauge charge minimally coupled to $C^1$, $C^2$, $\cdots$, $C^k$ fields respectively. $p_{1}$, $p_{2}$, $\cdots$ , $p_{k}$ in the superscript denote $p_{1}$, $p_{2}$, $\cdots$ , $p_{k}$ units of $\mathbb{Z} _{N_1}$, $\mathbb{Z} _{N_2}$, $\cdots$, $\mathbb{Z} _{N_k}$ gauge charge minimally coupled to $A^1$, $A^2$, $\cdots$, $A^k$ fields respectively. $m_{1}$, $m_{2}$, $\cdots$ , $m_{l}$ in the superscript denote $m_{1}$, $m_{2}$, $\cdots$ , $m_{l}$ units of $\mathbb{Z} _{N_1}$, $\mathbb{Z} _{N_2}$, $\cdots$, $\mathbb{Z} _{N_l}$ gauge charge minimally coupled to $B^1$, $B^2$, $\cdots$, $B^l$ fields respectively. $\tilde{m}_{1}$, $\tilde{m}_{2}$, $\cdots$ , $\tilde{m}_{l}$ in the superscript denote $\tilde{m}_{1}$, $\tilde{m}_{2}$, $\cdots$ , $\tilde{m}_{l}$ units of $\mathbb{Z} _{N_1}$, $\mathbb{Z} _{N_2}$, $\cdots$, $\mathbb{Z} _{N_l}$ gauge charge minimally coupled to $\tilde{B}^1$, $\tilde{B}^2$, $\cdots$, $\tilde{B}^l$ fields respectively. For simplicity, we omit the whole superscript if there is no decoration. We also write ``$p_1p_2\cdots;000,000$'' as ``$p_1p_2\cdots;$'' and ``$000;000,\tilde{m}_1\tilde{m}_2\cdots$''  as ``$,\tilde{m}_1\tilde{m}_2\cdots$'', respectively. In $BBA$ twisted term, a torus-like membrane carrying unit gauge charge minimally coupled to 3-form field $C^3$ is denoted as $\mathsf{M^T}_{001}$. The corresponding gauge invariant Wilson operator is given by
\begin{align}
	\mathcal{O}_{\mathsf{M^T}_{001}}=4\exp \left[ i\int_{\omega}{C^3+\frac{1}{2}\frac{2\pi q}{N_3}\left( d^{-1}B^1B^2+d^{-1}B^2B^1 \right)} \right] \delta \left( \int_{\sigma}{B^1} \right) \delta \left( \int_{\sigma}{B^2} \right) \,,
	\label{eq_mt001}
\end{align}
where $\omega=T^2\times S^1$ is the closed world volume of the membrane excitation. The nontrivial terms $\frac{1}{2}\frac{2\pi q}{N_3}\left( d^{-1}B^1B^2+d^{-1}B^2B^1 \right)$ are introduced to guarantee that the Wilson operator is gauge invariant. We define $d^{-1}B^1$ and $d^{-1}B^2$ as
$d^{-1}B^1=\int_{\mathcal{A} \in \omega}{B^1} $ and $ d^{-1}B^2=\int_{\mathcal{A} \in \omega}{B^2}$, where $\mathcal{A}$ is an open area on $\omega$. As 1-form fields, $d^{-1}B^1$ and $d^{-1}B^2$ are well defined on $\omega$ if and only if $B^1$ and $B^2$ are exact on $\omega$, i.e., $B^1$ and $B^2$ satisfy extra constraints:  $\int_{\sigma}{B^1}=0$ mod $2\pi$, $\int_{\sigma}{B^2}=0$ mod $2\pi$, where $\sigma$ is an arbitrary closed surface on $\omega$. To enforce these constraints, we introduce two  delta functionals in eq.~(\ref{eq_mt001}):
\begin{align}
	\delta \left( \int_{\sigma}{B^1} \right) =\begin{cases}
		1, \quad \int_{\sigma}{B^1}=0   \,\, \mathrm{mod}\,\, 2\pi\\
		0, \quad \mathrm{else}\\
	\end{cases}\,,\,\,
	\delta \left( \int_{\sigma}{B^2} \right) =\begin{cases}
		1, \quad \int_{\sigma}{B^2}=0 \,\,  \mathrm{mod} \,\, 2\pi\\
		0, \quad \mathrm{else}\\
	\end{cases}\,.
\end{align}

We can construct other Wilson operators for excitations by using the similar approach. In our previous papers~\cite{Zhang2023fusion,Huang2023}, we have exhausted all topologically distinct Wilson operators in $BF$ theory with $AAB$ and $BBA$ twisted terms respectively. We conclude that there are 19 topologically distinct excitations in $BF$ theory with $AAB$ twisted term and 29 topologically distinct excitations in $BF$ theory with $BBA$ twisted term.

\subsection{Fusion and shrinking rules in path integral formalism}\label{ss223}
Fusion rules of topological excitations are important topological data in topological orders. If we adiabatically bring two topological excitations together spatially, the combination of the two excitations behaves like another topological excitation. Within the framework of path integral, fusing two topological excitations $\mathsf{a}$ and $\mathsf{b}$ is represented by
\begin{align}
	\langle \mathsf{a}\otimes\mathsf{b} \rangle&=\frac{1}{\mathcal{Z}}\int\mathcal{D} \left[ ABC\right] \exp \left( iS \right) \times \left( \mathcal{O} _{\mathsf{a}}\times \mathcal{O} _{\mathsf{b}} \right) \,.
\end{align}
where $\mathcal{Z}$ and $S$ are partition function and action respectively. $\mathcal{D} \left[ ABC\right]$ represents the integration over all field configurations. Refs.~\cite{Zhang2023fusion,Huang2023} show that we can further calculate the fusion rules as
\begin{align}
	\langle \mathsf{a}\otimes\mathsf{b} \rangle&=\frac{1}{\mathcal{Z}}\int\mathcal{D} \left[ ABC\right] \exp \left( iS \right) \times \left( \mathcal{O} _{\mathsf{a}}\times \mathcal{O} _{\mathsf{b}} \right) \nonumber
	\\
	&=\frac{1}{\mathcal{Z}}\int \mathcal{D} \left[ ABC\right] \exp \left( iS \right) \times \left( \sum_\mathsf{c}{N_{\mathsf{c}}^{\mathsf{a}\mathsf{b}}}\mathcal{O} _{\mathsf{c}} \right)\nonumber
	\\
	&=\langle \oplus _{\mathsf{c}}N_{\mathsf{c}}^{\mathsf{a}\mathsf{b}}\mathsf{c} \rangle  \,,
	\label{eq_fusion_TQFT}
\end{align}
 We can simply rewrite eq.~(\ref{eq_fusion_TQFT}) as
\begin{align}
	\mathsf{a}\otimes \mathsf{b}=\oplus _{\mathsf{c}}N_{\mathsf{c}}^{\mathsf{a}\mathsf{b}}\mathsf{c}
	\label{eq_general_fusion}\,.
\end{align}
$\mathsf{a}$, $\mathsf{b}$ and $\mathsf{c}$ denote topological excitations and  $N_{\mathsf{c}}^{\mathsf{a}\mathsf{b}}\in\mathbb{Z}$ is fusion coefficient implying there are $N_{\mathsf{c}}^{\mathsf{a}\mathsf{b}}$ fusion channels to $\mathsf{c}$. Summation $\oplus _{\mathsf{c}}$ exhausts all topological excitations in the system. If $N_{\mathsf{c}}^{\mathsf{a}\mathsf{b}}=0$, it means that such fusion channel does not exist and fusing $\mathsf{a}$ and $\mathsf{b}$ to $\mathsf{c}$ is prohibited. A fusion process that only has a single fusion channel is called an Abelian fusion process. In contrast, a non-Abelian fusion process contains multiple fusion channels, indicating different possible fusion outputs.   For an excitation $\mathsf{a}$, if fusing $\mathsf{a}$ and $ \mathsf{b}$ is always Abelian for any $\mathsf{b}$, we call $\mathsf{a}$ an Abelian excitation. Otherwise, we call $\mathsf{a}$ a non-Abelian excitation. Fusion rules satisfy commutativity and associativity, i.e.,
\begin{gather}
	\mathsf{a}\otimes \mathsf{b}=\mathsf{a}\otimes \mathsf{b}\,,
	\\
	\left(\mathsf{a}\otimes \mathsf{b}\right)\otimes \mathsf{c}=\mathsf{a}\otimes\left(\mathsf{b}\otimes \mathsf{c}\right)\,.
\end{gather}

\begin{figure}
	\centering
	\includegraphics[scale=0.5,keepaspectratio]{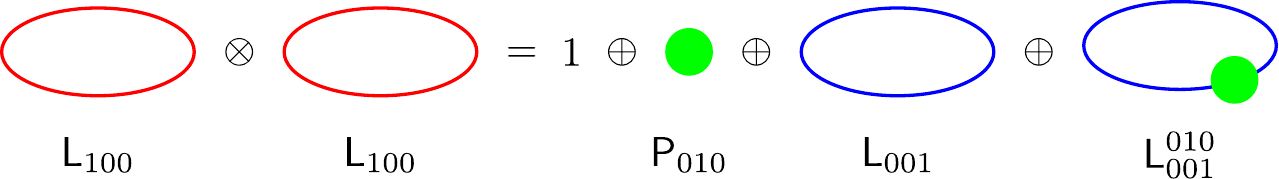}
	\caption{A typical example of fusion rules in $BF$ theory with $AAB$ twisted term. Fusing two loops $\mathsf{L}_{100}$ has four possible fusion channels:  the vacuum, the particle $\mathsf{P}_{010}$, the pure loop $\mathsf{L}_{001}$, and the decorated loop $\mathsf{L}_{001}^{010}$.}
	\label{fig_fusionAAB}
\end{figure}

Following the notations in ref.~\cite{Zhang2023fusion}, here we present an explicit example of fusing two loops. A loop in the $AAB$ twisted term carrying unit gauge charge minimally coupled to $B^1$ is denoted as $\mathsf{L}_{100}$, which can be represented by~\cite{Zhang2023fusion}
\begin{align}
	\mathcal{O}_{\mathsf{L}_{100}}=2\exp \left[ i\int_{\sigma}{B^1+\frac{1}{2}\frac{2\pi q}{N_1}\left( d^{-1}A^2B^3+d^{-1}B^3A^2 \right)} \right]\delta \left( \int_{\gamma}{A^2} \right) \delta \left( \int_{\sigma}{B^3} \right)\,.
\end{align}
We define $d^{-1}A^2$ and $d^{-1}B^3$ as
$d^{-1}A^2=\int_{\left[a,b\right] \in c}{A^2} $ and $ d^{-1}B^3=\int_{\mathcal{A} \in \sigma}{B^3}$ respectively, where $\left[a,b\right]$ is a segment of the closed curve $c$ and $\mathcal{A}$ is an open area on $\sigma$. The delta functionals are:
\begin{align}
	\delta \left( \int_{\sigma}{A^2} \right) =\begin{cases}
		1, \quad \int_{\gamma}{A^2}=0   \,\, \mathrm{mod}\,\, 2\pi\\
		0, \quad \mathrm{else}\\
	\end{cases}\,,\nonumber
	\\
	\delta \left( \int_{\sigma}{B^3} \right) =\begin{cases}
		1, \quad \int_{\sigma}{B^3}=0 \,\,  \mathrm{mod} \,\, 2\pi\\
		0, \quad \mathrm{else}\\
	\end{cases}\,.
\end{align}
Thus fusing two $\mathsf{L}_{100}$'s is given by:
\begin{align}
	&\langle \mathsf{L}_{100}\otimes \mathsf{L}_{100}\rangle \nonumber
	\\
	=&\frac{1}{\mathcal{Z}}\int{}\mathcal{D} \left[ AB \right] \exp \left( iS \right) \times \mathcal{O} _{\mathsf{L}_{100}}\times \mathcal{O} _{\mathsf{L}_{100}} \nonumber
	\\
	=&\frac{1}{\mathcal{Z}}\int{}\mathcal{D} \left[ AB \right] \exp \left( iS \right) \times\left\{2\exp \left[ i\int_{\sigma}{B^1+\frac{1}{2}\frac{2\pi q}{N_1}\left( d^{-1}A^2B^3+d^{-1}B^3A^2 \right)} \right]\right. \nonumber
	\\
	&\times\left. \delta \left( \int_{\gamma}{A^2} \right) \delta \left( \int_{\sigma}{B^3} \right)  \right\}^{2} \nonumber
	\\
	=&\frac{1}{\mathcal{Z}}\int{}\mathcal{D} \left[ AB \right] \exp \left( iS \right) \times \left[ 1+\exp \left( i\int_{\gamma}{A^2} \right)+\exp \left( i\int_{\sigma}{B^3} \right) +\exp \left( i\int_{\sigma}{A^2+B^3} \right) \right] \nonumber 
	\\
	=&\frac{1}{\mathcal{Z}}\int{}\mathcal{D} \left[ AB \right] \exp \left( iS \right) \times \left[ \mathcal{O} _1+\mathcal{O} _{\mathsf{P}_{010}}+\mathcal{O} _{\mathsf{L}_{001}}+\mathcal{O} _{\mathsf{L}_{001}^{010}} \right]  \nonumber
	\\
	=&\langle 1\oplus \mathsf{P}_{010}\oplus \mathsf{L}_{001}\oplus \mathsf{L}_{001}^{010}\rangle \,,
	\label{eq5}
\end{align}
where $\mathsf{1}$ is the vacuum, $\mathsf{P}_{010}$ is a particle carrying a gauge charge minimally coupled to $1$-form field $A^2$,  $\mathsf{L}_{001}$ is a loop carrying a gauge charge minimally coupled to $2$-form fields $B^3$, $\mathsf{L}_{001}^{010}$ is a decorated loop carrying both a gauge charge minimally coupled to $A^2$ and a gauge charge minimally coupled to $B^3$. Their corresponding gauge invariant Wilson operators are~\cite{Zhang2023fusion}
\begin{align}
	\mathcal{O} _{\mathsf{1}}=1\,,&\quad\mathcal{O} _{\mathsf{P}_{010}}=\exp \left( i\int_{\gamma}{A^2} \right)\,,\nonumber\\
	\quad\mathcal{O} _{\mathsf{L}_{001}}=\exp \left( i\int_{\sigma}{B^3} \right)\,,&\quad\mathcal{O} _{\mathsf{L}_{001}^{010}}=\exp \left( i\int_{\gamma}A^2+i\int_{\sigma} B^3 \right).
\end{align}
Here, $\gamma$ is the world line of particles and $\sigma$ is the world sheet of loops. 
Eq.~(\ref{eq5}) means that fusing two loops $\mathsf{L}_{100}$ has four possible outputs: the vacuum, the particle $\mathsf{P}_{010}$, the loop $\mathsf{L}_{001}$, and the loop $\mathsf{L}_{001}^{010}$. This is a non-Abelian fusion process as shown in figure~\ref{fig_fusionAAB}. One fusion channel can be changed to another by braiding processes.

In $3$D topological orders, topological excitations are point-like particles, i.e., anyons. However, in higher-dimensional topological orders, spatially extended topological excitations may occur. For instance, in $4$D, there exist loop excitations. By applying a series of local unitary operators, we can manipulate a particle to pass through the geometric hole of the loop, which forms a particle-loop braiding process. However, such a process can only happen when the spatial size of the loop is larger than the correlation length of the TQFT. If we shrink the loop to a size that is smaller than the correlation length, no excitation can pass through the hole of the loop, causing the loop to behave like a particle when braiding with other excitations. This motivates us to consider the shrinking rules: when we shrink a spatially higher dimensional topological excitation to a sufficiently small size, it behaves like a spatially lower dimensional excitation. In $4$D, there exist loop excitations and they can be shrunk to point-like particles. Keep lifting dimension to $5$D, spatially extended topological excitations include both loops and membranes. Some membranes can be shrunk to loops first and then shrunk to particles, and such shrinking processes are called \textit{hierarchical shrinking rules}. In the path integral representation, we define a shrinking operator $\mathcal{S}$, then shrinking an excitation $\mathsf{a}$ is~\cite{Zhang2023fusion,Huang2023} 
\begin{align}
	\langle \mathcal{S} \left( \mathsf{a} \right) \rangle =&\langle \underset{X_1\rightarrow X_2}{\lim}\mathsf{a} \rangle =\underset{X_1\rightarrow X_2}{\lim}\frac{1}{\mathcal{Z}}\int{\mathcal{D} \left[ ABC \right] \exp \left( iS \right) \mathcal{O} _{\mathsf{a}}} \nonumber
	\\
	=&\sum_{\mathsf{b}}\frac{1}{\mathcal{Z}}\int{\mathcal{D} \left[ ABC \right] \exp \left( iS \right) \mathrm{S}_{\mathsf{b}}^{\mathsf{a}}\mathcal{O} _{\mathsf{b}}}\nonumber
	\\
	=&\langle \oplus_{\mathsf{b}}{\mathrm{S}}_{\mathsf{b}}^{\mathsf{a}}\mathsf{b} \rangle \,,
	\label{eq_shrinking_TQFT}
\end{align}
where $\mathcal{Z}$ and $S$ are partition function and action respectively. $X_1$ and $X_2$ ($X_2\subset X_1$) are respectively  spacetime trajectories (manifold) of excitation $\mathsf{a}$ (Wilson operator $\mathcal{O} _{\mathsf{a}}$) before and after shrinking. We can simply rewrite eq.~(\ref{eq_shrinking_TQFT}) as
\begin{align}
	\mathcal{S}\left(\mathsf{a}\right)=\oplus_{\mathsf{b}} {\mathrm{S}}_{\mathsf{b}}^{\mathsf{a}} \mathsf{b}\,.\label{eq_general_shrinking}
\end{align}
${\mathrm{S}}_{\mathsf{b}}^{\mathsf{a}}\in\mathbb{Z}$ is the shrinking coefficient, and there are ${\mathrm{S}}_{\mathsf{b}}^{\mathsf{a}}$ shrinking channels to $\mathsf{b}$. Summation $\oplus _{\mathsf{b}}$ exhausts all topological excitations in the system. If ${\mathrm{S}}_{\mathsf{b}}^{\mathsf{a}}=0$, then  shrinking $\mathsf{a}$ to $\mathsf{b}$ is prohibited. A shrinking process is called Abelian if it has a single shrinking channel. Otherwise, it is deemed non-Abelian. Working in the TQFT paradigm, refs.~\cite{Huang2023,Zhang2023fusion} show that (hierarchical) shrinking rules respect fusion rules; that is, in $4$D, we have:
\begin{align}
	\mathcal{S} \left( \mathsf{a} \right) \otimes \mathcal{S} \left( \mathsf{b} \right) =\mathcal{S} \left( \mathsf{a}\otimes \mathsf{b} \right)\,.
	\label{eq_consistency_relation_4D}
\end{align}
Eq.~(\ref{eq_consistency_relation_4D}) only holds for $4$D because we only need one step to shrink loops to particles, which precludes the existence of non-trivial hierarchical shrinking structures. While in $5$D, topological excitations can be particles, loops, and membranes. For a sphere-like membrane, we can directly shrink it to particles. However, for a torus-like membrane, we may first shrink it to a loop and then continue to shrink the loop to a particle. If we need at least two steps to shrink such a membrane to a particle, we say there exist hierarchical shrinking rules. Ref.~\cite{Huang2023} shows that for $5$D topological order described by the $BF$ field theory, the $BBA$ and $AAAB$ twisted terms can lead to nontrivial hierarchical shrinking rules. For the $AAAAA$, $AAAdA$, and $AAC$ twisted terms, we find that they have non-Abelian shrinking rules, but their   shrinking rules are not hierarchical, i.e., torus-like membranes are shrunk to nontrivial particles and the trivial loop in the first step. Since the trivial loop is equivalent to the vacuum and thus can be ignored,   these torus-like membranes   are directly shrunk to particles in the first shrinking process. For the $AdAdA$, $AdAB$, and $AAdB$ twisted terms, they only admit Abelian shrinking rules and trivial hierarchical shrinking rules. Generally, in $5$D, hierarchical shrinking rules respect fusion rules, that is,
\begin{align}
	\mathcal{S} ^2\left( \mathsf{a} \right) \otimes \mathcal{S} ^2\left( \mathsf{b} \right) =\mathcal{S} ^2\left( \mathsf{a}\otimes \mathsf{b} \right) \,,
	\label{eq_consistency_relation_5D}
\end{align}
where $\mathcal{S} ^2\left(\cdot\right)=\mathcal{S}\left(\mathcal{S}\left(\cdot\right)\right)$ means shrinking twice.

\begin{figure}
	\centering
	\includegraphics[scale=0.12,keepaspectratio]{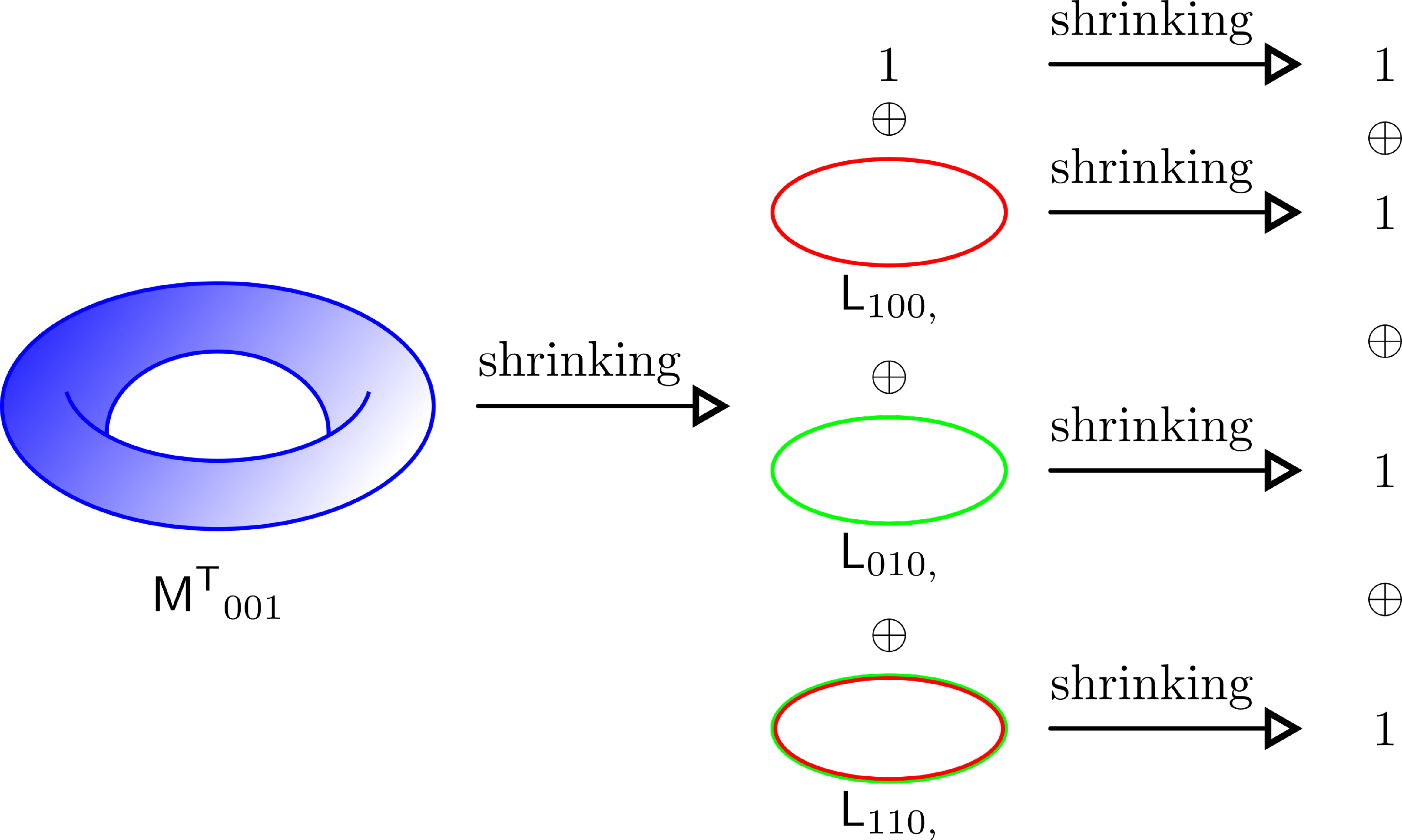}
	\caption{A typical example of hierarchical shrinking rules in $BF$ theory with $BBA$ twisted term. The torus membrane ${\mathsf{M}^{\mathsf{T}}}_{001}$ has four shrinking channels: the vacuum, the loop $\mathsf{L}_{100,}$, the loop $\mathsf{L}_{010,}$, and the loop $\mathsf{L}_{110,}$. Subsequently, these loops are shrunk to vacua.}
	\label{fig_shrinkingBBA}
\end{figure}

Following the notations in ref.~\cite{Huang2023}, here we calculate an explicit example of shrinking a torus-like membrane ${\mathsf{M}^{\mathsf{T}}}_{001}$ (whose Wilson operator is shown in eq.~(\ref{eq_mt001})) in the $BBA$ twisted term:
\begin{align}
	&\langle \mathcal{S} \left( {\mathsf{M}^{\mathsf{T}}}_{001} \right) \rangle =\langle \underset{{\omega}\rightarrow \sigma}{\lim}{\mathsf{M}^{\mathsf{T}}}_{001}\rangle =\underset{{\omega}\rightarrow \sigma}{\lim}\frac{1}{\mathcal{Z}}\int{\mathcal{D} \left[ ABC \right] \exp \left( iS \right)}\times\mathcal{O} _{{\mathsf{M}^{\mathsf{T}}}_{001}}\nonumber
	\\
	=&\underset{{\omega}\rightarrow \sigma}{\lim}\frac{1}{\mathcal{Z}}\int{\mathcal{D} \left[ ABC \right] \exp \left( iS \right)}\times4\exp \left[ i\int_{{\omega}}{C^3+\frac{1}{2}\frac{2\pi q}{N_3}\left( d^{-1}B^1B^2+d^{-1}B^2B^1 \right)} \right]\nonumber
	\\
	&\times \delta \left( \int_{\sigma}{B^1} \right) \delta \left( \int_{\sigma}{B^2} \right)  \nonumber
	\\
	=&\frac{1}{\mathcal{Z}}\int{\mathcal{D} \left[ ABC \right] \exp \left( iS \right)}  \left[ 1+\exp \left( i\int_{\sigma}{B^1} \right) +\exp \left( i\int_{\sigma}{B^2} \right) +\exp \left( i\int_{\sigma}B^1+i\int_{\sigma}B^2 \right) \right] \nonumber
	\\
	=&\frac{1}{\mathcal{Z}}\int{\mathcal{D} \left[ ABC \right] \exp \left( iS \right)} \left[ \mathcal{O} _1+\mathcal{O} _{\mathsf{L}_{100,}}+\mathcal{O} _{\mathsf{L}_{010,}}+\mathcal{O} _{\mathsf{L}_{110,}} \right] \nonumber
	\\
	=&\langle \mathsf{1}\oplus \mathsf{L}_{100,}\oplus \mathsf{L}_{010,}\oplus \mathsf{L}_{110,}\rangle \,.
\end{align}
Here, ${\mathsf{M}^{\mathsf{T}}}_{001}$ is a torus-like membrane carrying a gauge charge minimally coupled to $3$-form field $C^3$. $\mathsf{1}$ is the vacuum, $\mathsf{L}_{100,}$ and $\mathsf{L}_{010,}$ are loops carrying a gauge charge minimally coupled to $2$-form field $B^1$ and a gauge charge minimally coupled to $2$-form field $B^2$ respectively, $\mathsf{L}_{110,}$ is a loop carrying both a gauge charge minimally coupled to $B^1$ and a gauge charge minimally coupled to $B^2$. Their corresponding gauge invariant Wilson operators are~\cite{Huang2023}
\begin{align}
	&\mathcal{O} _1=1\,,\quad\mathcal{O} _{\mathsf{L}_{100,}}=\exp \left( i\int_{\sigma}{B^1} \right)\,,\quad\mathsf{L}_{010,}=\exp \left( i\int_{\sigma}{B^2} \right)\nonumber
	\\
	&\mathcal{O} _{\mathsf{L}_{110,}}=\exp \left( i\int_{\sigma}B^1+i\int_{\sigma}B^2 \right)\,.
\end{align}
This result means that shrinking the membrane ${\mathsf{M}^{\mathsf{T}}}_{001}$ has four shrinking channels: the vacuum, the loop $\mathsf{L}_{100,}$, the loop $\mathsf{L}_{010,}$, and the loop $\mathsf{L}_{110,}$. These loops can be further shrunk to particles, i.e., the membrane ${\mathsf{M}^{\mathsf{T}}}_{001}$ have hierarchical shrinking rules:
\begin{align}
	&\langle \mathcal{S} ^2\left( {\mathsf{M}^{\mathsf{T}}}_{001} \right) \rangle =\langle \mathcal{S} \left( 1\oplus \mathsf{L}_{100,}\oplus \mathsf{L}_{010,}\oplus \mathsf{L}_{110,} \right) \rangle =\langle \underset{\sigma \rightarrow \gamma}{\lim}\left( 1\oplus \mathsf{L}_{100,}\oplus \mathsf{L}_{010,}\oplus \mathsf{L}_{110,} \right) \rangle \nonumber
	\\
	=&\underset{\sigma \rightarrow \gamma}{\lim}\frac{1}{\mathcal{Z}}\int{\mathcal{D} \left[ ABC \right] \exp \left( iS \right)}\left[ 1+\exp \left( i\int_{\sigma}{B^1} \right) +\exp \left( i\int_{\sigma}{B^2} \right) +\exp \left( i\int_{\sigma}{B^1+B^2} \right) \right] \nonumber
	\\
	=&\frac{1}{\mathcal{Z}}\int{\mathcal{D} \left[ ABC \right] \exp \left( iS \right) } \left[ 1+\exp \left( i0 \right) +\exp \left( i0 \right) +\exp \left( i0 \right) \right]\nonumber
	\\
	=&\frac{1}{\mathcal{Z}}\int{\mathcal{D} \left[ ABC \right] \exp \left( iS \right) \times 4\cdot 1}=\langle 4\cdot \mathsf{1}\rangle \,.
\end{align}
This result indicates that, by performing shrinking twice, ${\mathsf{M}^{\mathsf{T}}}_{001}$ can be shrunk to the vacuum ultimately, as shown in figure~\ref{fig_shrinkingBBA}.

\section{Fusion and shrinking rules as mappings}\label{s22}
In this section, we will categorize excitations into different sets and investigate fusion and shrinking processes as mappings between these sets. Excitations from different sets need to be treated differently in our diagrammatic representations because they play distinct roles in (hierarchical) shrinking processes. Assuming only a finite number of topologically distinct excitations exist in a topological order, we can exhaust all   excitations and put them in a \textit{set of excitations}, denoted as $\Phi _{0}^{\mathcal{D}}$, where the superscript $\mathcal{D}$ denotes the spacetime dimension. In this paper, we mainly focus on $\mathcal{D}=4,5$. In the following discussion, we interpret shrinking as a mapping from the set $\Phi _{0}^{\mathcal{D}}$ to one of its subsets, denoted as $\Phi _{1}^{\mathcal{D}}$. If nontrivial hierarchical shrinking rules exist, we are allowed to   map the subset $\Phi _{i}^{\mathcal{D}}$ to a smaller subset $\Phi _{i+1}^{\mathcal{D}}$, where $i=0,1,\cdots,\mathcal{D}-4$.  

\subsection{4D topological orders}\label{s222}
In a $4$D topological order, we generally list all excitations in a set $\Phi _{0}^{4}$ as follows: 
\begin{align}
	\Phi _{0}^{4}=\left\{ \mathsf{1},\mathsf{P}_1,\mathsf{P}_2,\cdots \mathsf{P}_n,\mathsf{L}_1,\mathsf{L}_2,\cdots ,\mathsf{L}_m \right\}\,,
	\label{set_4D_U0}
\end{align}
where $\mathsf{1}$ denotes the vacuum (both the trivial particle $\mathsf{P}_0$ and trivial loop  $\mathsf{L}_0$ are equivalent to vacuum, i.e., $\mathsf{1}=\mathsf{P}_0=\mathsf{L}_0$). $\mathsf{P}_i$ and $\mathsf{L}_i$ denote particles and loops respectively. The possible outputs of fusing two excitations still form the set $\Phi _{0}^{4}$, thus we can regard fusion as a mapping:
\begin{align}
	\otimes :\quad \Phi _{0}^{4}\times \Phi _{0}^{4}\rightarrow \Phi _{0}^{4}\,,\label{equation_fusion_3+1}
\end{align}
where  $\Phi _{0}^{4}\times \Phi _{0}^{4}=\left\{ \left( \mathsf{E}_i,\mathsf{E}_j \right) \mid \mathsf{E}_i\in \Phi _{0}^{4},\mathsf{E}_j\in \Phi _{0}^{4} \right\} $ and the symbol $\otimes$ means fusion.

A shrinking operator can shrink loops into particles, but it will not alter particles (which means the shrinking operator acts on a particle will give the particle itself) as, evidently, there is no way to further shrink a point. Thus, shrinking can be considered as a mapping from the original set $\Phi _{0}^{4}$ to a subset $\Phi _{1}^{4}$: 
\begin{gather}
	\mathcal{S} :\quad \Phi _{0}^{4}\rightarrow \Phi _{1}^{4}\,.
	\\
	\Phi _{1}^{4}=\left\{ \mathsf{P}_0,\mathsf{P}_1,\mathsf{P}_2,\cdots \mathsf{P}_n \right\}\,.
	\label{set_4D_U1}
\end{gather}
Ref.~\cite{Zhang2023fusion} shows that such a subset $\Phi _{1}^{4}$ also has closed fusion rules, i.e.,  the possible outputs of fusing two excitations in $\Phi _{1}^{4}$ still form the set $\Phi _{1}^{4}$. As a result, eq.~(\ref{equation_fusion_3+1}) can be generalized to encompass the following fusion processes:
\begin{align}
	\otimes :\,\, \Phi _{i}^{4}\times \Phi _{i}^{4}\rightarrow \Phi _{i}^{4}\,,\quad i=0,1\,.
\end{align}
Now consider two copies of the set $\Phi _{0}^{4}$, we have two ways to map $\Phi _{0}^{4}\times \Phi _{0}^{4}$ to the subset $\Phi _{1}^{4}$. The first approach involves mapping $\Phi _{0}^{4}\times \Phi _{0}^{4}$ to $\Phi _{0}^{4}$ by fusion, and then mapping $\Phi _{0}^{4}$ to $\Phi _{1}^{4}$ by shrinking. Alternatively, we can map $\Phi _{0}^{4}\times \Phi _{0}^{4}$ to $\Phi _{1}^{4}\times \Phi _{1}^{4}$ by shrinking, and subsequently map $\Phi _{1}^{4}\times \Phi _{1}^{4}$ to $\Phi _{1}^{4}$ by fusion. These two procedures can be represented as
\begin{align}
	&\text{(a)}\quad\otimes :\,\,\, \Phi _{0}^{4}\times \Phi _{0}^{4}\rightarrow \Phi _{0}^{4}\,,\quad\quad\,\,\,\,\,\quad \mathcal{S} :\,\,\, \Phi _{0}^{4}\rightarrow \Phi _{1}^{4}\,,
	\\
	&\text{(b)}\quad\mathcal{S} :\,\,\, \Phi _{0}^{4}\times \Phi _{0}^{4}\rightarrow \Phi _{1}^{4}\times \Phi _{1}^{4}\,,\,\,\,\quad \otimes :\,\,\, \Phi _{1}^{4}\times \Phi _{1}^{4}\rightarrow \Phi _{1}^{4}\,,
\end{align}
respectively. Recall eq.~(\ref{eq_consistency_relation_4D}) in $4$D, shrinking rules respect fusion rules. We conclude that for $\mathsf{a},\mathsf{b}\in\Phi _{0}^{4}$, $\mathcal{S} \left( \mathsf{a}\otimes \mathsf{b} \right) $ and $\mathcal{S} \left( \mathsf{a} \right) \otimes \mathcal{S} \left( \mathsf{b} \right) $ only differ by intermediate states and they give the same outputs.

In the following, we take the $AAB$ twisted term in 4D as an example. We consider the gauge group as $G=\prod_{i=1}^n{\mathbb{Z} _{N_i}}=\left( \mathbb{Z} _2 \right) ^3$ and the action given by eq.~(\ref{eq_aab}). Ref.~\cite{Zhang2023fusion} shows that there are 19 distinct excitations, which form the set $\Phi _{0}^{4}$
\begin{align}
	\Phi _{0}^{4}=&\left\{ \mathsf{1},\mathsf{P}_{100},\mathsf{P}_{010},\mathsf{P}_{001},\mathsf{P}_{110},\mathsf{L}_{100},\mathsf{L}_{010},\mathsf{L}_{001},\mathsf{L}_{110},\right.\nonumber
	\\
	&\left.\mathsf{L}_{100}^{001},\mathsf{L}_{100}^{100},\mathsf{L}_{010}^{010},\mathsf{L}_{010}^{001},\mathsf{L}_{001}^{100},\mathsf{L}_{001}^{010},\mathsf{L}_{001}^{110},\mathsf{L}_{001}^{001},\mathsf{L}_{110}^{100},\mathsf{L}_{110}^{001} \right\} .
\end{align}
Here, $\mathsf{1}$, $\mathsf{P}$, and $\mathsf{L}$ denote the vacuum, particles, and loops respectively. Consider $n_i,m_i=0,1$, the subscript ``$n_1n_2n_3$'' of a particle denotes that the particle carries $n_{1}$, $n_{2}$, and $n_{3}$ gauge charges minimally coupled to the 1-form gauge  fields $A^{1}$, $A^{2}$, and $A^{3}$, respectively. The subscript ``$n_1n_2n_3$'' of a loop denotes that the loop carries $n_{1}$, $n_{2}$, and $n_{3}$ gauge charges minimally coupled to 2-form gauge fields $B^{1}$, $B^{2}$, and $B^{3}$, respectively. The superscript ``$m_1m_2m_3$'' of a loop denotes that the loop is decorated by $m_{1}$, $m_{2}$, and $m_{3}$ gauge charges minimally coupled to the 1-form gauge  fields $A^{1}$, $A^{2}$, and $A^{3}$, respectively. From ref.~\cite{Zhang2023fusion}, we conclude that the subset $\Phi _{1}^{4}$ is given by
\begin{align}
	\Phi _{1}^{4}=\left\{ \mathsf{1},\,\mathsf{P}_{100},\,\mathsf{P}_{010},\,\mathsf{P}_{001},\,\mathsf{P}_{110} \right\} .
\end{align}

\subsection{$5$D topological orders}\label{s223}
We can generalize the above prescription to $5$D, where excitations   include particles, loops, and membranes. We can generally list the set $\Phi _{0}^{5}$ as:
\begin{align}
	\Phi _{0}^{5}=\left\{ 1,\mathsf{P}_1,\mathsf{P}_2,\cdots \mathsf{P}_n,\mathsf{L}_1,\mathsf{L}_2,\cdots ,\mathsf{L}_m,\mathsf{M}_1,\mathsf{M}_1,\cdots ,\mathsf{M}_k \right\} ,
	\label{set_5D_W0}
\end{align}
where $\mathsf{1}$ is the vacuum (trivial particle $\mathsf{P}_0$, trivial loop  $\mathsf{L}_0$, and trivial membrane $\mathsf{M}_0$ are equivalent to vacuum), $\mathsf{P}_i$, $\mathsf{L}_i$, and $\mathsf{M}_i$ denote particles, loops, and membranes respectively. If we consider nontrivial hierarchical shrinking rules, loops and membranes are shrunk to particles and loops respectively in the first shrinking process, which can be considered as a mapping from the original set $\Phi _{0}^{5}$ to a subset $\Phi _{1}^{5}$:
\begin{gather}
	\Phi _{1}^{5}=\left\{ 1,\mathsf{P}_1,\mathsf{P}_2,\cdots \mathsf{P}_n,\mathsf{L}_1,\mathsf{L}_2,\cdots ,\mathsf{L}_p \right\} \,,
	\label{set_5D_W1}
\end{gather}
where $p\leqslant m$ because some loops may not appear as shrinking outputs. In fact, ref.~\cite{Huang2023} shows that we have $p\textless m$ for the $BBA$ and $AAAB$ twisted terms. (If the topological order does not have nontrivial hierarchical shrinking rules, then $\mathsf{L}_1,\mathsf{L}_2,\cdots ,\mathsf{L}_p$ vanish, which means that the subset $\Phi _{1}^{5}$ and the subset $\Phi _{2}^{5}$ [to be defined below] contain the same excitations). Continue to shrink the excitations in subset $\Phi _{1}^{5}$, we get a smaller subset:
\begin{gather}
	\Phi _{2}^{5}=\left\{ 1,\mathsf{P}_1,\mathsf{P}_2,\cdots \mathsf{P}_n \right\} \,.
	\label{set_5D_W2}
\end{gather}
Given the result in ref.~\cite{Huang2023}, we can verify that the sets $\Phi _{0}^{5}$, $\Phi _{1}^{5}$, and $\Phi _{2}^{5}$ all have closed fusion rules. Thus we can consider shrinking and fusion as mappings:
\begin{align}
	&\otimes :\,\,\,\, \Phi _{i}^{5}\times \Phi _{i}^{5}\rightarrow \Phi _{i}^{5}\,,\,\, i=0,1,2\nonumber
	\\
	&\mathcal{S} :\,\,\,\, \Phi _{j}^{5}\rightarrow \Phi _{j+1}^{5}\,,\,\, j=0,1
\end{align}
 
If we consider fusion and shrinking simultaneously, we will find that the original set $\Phi _{0}^{5}$ only admits 
\begin{align}
	\mathcal{S} ^2\left( \mathsf{a} \right) \otimes \mathcal{S} ^2\left( \mathsf{b} \right) =\mathcal{S} ^2\left( \mathsf{a}\otimes \mathsf{b} \right), \quad \mathsf{a},\,\mathsf{b}\in \Phi _{0}^{5}\,.
\end{align}
While the subset $\Phi _{1}^{5}$ and $\Phi _{2}^{5}$ admit
\begin{align}
	\mathcal{S} \left( \mathsf{c} \right) \otimes \mathcal{S} \left( \mathsf{d} \right) =\mathcal{S} \left( \mathsf{c}\otimes \mathsf{d} \right), \quad \mathsf{c},\,\mathsf{d}\in \Phi _{1,2}^{5}\,.
	\label{eq_consistency_relation_5D2}
\end{align}
Eq.~(\ref{eq_consistency_relation_5D2}) is not the most general relation in $5$D because  $\mathsf{c}$ and $\mathsf{d}$ can only be particles and loops. Since the subset $\Phi _{1}^{5}$ and $\Phi _{2}^{5}$ can be obtained by acting a shrinking process on $\Phi _{0}^{5}$, i.e., we can always find $\mathsf{a}\in \Phi _{0}^{5}$ and $\mathsf{b}\in \Phi _{0}^{5}$ such that $\mathcal{S} \left( \mathsf{a} \right)$ and $\mathcal{S} \left( \mathsf{b} \right)$ have shrinking channels to $\mathsf{c}\in \Phi _{1,2}^{5}$ and $\mathsf{d}\in \Phi _{1,2}^{5}$ respectively, we have
\begin{align}
	\mathcal{S} ^2\left( \mathsf{a} \right) \otimes \mathcal{S} ^2\left( \mathsf{b} \right) =\mathcal{S}\left(\mathcal{S} \left( \mathsf{a}\right)\otimes\mathcal{S} \left( \mathsf{b}\right)\right) 
\end{align}
holds for the original set $\Phi _{0}^{5}$. Now consider two copies of the set $\Phi _{0}^{5}$, we have three ways to map $\Phi _{0}^{5}\times\Phi _{0}^{5}$ to $\Phi _{2}^{5}$:
\begin{align}
	\text{(a)}\quad\,\mathcal{S} :\quad&\Phi _{0}^{5}\times \Phi _{0}^{5}\rightarrow \Phi _{1}^{5}\times \Phi _{1}^{5}\,,\nonumber
	\\
	\mathcal{S} :\quad&\Phi _{1}^{5}\times \Phi _{1}^{5}\rightarrow \Phi _{2}^{5}\times \Phi _{2}^{5}\,,\nonumber
	\\
	\otimes :\quad&\Phi _{2}^{5}\times \Phi _{2}^{5}\rightarrow \Phi _{2}^{5}\,,
	\\
	\text{(b)}\quad\otimes :\quad&\Phi _{0}^{5}\times \Phi _{0}^{5}\rightarrow \Phi _{0}^{5}\,,\nonumber
	\\
	\mathcal{S} :\quad&\Phi _{0}^{5}\rightarrow \Phi _{1}^{5}\,,\nonumber
	\\
	\mathcal{S} :\quad&\Phi _{1}^{5}\rightarrow \Phi _{2}^{5}\,,
	\\
	\text{(c)}\quad\mathcal{S} :\quad&\Phi _{0}^{5}\times \Phi _{0}^{5}\rightarrow \Phi _{1}^{5}\times \Phi _{1}^{5}\,,\nonumber
	\\
	\otimes :\quad&\Phi _{1}^{5}\times \Phi _{1}^{5}\rightarrow \Phi _{1}^{5}\,,\nonumber
	\\
	\mathcal{S} :\quad &\Phi _{1}^{5}\rightarrow \Phi _{2}^{5}\,.
\end{align}
For $\mathsf{a},\mathsf{b}\in\Phi _{0}^{5}$, the condition $\mathcal{S} ^2\left( \mathsf{a} \right) \otimes \mathcal{S} ^2\left( \mathsf{b} \right) =\mathcal{S} ^2\left( \mathsf{a}\otimes \mathsf{b} \right) =\mathcal{S} \left( \mathcal{S} \left( \mathsf{a} \right) \otimes \mathcal{S} \left( \mathsf{b} \right) \right) $ indicates that the three processes above only differ by intermediate states and they should give the same outputs.

Here, we take the $BBA$ twisted term in 5D as an example. We consider the gauge group as $G=\prod_{i=1}^n{\mathbb{Z} _{N_i}}=\left( \mathbb{Z} _2 \right) ^3$ and the action given by eq.~(\ref{eq_bba}). Ref.~\cite{Huang2023} shows that there are 29 distinct excitations, which form the set $\Phi _{0}^{5}$ (pay attention to commas and semicolons in sub- and super-scripts):
\begin{align}
	\Phi _{0}^{5}=&\left\{ 1,\,\mathsf{P}_{001},\,\mathsf{L}_{100,},\,\mathsf{L}_{010,},\,\mathsf{L}_{110,},\,\mathsf{L}_{100,}^{001},\,\mathsf{L}_{010,}^{001},\,\mathsf{L}_{110,}^{001},\, \mathsf{L}_{,100},\,\mathsf{L}_{,010},\,\mathsf{L}_{,110},\,\mathsf{L}_{100,100},\,\mathsf{L}_{010,010},\,\mathsf{L}_{100,110},\right.\nonumber
	\\
	&\left.{\mathsf{M}^{\mathsf{S}}}_{001},\,{\mathsf{M}^{\mathsf{S}}}_{001}^{001;},\,{\mathsf{M}^{\mathsf{S}}}_{001}^{,100},\,{\mathsf{M}^{\mathsf{S}}}_{001}^{,010},\,{\mathsf{M}^{\mathsf{S}}}_{001}^{,110},\,{\mathsf{M}^{\mathsf{T}}}_{001},\,{\mathsf{M}^{\mathsf{T}}}_{001}^{001;},{\mathsf{M}^{\mathsf{T}}}_{001}^{,100},\,{\mathsf{M}^{\mathsf{T}}}_{001}^{,010},\,{\mathsf{M}^{\mathsf{T}}}_{001}^{,110},\,\mathsf{M}^{\mathsf{ST}},\right.\nonumber
	\\
	&\left.{\mathsf{M}^{\mathsf{ST}}}^{001;},\, {\mathsf{M}^{\mathsf{ST}}}^{,100},\,{\mathsf{M}^{\mathsf{ST}}}^{,010},\,{\mathsf{M}^{\mathsf{ST}}}^{,110} \right\} \,.
\end{align}
Here, $\mathsf{1}$, $\mathsf{P}$, $\mathsf{L}$, ${\mathsf{M}^{\mathsf{S}}}$, and ${\mathsf{M}^{\mathsf{T}}}$ denote the vacuum, particles, loops, sphere-like membranes, and torus-like membranes respectively.  The superscripts and subscripts are defined as  follows:
\begin{itemize}
\item For particles, the subscript ``$n_1n_2n_3$'' denotes $n_1$, $n_2$, and $n_3$  gauge charges minimally coupled to $A^{1}$, $A^{2}$, and $A^{3}$, respectively. 
\item For loops, the superscript ``$m_1m_2m_3$'' denotes $m_1$, $m_2$, and $m_3$  gauge charges minimally coupled to $A^{1}$, $A^{2}$, and $A^{3}$, respectively.  The subscript ``$n_1n_2n_3,\tilde{n}_1\tilde{n}_2\tilde{n}_3$'' of loops denotes (i) $n_1$, $n_2$, and $n_3$  gauge charges minimally coupled to $B^{1}$, $B^{2}$, and $B^{3}$, respectively, and (ii)   $\tilde{n}_1$, $\tilde{n}_2$, and $\tilde{n}_3$  gauge charges minimally coupled to $\tilde{B}^{1}$, $\tilde{B}^{2}$, and $\tilde{B}^{3}$, respectively. 
\item For the sake of convenience, we simply write ``$n_1n_2n_3,000$'' and ``$000,n_1n_2n_3$'' as ``$n_1n_2n_3,$'' and ``$,n_1n_2n_3$'' respectively. 
\item For sphere-like and torus-like membranes, the superscript ``$p_1p_2p_3;m_1m_2m_3,\tilde{m}_1\tilde{m}_2\tilde{m}_3$'' denotes (i) $p_1$, $p_2$, and $p_3$ gauge charges minimally coupled to $A^{1}$, $A^{2}$, and $A^{3}$, respectively, (ii) $m_1$, $m_2$, and $m_3$ gauge charges minimally coupled to $\tilde{B}^{1}$, $\tilde{B}^{2}$, and $\tilde{B}^{3}$, respectively, and (iii)  $\tilde{m}_1$, $\tilde{m}_2$, and $\tilde{m}_3$ gauge charges minimally coupled to $\tilde{B}^{1}$, $\tilde{B}^{2}$, and $\tilde{B}^{3}$, respectively. The subscript $n_1n_2n_3$ denotes $n_1$, $n_2$, and $n_3$ gauge charges minimally coupled to the 3-form gauge fields $C^{1}$, $C^{2}$, and $C^{3}$, respectively. 
\item For the sake of convenience, we simply express ``$p_1p_2p_3;000,000$'' and ``$000;000,\tilde{m}_1\tilde{m}_2\tilde{m}_3$'' as ``$p_1p_2p_3;$'' and ``$,\tilde{m}_1\tilde{m}_2\tilde{m}_3$'' respectively. 
\end{itemize}

${\mathsf{M}^{\mathsf{ST}}}^{n_1n_2n_3;m_1m_2m_3,p_1p_2p_3}$ can be generated by fusing a sphere-like ${\mathsf{M}^{\mathsf{S}}}_{001}$ membrane and a torus-like membrane ${\mathsf{M}^{\mathsf{T}}}_{001}^{n_1n_2n_3;m_1m_2m_3,p_1p_2p_3}$, i.e.,
\begin{align}
	{\mathsf{M}^{\mathsf{ST}}}^{n_1n_2n_3;m_1m_2m_3,p_1p_2p_3}={\mathsf{M}^{\mathsf{S}}}_{001}\otimes{\mathsf{M}^{\mathsf{T}}}_{001}^{n_1n_2n_3;m_1m_2m_3,p_1p_2p_3}\,.
\end{align}
The subsets $\Phi _{1}^{5}$ and $\Phi _{2}^{5}$ are given by
\begin{align}
	&\Phi _{1}^{5}=\left\{ 1,\,\mathsf{P}_{001},\,\mathsf{L}_{100,},\,\mathsf{L}_{010,},\,\mathsf{L}_{110,},\,\mathsf{L}_{100,}^{001},\,\mathsf{L}_{010,}^{001},\,\mathsf{L}_{110,}^{001} \right\} 
	\\
	&\Phi _{2}^{5}=\left\{ 1,\,\mathsf{P}_{001} \right\} \,.
\end{align}
By means of these definitions of sets, we will construct diagrammatic representations of higher-dimensional topological orders in section~\ref{s3} and section~\ref{s4}.

\section{Diagrammatics of $4$D topological orders}\label{s3} 

In this section, inspired by several results from TQFT, we construct diagrammatic representations of $4$D topological orders that can consistently describe fusion and shrinking rules, especially focusing on unitary operators during the processes of fusion and  shrinking.  We obtain pentagon equations and  shrinking-fusion hexagon equations for anomalous-free topological orders in $4$D, which enforces constraints on legitimate unitary operators. We will generalize our construction to $5$D in section~\ref{s4}, in which the existence of membrane excitations makes   hierarchical shrinking processes possible. 

\subsection{Fusion diagrams: fusion space,  $F$-symbols, and pentagon relation}\label{s31}
\begin{figure}
	\centering
	\includegraphics[scale=0.6,keepaspectratio]{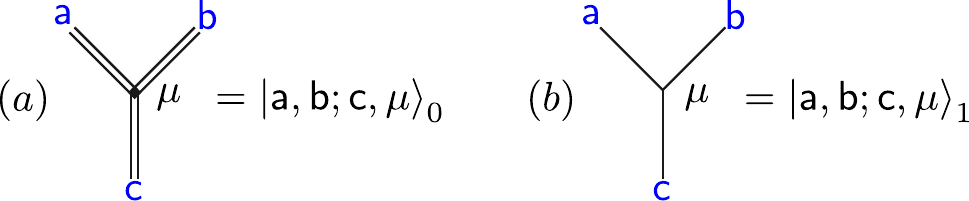}
	\caption{Elementary fusion diagrams. $\mathsf{a}$, $\mathsf{b}$ and $\mathsf{c}$ denote excitations and we highlight them in blue. $\mu=\left\{1,2,\cdots,N_{\mathsf{c}}^{\mathsf{a}\mathsf{b}}\right\}$ labels different fusion channels with the same output $\mathsf{c}$. (a) Double-lines mean that $\mathsf{a},\mathsf{b},\mathsf{c}\in \Phi _{0}^{4}$ and this diagram represents fusion for set $\Phi _{0}^{4}$. (b) Single-lines mean that $\mathsf{a},\mathsf{b},\mathsf{c}\in \Phi _{1}^{4}$ and this diagram represents fusion for set $\Phi _{1}^{4}$.}
	\label{fig_fusion4d}
\end{figure}

First, we explore the construction of fusion diagrams in $4$D. In the following discussion, Latin and Greek letters denote excitations and channels respectively. Suppose the fusion process $\mathsf{a}\otimes \mathsf{b}$ has fusion channels to $\mathsf{c}$, we can represent these fusion channels diagrammatically in figure~\ref{fig_fusion4d}. The solid lines can be understood as the spacetime trajectories of excitations. As mentioned in section~\ref{s222}, we can either fuse two excitations in the set $\Phi _{0}^{4}$ or two excitations in the subset $\Phi _{1}^{4}$. Hence, we use double-line and single-line to represent excitations from the sets $\Phi _{0}^{4}$ and $\Phi _{1}^{4}$ respectively.  $\mu=\left\{1,2,\cdots,N_{\mathsf{c}}^{\mathsf{a}\mathsf{b}}\right\}$ labels different fusion channels to $\mathsf{c}$. Notice that if $N_{\mathsf{c}}^{\mathsf{a}\mathsf{b}}=0$, we say diagrams in figure~\ref{fig_fusion4d} cannot happen. The left diagrams in figure~\ref{fig_fusion4d} can be defined as a vector $\ket{\mathsf{a},\mathsf{b};\mathsf{c},\mu}_0 $, where different $\mu$ represent orthogonal vectors. This set of vectors $\left\{\ket{\mathsf{a},\mathsf{b};\mathsf{c},\mu}_0,\mu=1,2,\cdots,N_{\mathsf{c}}^{\mathsf{a}\mathsf{b}}\right\} $ spans a fusion space ${V_0}_{\mathsf{c}}^{\mathsf{ab}}$ with $\text{dim}({V_0}_{\mathsf{c}}^{\mathsf{ab}})=N_{\mathsf{c}}^{\mathsf{a}\mathsf{b}}$. Similarly, we can define the right diagram in figure~\ref{fig_fusion4d} as a vector $\ket{\mathsf{a},\mathsf{b};\mathsf{c},\mu}_1 $, and the corresponding fusion space is ${V_1}_{\mathsf{c}}^{\mathsf{ab}}$. Since these two diagrams only differ by single-lines and double-lines, we only draw the fusion diagrams for the set $\Phi _{0}^{4}$ in the following discussion and omit the index ``$0$'' of vectors and spaces. \textit{One can straightforwardly obtain the diagrams for   set $\Phi _{1}^{4}$ by simply replacing double-lines with single-lines.} 

\begin{figure}
	\centering
	\includegraphics[scale=0.6,keepaspectratio]{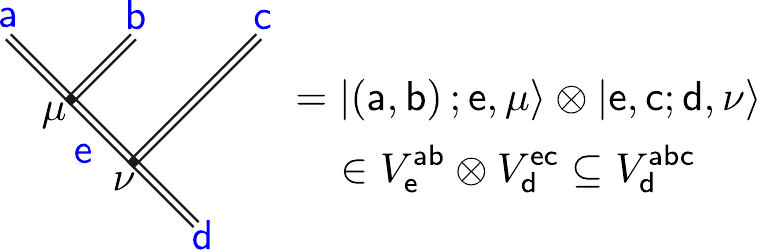}
	\caption{Fusion diagram of three excitations. $\mathsf{a}$, $\mathsf{b}$, and $\mathsf{c}$ are input excitations, ultimately  being fused  into $\mathsf{d}$. We highlight excitations in blue. The bracket ``$\left(\mathsf{a},\mathsf{b}\right)$'' is added to emphasize that  $\mathsf{a}$ and $\mathsf{b}$ fuse together first in the whole three-excitation fusion process. Note that, the bracket does not change the vector at all. $\otimes$ means   tensor product operation. Vectors in set $\left\{\ket{\left( \mathsf{a},\mathsf{b} \right) ;\mathsf{e},\mu} \otimes\ket{\mathsf{e},\mathsf{c};\mathsf{d},\nu}\right\} $ span the whole  space $V_{\mathsf{d}}^{\mathsf{abc}}$, in which different $\mu$, $\nu$ and $\mathsf{e}$ label different orthogonal vectors. The space $V_{\mathsf{d}}^{\mathsf{abc}}$ is isomorphic to $\oplus _{\mathsf{e}}(V_{\mathsf{e}}^{\mathsf{ab}}\otimes V_{\mathsf{d}}^{\mathsf{ec}})$, where $\oplus$ means direct sum. The dimension of $V_{\mathsf{d}}^{\mathsf{a}\mathsf{b}\mathsf{c}}$ is given by $\text{dim}V_{\mathsf{d}}^{\mathsf{a}\mathsf{b}\mathsf{c}}=\sum_{\mathsf{e}}{N_{\mathsf{e}}^{\mathsf{ab}}}N_{\mathsf{d}}^{\mathsf{ec}}$. Here we only draw the diagram for fusion process in the set $\Phi _{0}^{4}$ and simply write $\ket{\left( \mathsf{a},\mathsf{b} \right) ;\mathsf{e},\mu}_0$ and ${V_0}_{\mathsf{e}}^{\mathsf{ab}}$ as $\ket{\left( \mathsf{a},\mathsf{b} \right) ;\mathsf{e},\mu}$ and ${V}_{\mathsf{e}}^{\mathsf{ab}}$ respectively. One can obtain the diagram for subset $\Phi _{1}^{4}$ by simply replacing all double-lines with single-lines.
	}
	\label{fig_3fusion4d_1}
\end{figure}

Now we further consider diagrams that involve more excitations. Suppose the fusion process $\left(\mathsf{a}\otimes \mathsf{b}\right)\otimes \mathsf{c}$ has fusion channels to $\mathsf{d}$, diagrammatically we have figure~\ref{fig_3fusion4d_1}. Such a diagram can be constructed by stacking two fusion diagrams shown in figure~\ref{fig_fusion4d}. The diagram in figure~\ref{fig_3fusion4d_1} can also be defined as a vector $\ket{\left( \mathsf{a},\mathsf{b} \right) ;\mathsf{e},\mu} \otimes\ket{\mathsf{e},\mathsf{c};\mathsf{d},\nu} $, where $\otimes$ means tensor product. The corresponding space is denoted as $V_{\mathsf{d}}^{\mathsf{a}\mathsf{b}\mathsf{c}}$, whose dimension is $\text{dim}(V_{\mathsf{d}}^{\mathsf{a}\mathsf{b}\mathsf{c}})=\sum_{\mathsf{e}}{N_{\mathsf{e}}^{\mathsf{ab}}}N_{\mathsf{d}}^{\mathsf{ec}}$ because $V_{\mathsf{d}}^{\mathsf{a}\mathsf{b}\mathsf{c}}$ is isomorphic to $\oplus _{\mathsf{e}}V_{\mathsf{e}}^{\mathsf{ab}}\otimes V_{\mathsf{d}}^{\mathsf{ec}}$. 
\begin{figure}
	\centering
	\includegraphics[scale=0.6,keepaspectratio]{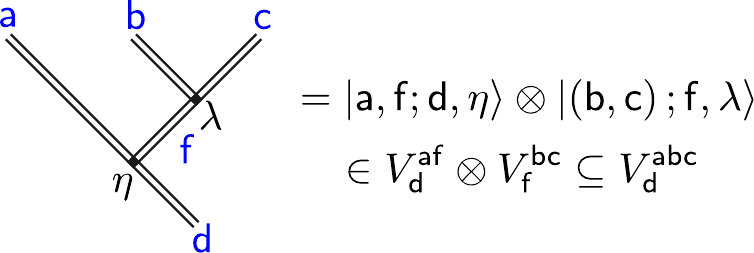}
	\caption{A different diagram of fusing three excitations. The associativity of fusion rules guarantees that this diagram actually describes the same physics as the diagram shown in figure~\ref{fig_3fusion4d_1}. The only difference between them is just a change of basis.}
	\label{fig_3fusion4d_2}
\end{figure}

\begin{figure}
	\centering
	\includegraphics[scale=0.6,keepaspectratio]{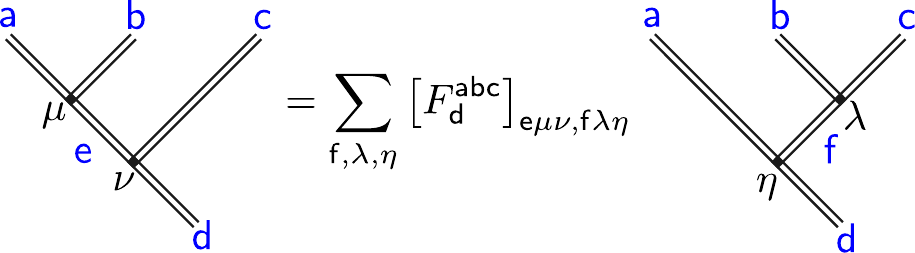}
	\caption{Definition of the $F$-symbol. The left and right diagrams describe the same physics in different bases. We use the unitary $F$-symbol to change the basis. $\mathsf{a}$, $\mathsf{b}$, and $\mathsf{c}$ denote the inputs. $\mathsf{d}$ denotes the output. $\mathsf{e}$, $\mu$, $\nu$, $\mathsf{f}$, $\lambda$, and $\eta$ denote different channels.}
	\label{fig_Fmatrix4d_2}
\end{figure}

\begin{figure}
	\centering
	\includegraphics[scale=0.6,keepaspectratio]{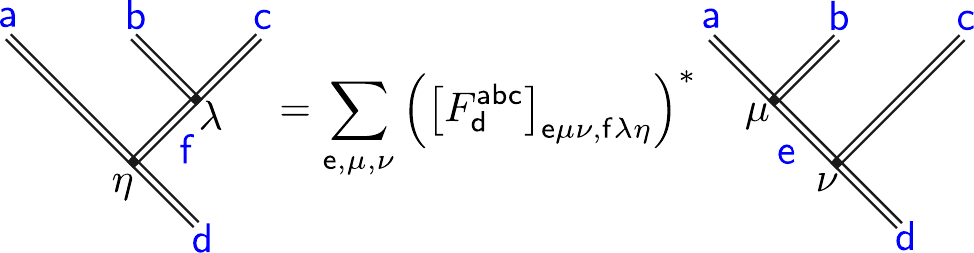}
	\caption{Inverse transformation. This figure comes from the unitarity of the  $F$-symbols. We use the star to denote complex conjugate.}
	\label{fig_Fmatrix4d_1}
\end{figure}

In section~\ref{s22}, we have discussed the crucial physical property of fusion rules known as associativity, i.e., $\left(\mathsf{a}\otimes\mathsf{b}\right)\otimes\mathsf{c}=\mathsf{a}\otimes\left(\mathsf{b}\otimes\mathsf{c}\right)$. This means that fusing $\mathsf{b}$ and $\mathsf{c}$ first should gives the same final output as fusing $\mathsf{a}$ and $\mathsf{b}$ first. Therefore, the diagram shown in figure~\ref{fig_3fusion4d_2} also represents a set of basis vectors in $V_{\mathsf{d}}^{\mathsf{abc}}$. Actually, figure~\ref{fig_3fusion4d_1} and figure~\ref{fig_3fusion4d_2} represent different bases and we can use a unitary matrix, known as the $F$-symbol, to change the basis. The definition of the $F$-symbol is given by figure~\ref{fig_Fmatrix4d_2}. We can express the equation in figure~\ref{fig_Fmatrix4d_2} as:
\begin{align}
	&\ket{\left( \mathsf{a},\mathsf{b} \right) ;\mathsf{e},\mu} \otimes \ket{\mathsf{e},\mathsf{c};\mathsf{d},\nu} =\sum_{\mathsf{f},\lambda ,\eta}{\left[ F_{\mathsf{d}}^{\mathsf{abc}} \right] _{\mathsf{e}\mu \nu ,\mathsf{f}\lambda \eta}}\ket{\mathsf{a},\mathsf{f};\mathsf{d},\eta} \otimes \ket{\left( \mathsf{b},\mathsf{c} \right) ;\mathsf{f},\lambda}\,. 
\end{align}
The summation over $\mathsf{f}$ exhausts all excitations in the system, $\lambda =\left\{ 1,2,\cdots ,N_{\mathsf{f}}^{\mathsf{bc}} \right\} $ and $\eta=\left\{ 1,2,\cdots ,N_{\mathsf{d}}^{\mathsf{af}} \right\}$. Since the $F$-symbol is unitary, we have
\begin{gather}
	\sum_{\mathsf{f},\lambda, \eta}{\left[ F_{\mathsf{d}}^{\mathsf{abc}} \right] _{\mathsf{e}\mu \nu ,\mathsf{f}\lambda \eta}}\left( \left[ F_{\mathsf{d}}^{\mathsf{abc}} \right] _{\mathsf{e}^{\prime}\mu^{\prime}\nu ^{\prime},\mathsf{f}\lambda \eta} \right) ^{\ast}=\delta _{\mathsf{ee}^{\prime}}\delta _{\mu \mu ^{\prime}}\delta _{\nu \nu ^{\prime}}\,
\end{gather}
which implies figure~\ref{fig_Fmatrix4d_1}. We can also derive a constraint on fusion coefficients by comparing the dimension of $V_{\mathsf{d}}^{\mathsf{a}\mathsf{b}\mathsf{c}}$ calculated from figure~\ref{fig_3fusion4d_1} and figure~\ref{fig_3fusion4d_2}, where the total space $V_{\mathsf{d}}^{\mathsf{abc}}$ is constructed by $\oplus _{\mathsf{e}}V_{\mathsf{e}}^{\mathsf{ab}}\otimes V_{\mathsf{d}}^{\mathsf{ec}}$ and $\oplus _{\mathsf{f}}V_{\mathsf{d}}^{\mathsf{af}}\otimes V_{\mathsf{f}}^{\mathsf{bc}}$ respectively. Thus
\begin{align}
	\text{dim}(V_{\mathsf{d}}^{\mathsf{a}\mathsf{b}\mathsf{c}})=\sum_{\mathsf{e}}{N_{\mathsf{e}}^{\mathsf{ab}}}N_{\mathsf{d}}^{\mathsf{ec}}=\sum_{\mathsf{f}}{N_{\mathsf{d}}^{\mathsf{af}}}N_{\mathsf{f}}^{\mathsf{bc}}\,.
\end{align}
This formula can also be derived by directly using associativity and eq.~(\ref{eq_general_fusion}):
\begin{gather}
	\left( \mathsf{a}\otimes \mathsf{b} \right) \otimes \mathsf{c}=\left( \oplus _{\mathsf{e}}N_{\mathsf{e}}^{\mathsf{ab}}\mathsf{e} \right) \otimes \mathsf{c}=\oplus _{\mathsf{ed}}N_{\mathsf{e}}^{\mathsf{ab}}N_{\mathsf{d}}^{\mathsf{ec}}\mathsf{d}\,,
	\\
	\mathsf{a}\otimes \left( \mathsf{b}\otimes \mathsf{c} \right) =\mathsf{a}\otimes \left( \oplus _{\mathsf{f}}N_{\mathsf{f}}^{\mathsf{bc}}\mathsf{f} \right) =\oplus _{\mathsf{fd}}N_{\mathsf{f}}^{\mathsf{bc}}N_{\mathsf{d}}^{\mathsf{af}}\mathsf{d}\,.
\end{gather}
Comparing the fusion coefficients, we have:
\begin{align}
	\sum_{\mathsf{e}}{N_{\mathsf{e}}^{\mathsf{ab}}}N_{\mathsf{d}}^{\mathsf{ec}}=\sum_{\mathsf{f}}{N_{\mathsf{d}}^{\mathsf{af}}}N_{\mathsf{f}}^{\mathsf{bc}}\,.
\end{align}


\begin{figure*}
	\centering
	\includegraphics[scale=0.6,keepaspectratio]{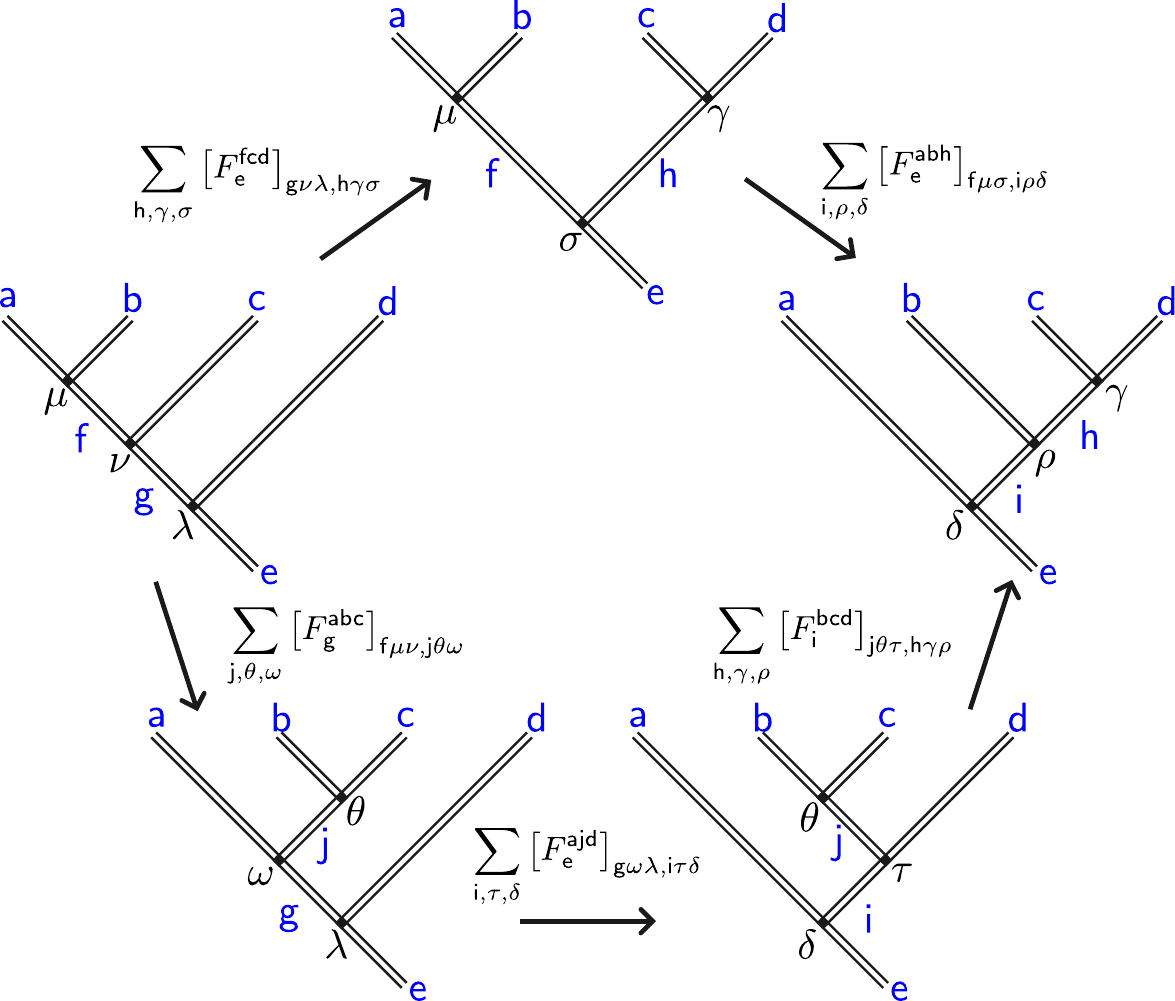}
	\caption{Diagrammatic representation of the pentagon equation (\ref{3eq_pentagon}). We highlight the excitations in blue. Starting from the far left diagram, we can go to the far right diagram through either the upper path or the lower path. Comparing these two different paths, we can derive the pentagon equation.}
	\label{fig_pentagon4d}
\end{figure*}

For other diagrams involving more excitations, we employ a similar tensor product approach to write down the corresponding vectors. Also, the $F$-symbols can be applied inside more complicated diagrams. For instance, consider fusing four excitations, we have different ways to transform one diagram into another. As shown in figure~\ref{fig_pentagon4d}, we have two paths to transform the far left diagram into the far right diagram, imposing a very strong constraint on the $F$-symbols, known as the pentagon equation:
\begin{align}
	&\sum_{\sigma =1}^{N_{\mathsf{e}}^{\mathsf{fh}}}{\left[ F_{\mathsf{e}}^{\mathsf{fcd}} \right] _{\mathsf{g}\nu \lambda ,\mathsf{h} \gamma \sigma}}\left[ F_{\mathsf{e}}^{\mathsf{abh}} \right] _{\mathsf{f}\mu \sigma ,\mathsf{i} \rho \delta}=\sum_{\mathsf{j}}{\sum_{\omega =1}^{N_{\mathsf{g}}^{\mathsf{aj}}}{\sum_{\theta =1}^{N_{\mathsf{j}}^{\mathsf{bc}}}{\sum_{\tau =1}^{N_{\mathsf{i}}^{\mathsf{jd}}}{\left[ F_{\mathsf{g}}^{\mathsf{abc}} \right] _{\mathsf{f}\mu \nu ,\mathsf{j} \theta \omega}\left[ F_{\mathsf{e}}^{\mathsf{ajd}} \right] _{\mathsf{g}\omega \lambda ,\mathsf{i} \tau \delta}\left[ F_{\mathsf{i}}^{\mathsf{bcd}} \right] _{\mathsf{j}\theta \tau ,\mathsf{h} \gamma \rho}}}}}.
	\label{3eq_pentagon}
\end{align}
 Such pentagon equation also exists in diagrammatic representations of $3$D anyons. It has been proven that no more identities beyond the pentagon equation can be derived by drawing more complicated fusion diagrams in $3$D. Essentially, the pentagon equation in any dimension arises from the associativity of the fusion rules,  leading to the conclusion that no more identities can be derived from fusion diagrams in any dimension. As a side note, if we draw all previous fusion diagrams in a single-line fashion, our fusion diagrams then reduce to $3$D anyonic fusion diagrams. A brief review of $3$D anyon diagrams is shown in appendix~\ref{ap1}.

\subsection{Shrinking diagrams: shrinking space and $\Delta$-symbol\label{ss32}}
\begin{figure}
	\centering
	\includegraphics[scale=0.6,keepaspectratio]{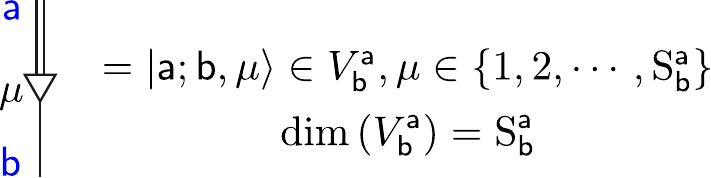}
	\caption{Shrinking diagram. According to the definition at the beginning of section~\ref{s31}, the double-line and single-line represent excitations from  $\Phi _{0}^{4}$ and $\Phi _{1}^{4}$ respectively. Here, $\mathsf{a}\in\Phi _{0}^{4}$ is the input of the shrinking process and thus $\mathsf{a}$ can be a loop or a particle. $\mathsf{b}\in\Phi _{1}^{4}$ is the output of the shrinking process and thus  $\mathsf{b}$ can only be a particle. We use a triangle to represent the shrinking process. This diagram describes shrinking $\mathsf{a}$ to $\mathsf{b}$ in the $\mu$ channel and we define it as a vector. The orthogonal set $\left\{ \ket{\mathsf{a};\mathsf{b},\mu} ,\mu =1,2,\cdots ,\mathrm{S}_{\mathsf{b}}^{\mathsf{a}} \right\}$ spans the shrinking space  $V_{\mathsf{b}}^{\mathsf{a}}$ with $\mathrm{dim}\left( V_{\mathsf{b}}^{\mathsf{a}} \right) =\mathrm{S}_{\mathsf{b}}^{\mathsf{a}}$.
	}
	\label{fig_shrinking}
\end{figure}

Suppose for $\mathsf{a}\in \Phi _{0}^{4}$, $\mathcal{S}\left(\mathsf{a}\right)$ has shrinking channels to $\mathsf{b}\in \Phi _{1}^{4}$, then we define the corresponding shrinking diagram in figure~\ref{fig_shrinking}. We use a triangle to represent the shrinking process.  Note that in a specific diagram, a double-line and a single-line may actually represent the same particle. However, they carry distinct meanings: the double-line means that this particle from $\Phi _{0}^{4}$ is treated as the input of the shrinking process while the single-line means that this particle from $\Phi _{1}^{4}$ is the output. Similar to the fusion diagram, the shrinking diagram in figure~\ref{fig_shrinking} can be understood as a vector $\ket{\mathsf{a};\mathsf{b},\mu}$. The orthogonal set $\left\{ \ket{\mathsf{a};\mathsf{b},\mu} ,\mu =1,2,\cdots ,\mathrm{S}_{\mathsf{b}}^{\mathsf{a}} \right\}$ spans a shrinking space $V_{\mathsf{b}}^{\mathsf{a}}$, whose dimension is $\mathrm{dim}\left( V_{\mathsf{b}}^{\mathsf{a}} \right) =\mathrm{S}_{\mathsf{b}}^{\mathsf{a}}$. 

\begin{figure}
	\centering
	\includegraphics[scale=0.6,keepaspectratio]{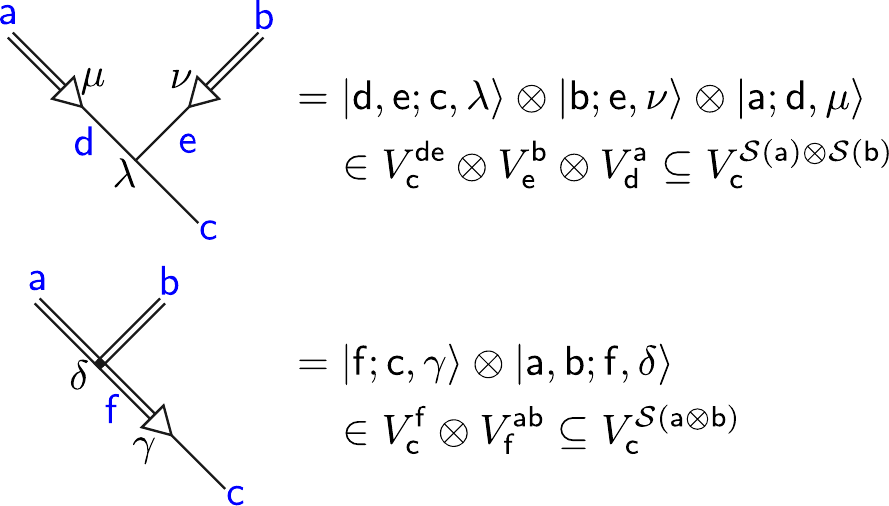}
	\caption{Elementary diagrams that incorporate both fusion and shrinking processes. The upper diagram describes $\mathcal{S} \left( \mathsf{a} \right) \otimes \mathcal{S} \left( \mathsf{b} \right) $, while the lower diagram describes $\mathcal{S} \left( \mathsf{a}\otimes \mathsf{b} \right) $. These diagrams can be constructed by tensoring smaller diagrams. Thus, we have $V_{\mathsf{c}}^{\mathcal{S} \left( \mathsf{a} \right) \otimes \mathcal{S} \left( \mathsf{b} \right)}\cong \oplus _{\mathsf{de}}V_{\mathsf{c}}^{\mathsf{de}}\otimes V_{\mathsf{e}}^{\mathsf{b}}\otimes V_{\mathsf{d}}^{\mathsf{a}}$ and $V_{\mathsf{c}}^{\mathcal{S} \left( \mathsf{a}\otimes \mathsf{b} \right)}\cong \oplus _{\mathsf{f}}V_{\mathsf{c}}^{\mathsf{f}}\otimes V_{\mathsf{f}}^{\mathsf{ab}}$. Here, $\otimes$ means tensor product, $\oplus$ means direct sum, and $\cong $ means isomorphism. We can verify that $\mathrm{dim}\left( V_{\mathsf{c}}^{\mathcal{S} \left( \mathsf{a} \right) \otimes \mathcal{S} \left( \mathsf{b} \right)} \right) =\sum_{\mathsf{de}}{N_{\mathsf{c}}^{\mathsf{de}}\mathrm{S}_{\mathsf{e}}^{\mathsf{b}}\mathrm{S}_{\mathsf{d}}^{\mathsf{a}}}$ and $\mathrm{dim}\left( V_{\mathsf{c}}^{\mathcal{S} \left( \mathsf{a}\otimes \mathsf{b} \right)} \right) =\sum_{\mathsf{f}}{\mathrm{S}_{\mathsf{c}}^{\mathsf{f}}N_{\mathsf{f}}^{\mathsf{ab}}}$. 
	}
	\label{fig_shrinking_fusion}
\end{figure}

Now we use tensor product to incorporate fusion and shrinking processes in a diagram. Consider $\mathcal{S} \left( \mathsf{a} \right) \otimes \mathcal{S} \left( \mathsf{b} \right) $ and $\mathcal{S} \left( \mathsf{a}\otimes \mathsf{b} \right) $, their corresponding diagrams are shown as the upper diagram and the lower diagram in figure~\ref{fig_shrinking_fusion} respectively. Recall eq.~(\ref{eq_consistency_relation_4D}), shrinking rules respect fusion rules. Thus, if $\mathcal{S} \left( \mathsf{a} \right) \otimes \mathcal{S} \left( \mathsf{b} \right) $ finally gives $\mathsf{c}$, then $\mathcal{S} \left( \mathsf{a}\otimes \mathsf{b} \right) $ must also give $\mathsf{c}$. Although these two processes have the same final output, they experience different intermediate steps. For the upper diagram in figure~\ref{fig_shrinking_fusion}, $\mathsf{a}$ and $\mathsf{b}$ shrink to $\mathsf{d}$ and $\mathsf{e}$ in the $\mu$ and $\nu$ channels respectively first, then $\mathsf{d}$ and $\mathsf{e}$ fuse to $\mathsf{c}$ in the $\lambda$ channel. In a non-Abelian case, different legitimate choices of $\mathsf{d}$, $\mathsf{e}$, $\mu$, $\nu$, and $\lambda$ may exist, i.e. they can finally produce $\mathsf{c}$. We define such diagram as $\ket{\mathsf{d},\mathsf{e};\mathsf{c},\lambda}\otimes\ket{\mathsf{b};\mathsf{e},\nu}\otimes\ket{\mathsf{a};\mathsf{d},\mu}$. Since $\mathsf{a}$, $\mathsf{b}$ and $\mathsf{c}$ are already fixed, different $\mathsf{d}$, $\mathsf{e}$, $\mu$, $\nu$, and $\lambda$ label different vectors. By exhausting all possible $\mathsf{d}$, $\mathsf{e}$, $\mu$, $\nu$, and $\lambda$, we obtain a set of orthogonal vectors and they span a space denoted as $V_{\mathsf{c}}^{\mathcal{S} \left( \mathsf{a} \right) \otimes \mathcal{S} \left( \mathsf{b} \right)}$, which is isomorphic to $\oplus _{\mathsf{de}}V_{\mathsf{c}}^{\mathsf{de}}\otimes V_{\mathsf{e}}^{\mathsf{b}}\otimes V_{\mathsf{d}}^{\mathsf{a}}$ due to our tensor product construction. For the lower diagram in figure~\ref{fig_shrinking_fusion}, $\mathsf{a}$ and $\mathsf{b}$ fuse to $\mathsf{f}$ in the $\delta$ channel first, then $\mathsf{f}$ shrinks to $\mathsf{c}$ in the $\gamma$ channel. We define such diagram as $\ket{\mathsf{f};\mathsf{c},\gamma}\otimes\ket{\mathsf{a},\mathsf{b};\mathsf{f},\delta}$. Similarly, $\mathsf{a}$, $\mathsf{b}$, and $\mathsf{c}$ are fixed. Different legitimate choices of $\mathsf{f}$, $\delta$, and $\gamma$ may exist, which label different orthogonal vectors. By exhausting all possible $\mathsf{f}$, $\delta$, and $\gamma$, we obtain a set of vectors spanning the space $V_{\mathsf{c}}^{\mathcal{S} \left( \mathsf{a}\otimes \mathsf{b} \right)}$, which is isomorphic to $\oplus _{\mathsf{f}}V_{\mathsf{c}}^{\mathsf{f}}\otimes V_{\mathsf{f}}^{\mathsf{ab}}$. The conclusion that shrinking rules respect fusion rules also implies that the space $V_{\mathsf{c}}^{\mathcal{S} \left( \mathsf{a} \right) \otimes \mathcal{S} \left( \mathsf{b} \right)}$ is isomorphic to $V_{\mathsf{c}}^{\mathcal{S} \left( \mathsf{a}\otimes \mathsf{b} \right)}$, leading to:
\begin{align}
	&\mathrm{dim}\left( V_{\mathsf{c}}^{\mathcal{S} \left( \mathsf{a} \right) \otimes \mathcal{S} \left( \mathsf{b} \right)} \right) =\mathrm{dim}\left( V_{\mathsf{c}}^{\mathcal{S} \left( \mathsf{a}\otimes \mathsf{b} \right)} \right)=\sum_{\mathsf{de}}{N_{\mathsf{c}}^{\mathsf{de}}\mathrm{S}_{\mathsf{e}}^{\mathsf{b}}\mathrm{S}_{\mathsf{d}}^{\mathsf{a}}}=\sum_{\mathsf{f}}{\mathrm{S}_{\mathsf{c}}^{\mathsf{f}}N_{\mathsf{f}}^{\mathsf{ab}}}\,.
	\label{eq1}
\end{align}
Eq.~(\ref{eq1}) can also be found in ref.~\cite{Zhang2023fusion} where all fusion and shrinking rules in the $BF$ theory with the $AAB$ twisted term are obtained. Also, eq.~(\ref{eq1}) is numerically verified. This constraint on fusion coefficients and shrinking coefficients can be considered as a consistency condition for fusion rules and shrinking rules in anomaly-free $4$D topological orders. 

\begin{figure}
	\centering
	\includegraphics[scale=0.6,keepaspectratio]{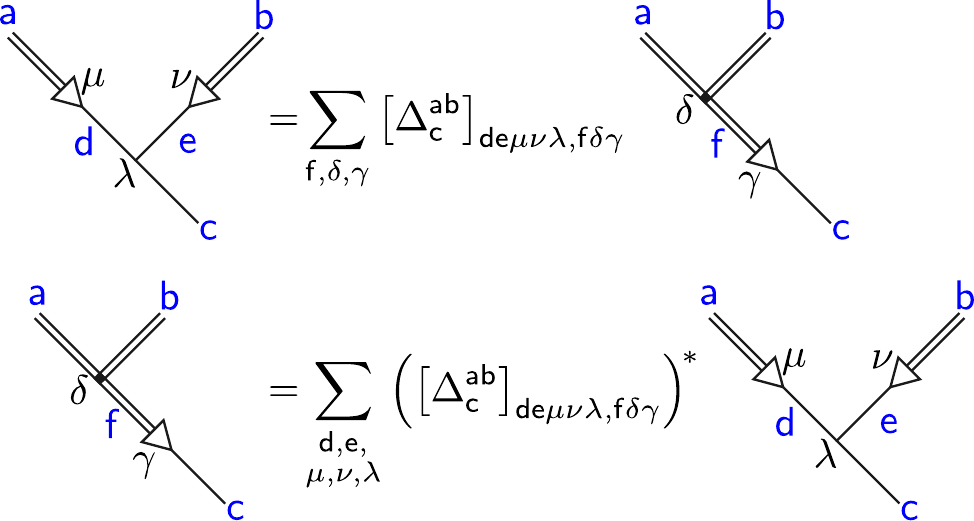}
	\caption{Definition of the $\Delta$-symbol. The left diagram and the right diagram describe the same physics in different bases. Similar to the $F$-symbol, the $\Delta$-symbol is also unitary and we can use it to change bases. $\mathsf{a}$ and $\mathsf{b}$ denote the inputs. $\mathsf{c}$ denotes the output. $\mathsf{d}$, $\mathsf{e}$, $\mu$, $\nu$, $\lambda$, $\mathsf{f}$, $\delta$, and $\gamma$ denote different shrinking channels.}
	\label{fig_delta_matrix}
\end{figure}

The two sets of vectors shown in figure~\ref{fig_shrinking_fusion} are   two different bases of the total space $V_{\mathsf{c}}^{\mathcal{S} \left( \mathsf{a} \right) \otimes \mathcal{S} \left( \mathsf{b} \right)}$ (or equivalently, we can write $V_{\mathsf{c}}^{\mathcal{S} \left( \mathsf{a}\otimes \mathsf{b} \right)}$ here). We expect that these two bases can transform to each other by a unitary matrix, called the $\Delta$-symbol. The definition of the $\Delta$-symbol is shown in figure~\ref{fig_delta_matrix}. We can explicitly write down the transformations in figure~\ref{fig_delta_matrix} as:
\begin{align}
	&\ket{\mathsf{d},\mathsf{e};\mathsf{c},\lambda}\otimes\ket{\mathsf{b};\mathsf{e},\nu}\otimes\ket{\mathsf{a};\mathsf{d},\mu}=\sum_{\mathsf{f},\delta ,\gamma}{\left[ \Delta _{\mathsf{c}}^{\mathsf{ab}} \right] _{\mathsf{de}\mu \nu \lambda ,\mathsf{f}\delta \gamma}\ket{\mathsf{f};\mathsf{c},\gamma}\otimes\ket{\mathsf{a},\mathsf{b};\mathsf{f},\delta}}\,,
	\\
	&\ket{\mathsf{f};\mathsf{c},\gamma}\otimes\ket{\mathsf{a},\mathsf{b};\mathsf{f},\delta}=\sum_{\substack{\mathsf{d},\mathsf{e},\\ \mu ,\nu ,\lambda}}{\left( \left[ \Delta _{\mathsf{c}}^{\mathsf{ab}} \right] _{\mathsf{de}\mu \nu \lambda ,\mathsf{f}\delta \gamma } \right) ^{\ast}\ket{\mathsf{d},\mathsf{e};\mathsf{c},\lambda}\otimes\ket{\mathsf{b};\mathsf{e},\nu}\otimes\ket{\mathsf{a};\mathsf{d},\mu}}\,.
\end{align}
Unitarity of the $\Delta$-symbol demands:
\begin{align}
	&\sum_{\mathsf{f},\delta ,\gamma}{\left[ \Delta _{\mathsf{c}}^{\mathsf{ab}} \right] _{\mathsf{de}\mu \nu \lambda ,\mathsf{f}\delta \gamma}\left( \left[ \Delta _{\mathsf{c}}^{\mathsf{ab}} \right] _{\mathsf{d}^{\prime}\mathsf{e}^{\prime}\mu^{\prime}\nu^{\prime}\lambda^{\prime},\mathsf{f}\delta \gamma} \right)}^{\ast}=\delta _{\mathsf{dd}^{\prime}}\delta _{\mathsf{ee}^{\prime}}\delta _{\mu \mu^{\prime}}\delta _{\nu \nu^{\prime}}\delta _{\lambda \lambda^{\prime}}\,,
	\\
	&\sum_{\substack{\mathsf{d},\mathsf{e},\\ \mu ,\nu ,\lambda}}{\left[ \Delta _{\mathsf{c}}^{\mathsf{ab}} \right] _{\mathsf{de}\mu \nu \lambda ,\mathsf{f}\delta \gamma}\left( \left[ \Delta _{\mathsf{c}}^{\mathsf{ab}} \right] _{\mathsf{de}\mu \nu \lambda ,\mathsf{f}^{\prime}\delta ^{\prime}\gamma ^{\prime}} \right) ^{\ast}}=\delta _{\mathsf{ff}^{\prime}}\delta _{\delta \delta ^{\prime}}\delta _{\gamma \gamma ^{\prime}}\,.
\end{align}
The element of the $\Delta$-symbol $\left[ \Delta _{\mathsf{c}}^{\mathsf{ab}} \right] _{\mathsf{de}\mu \nu \lambda ,\mathsf{f}\delta \gamma}$ is zero if the corresponding diagram is not allowed. Although we only use diagrams that involve shrinking and fusing two excitations to define the $\Delta$-symbol, we are allowed to act a $\Delta$-symbol on a part of a larger diagram, such as the diagram shown in figure~\ref{fig_shrinking_fusion_hexagon}.

\subsection{Shrinking-fusion hexagon equation}\label{ss33}
\begin{figure*}
	\centering
	\includegraphics[scale=0.57,keepaspectratio]{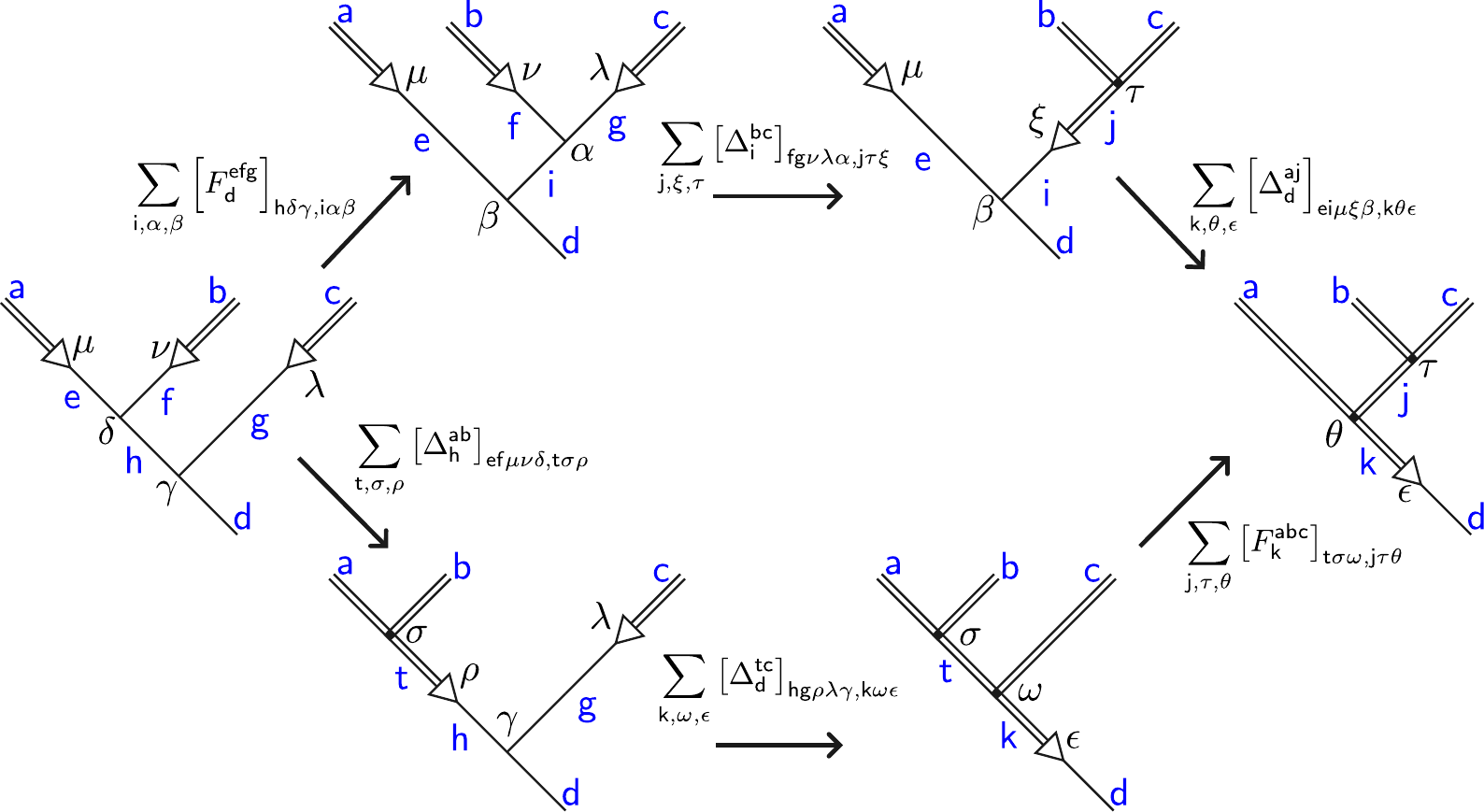}
	\caption{Diagrammatic representations of the shrinking-fusion hexagon equation (\ref{eq_shrinking_fusion_hexagon}). Two different paths can be constructed to transform the far left diagram to the far right diagram. Comparing these two different paths, we obtain the shrinking-fusion hexagon equation.}
	\label{fig_shrinking_fusion_hexagon}
\end{figure*}

In $4$D, fusion rules in both set $\Phi _{0}^{4}$ and subset $\Phi _{1}^{4}$ are closed and satisfy the associativity: $\left(\mathsf{a}\otimes\mathsf{b}\right)\otimes\mathsf{c}=\mathsf{a}\otimes\left(\mathsf{b}\otimes\mathsf{c}\right)$, which means we can use the $F$-symbols to change the bases when we consider diagrams involve fusing three excitations, regardless of which set do these excitations come from. We conclude that by applying the $F$-symbols and $\Delta$-symbols inside of diagrams involving three excitations, we can obtain another consistency relation called the shrinking-fusion hexagon equation besides the pentagon equation. As shown in figure~\ref{fig_shrinking_fusion_hexagon}, both the upper and lower paths can transform the far left diagram (i.e., $\left( \mathcal{S} \left( \mathsf{a} \right) \otimes \mathcal{S} \left( \mathsf{b} \right) \right) \otimes \mathcal{S} \left( \mathsf{c} \right) $) to the far right diagram (i.e., $\mathcal{S} \left( \mathsf{a}\otimes \left( \mathsf{b}\otimes \mathsf{c} \right) \right) $), these two paths correspond to:
	\begin{align}
		&\mathrm{far}\, \mathrm{left}=\sum_{\substack{\mathsf{i},\alpha ,\beta , \mathsf{j}, \\ \tau ,\xi ,  \mathsf{k},\theta ,\epsilon}}\!{\left[ F_{\mathsf{d}}^{\mathsf{efg}} \right] _{\mathsf{h}\delta \gamma ,\mathsf{i}\alpha \beta}\!\left[ \Delta _{\mathsf{i}}^{\mathsf{bc}} \right] _{\mathsf{fg}\nu \lambda \alpha ,\mathsf{j}\tau \xi}\!\left[ \Delta _{\mathsf{d}}^{\mathsf{aj}} \right] _{\mathsf{ei}\mu \xi \beta ,\mathsf{k}\theta \epsilon}\mathrm{far}\, \mathrm{right}}\,,\label{eq2}
		\\
		&\mathrm{far}\, \mathrm{left}=\sum_{\substack{\mathsf{t},\sigma ,\rho , \mathsf{k},\\ \omega ,\epsilon ,  \mathsf{j},\tau ,\theta}}\!{\left[ \Delta _{\mathsf{h}}^{\mathsf{ab}} \right] _{\mathsf{ef}\mu \nu \delta ,\mathsf{t}\sigma \rho}\!\left[ \Delta _{\mathsf{d}}^{\mathsf{tc}} \right] _{\mathsf{hg}\rho \lambda \gamma ,\mathsf{k}\omega \epsilon}\!\left[ F_{\mathsf{k}}^{\mathsf{abc}} \right] _{\mathsf{t}\sigma \omega ,\mathsf{j}\tau \theta}\mathrm{far}\, \mathrm{right}}\,,\label{eq3}
	\end{align}
respectively. Notice that the far right diagram is labeled by $\mathsf{j},\mathsf{k},\tau,
\theta$, and $\epsilon$. By comparing the coefficients of the far right diagrams in eq.~(\ref{eq2}) and eq.~(\ref{eq3}), we obtain a constraint on both the $F$-symbols and $\Delta$-symbols:
\begin{align}
	&\sum_{\mathsf{i}}{\sum_{\alpha =1}^{N_{\mathsf{i}}^{\mathsf{fg}}}{\sum_{\beta =1}^{N_{\mathsf{d}}^{\mathsf{ei}}}{\sum_{\xi =1}^{\mathrm{S}_{\mathsf{i}}^{\mathsf{j}}}{}}}}\left[ F_{\mathsf{d}}^{\mathsf{efg}} \right] _{\mathsf{h}\delta \gamma ,\mathsf{i}\alpha \beta}\left[ \Delta _{\mathsf{i}}^{\mathsf{bc}} \right] _{\mathsf{fg}\nu \lambda \alpha ,\mathsf{j}\tau \xi}\left[ \Delta _{\mathsf{d}}^{\mathsf{aj}} \right] _{\mathsf{ei}\mu \xi \beta ,\mathsf{k}\theta \epsilon}\nonumber
	\\
	=&\sum_{\mathsf{t}}{\sum_{\sigma =1}^{N_{\mathsf{t}}^{\mathsf{ab}}}{\sum_{\rho =1}^{\mathrm{S}_{\mathsf{h}}^{\mathsf{t}}}{\sum_{\omega =1}^{N_{\mathsf{k}}^{\mathsf{tc}}}{}}}}\left[ \Delta _{\mathsf{h}}^{\mathsf{ab}} \right] _{\mathsf{ef}\mu \nu \delta ,\mathsf{t}\sigma \rho}\left[ \Delta _{\mathsf{d}}^{\mathsf{tc}} \right] _{\mathsf{hg}\rho \lambda \gamma ,\mathsf{k}\omega \epsilon}\left[ F_{\mathsf{k}}^{\mathsf{abc}} \right] _{\mathsf{t}\sigma \omega ,\mathsf{j}\tau \theta}\,.
	\label{eq_shrinking_fusion_hexagon}
\end{align}
Diagrammatically represented by figure~\ref{fig_shrinking_fusion_hexagon}, this shrinking-fusion hexagon equation   can be understood as the consistency relation between the $F$-symbols and $\Delta$-symbols. Since being able to use the $\Delta$-symbols to change basis requires the consistency relation between fusion and shrinking coefficients (i.e., eq.~(\ref{eq1}) holds), we conclude that the shrinking-fusion hexagon equation actually implies eq.~(\ref{eq1}). Therefore, the shrinking-fusion hexagon equation is a stronger constraint on consistent fusion and shrinking rules, describing not only the behavior of fusion and shrinking coefficients but also the transformations of bases.  {We conjecture that such a shrinking-fusion hexagon is universal. All anomaly-free $4$D topological orders should have consistent fusion and shrinking rules, thus satisfying this equation.}

We can further consider applying the $F$-symbols and $\Delta$-symbols inside of diagrams involving more excitations, such as starting from the diagram for $\mathcal{S} \left( \mathsf{a} \right) \otimes \mathcal{S} \left( \mathsf{b} \right) \otimes \mathcal{S} \left( \mathsf{c} \right) \otimes \mathcal{S} \left( \mathsf{d} \right) $ and then implementing basis transformations. However, we can verify that no more independent equations can be obtained from the diagram involving four excitations. Details  
can be found in appendix~\ref{ap2}.  {We further conjecture that no more independent equations can be obtained from diagrams involving more excitations.}

\section{Diagrammatics of $5$D topological orders\label{s4}}
The idea of constructing the diagrammatic representations of $4$D topological orders can be generalized to $5$D, where nontrivial hierarchical shrinking rules may appear. An anomaly-free $5$D topological order should have consistent fusion and shrinking rules. If there exist nontrivial hierarchical shrinking rules, we further demand that such hierarchical shrinking rules are also consistent with fusion and shrinking rules, which leads to the hierarchical shrinking-fusion hexagon. 

\subsection{Fusion diagrams}\label{ss41}
The construction of fusion diagrams in section~\ref{s31} can be easily generalized to $5$D. We can simply replace double-lines with triple-lines in the previous fusion diagrams to represent fusing two excitations in the set $\Phi _{0}^{5}$ (see eq.~(\ref{set_5D_W0})). Double-lines and single-lines in $5$D represent excitations from the subset $\Phi _{1}^{5}$ and $\Phi _{2}^{5}$ respectively. The definitions of the $F$-symbols and the pentagon equation in section~\ref{s31} remain valid. Figure~\ref{fig_Fmatrix4d_2}, ~\ref{fig_Fmatrix4d_1}, and~\ref{fig_pentagon4d} can be drawn in triple-, double- and single-line fashions.
\begin{figure}
	\centering
	\includegraphics[scale=0.6,keepaspectratio]{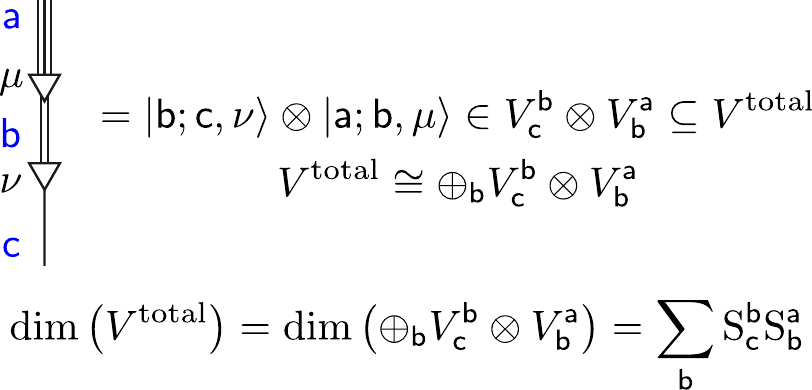}
	\caption{Hierarchical shrinking diagram. The triangle represents the shrinking process. The triple-line, double-line, and single-line represent excitations in the sets $\Phi _{0}^{5}$, $\Phi _{1}^{5}$, and $\Phi _{2}^{5}$ respectively. This diagram describes the process where the excitation $\mathsf{a}$ is shrunk to $\mathsf{b}$ in the $\mu$ channel first, followed by the further shrinking of $\mathsf{b}$ to $\mathsf{c}$ in the $\nu$ channel. The whole diagram can be considered as a vector in the total shrinking space, which is isomorphic to $\oplus _{\mathsf{b}}V_{\mathsf{c}}^{\mathsf{b}}\otimes V_{\mathsf{b}}^{\mathsf{a}}$. Different ways for shrinking $\mathsf{a}$ to $\mathsf{c}$ correspond to different $\mathsf{b}$, $\mu$, and $\nu$, thus correspond to different orthogonal vectors.}
	\label{fig_hierarchical_shrinking}
\end{figure}

\subsection{Hierarchical shrinking diagrams and $\Delta^2$-symbols}
Suppose $\mathsf{a}$ has nontrivial hierarchical shrinking rules and $\mathcal{S}^2\left(\mathsf{a}\right)$ gives $\mathsf{c}$ finally. We can define the corresponding hierarchical shrinking diagram in figure~\ref{fig_hierarchical_shrinking}. Here, the triangle denotes the shrinking process. The triple-line, double-line, and single-line represent excitations in the sets $\Phi _{0}^{5}$, $\Phi _{1}^{5}$, and $\Phi _{2}^{5}$ respectively. Similar to the case in $4$D, although these three kinds of lines may represent the same excitation in a specific diagram, they carry different meanings, implying different roles (input or output) in a hierarchical shrinking process. The diagram in figure~\ref{fig_hierarchical_shrinking} can be understood as a vector $\ket{\mathsf{b};\mathsf{c},\nu}\otimes\ket{\mathsf{a};\mathsf{b},\mu}$. Generally, there are various legitimate choices for $\mathsf{b}$, $\mu$, and $\nu$ that can finally give the desired output $\mathsf{c}$. Thus, different $\mathsf{b}$, $\mu$, and $\nu$ label different orthogonal vectors. The orthogonal set $\left\{ \ket{\mathsf{b};\mathsf{c},\nu}\otimes\ket{\mathsf{a};\mathsf{b},\mu} \right\}$ spans the total shrinking space $V^{\text{total}}$. This total shrinking space is isomorphic to $\oplus _{\mathsf{b}}V_{\mathsf{c}}^{\mathsf{b}}\otimes V_{\mathsf{b}}^{\mathsf{a}}$. Therefore, we can calculate its dimension as:
\begin{align}
	\mathrm{dim}\left( V^{\mathrm{total}} \right) =\mathrm{dim}\left( \oplus _{\mathsf{b}}V_{\mathsf{c}}^{\mathsf{b}}\otimes V_{\mathsf{b}}^{\mathsf{a}} \right) =\sum_{\mathsf{b}}{\mathrm{S}_{\mathsf{c}}^{\mathsf{b}}\mathrm{S}_{\mathsf{b}}^{\mathsf{a}}}\,.
\end{align}

\begin{figure}
	\centering
	\includegraphics[scale=0.6,keepaspectratio]{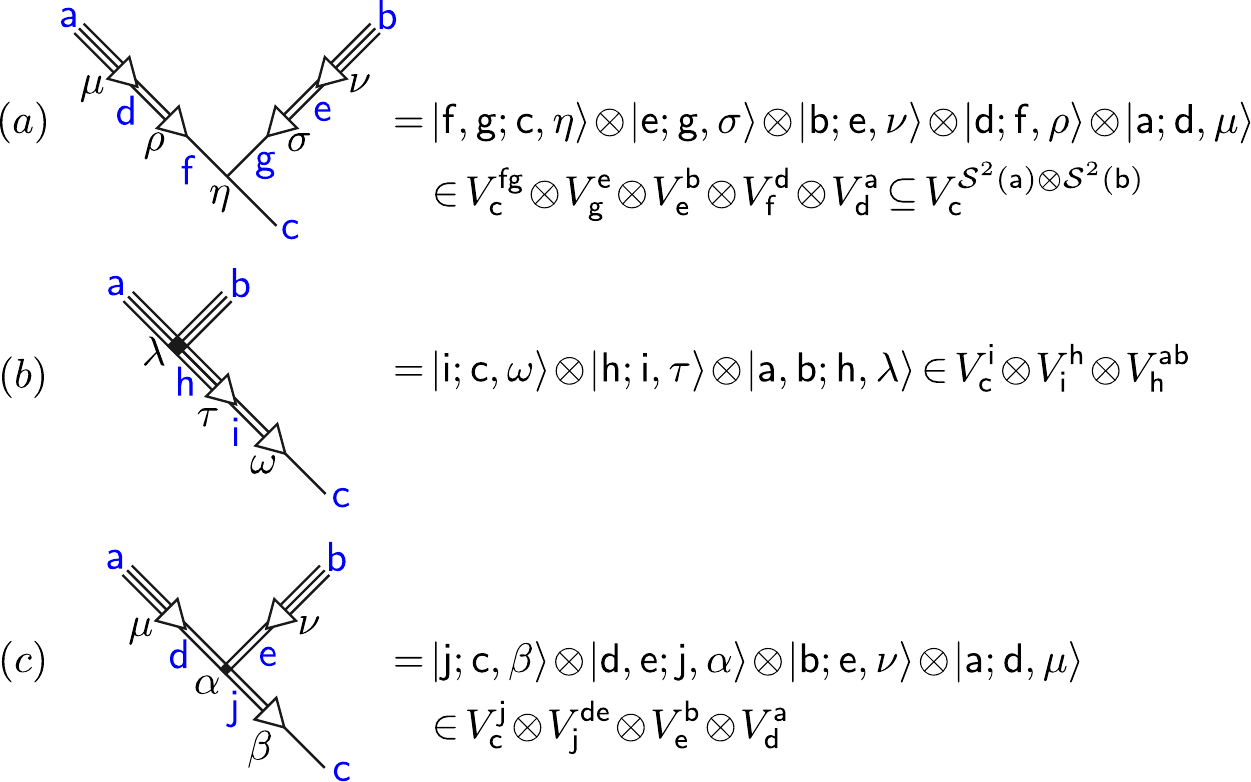}
	\caption{Diagrams for $\mathcal{S} ^2\left( \mathsf{a} \right) \otimes \mathcal{S} ^2\left( \mathsf{b} \right) $, $\mathcal{S} ^2\left( \mathsf{a}\otimes \mathsf{b} \right) $ and $\mathcal{S} \left( \mathcal{S} \left( \mathsf{a} \right) \otimes \mathcal{S} \left( \mathsf{b} \right) \right) $. These diagrams are understood as vectors. (a) Diagram for $\mathcal{S} ^2\left( \mathsf{a} \right) \otimes \mathcal{S} ^2\left( \mathsf{b} \right) $. Different $\mathsf{d}$, $\mathsf{e}$, $\mathsf{f}$, $\mathsf{g}$, $\mu$, $\nu$, $\rho$, $\sigma$, and $\eta$ label different orthogonal vectors in the whole space $V_{\mathsf{c}}^{\mathcal{S} ^2\left( \mathsf{a} \right) \otimes \mathcal{S} ^2\left( \mathsf{b} \right)}$, which is isomorphic to  $\oplus _{\mathsf{defg}}V_{\mathsf{c}}^{\mathsf{fg}}\otimes V_{\mathsf{g}}^{\mathsf{e}}\otimes V_{\mathsf{e}}^{\mathsf{b}}\otimes V_{\mathsf{f}}^{\mathsf{d}}\otimes V_{\mathsf{d}}^{\mathsf{a}}$. (b) Diagram for $\mathcal{S} ^2\left( \mathsf{a}\otimes \mathsf{b} \right) $. Different $\mathsf{h}$, $\mathsf{i}$, $\lambda$, $\tau$, and $\omega$ label different orthogonal vectors. The tensor product construction in this diagram indicates that the whole space $V_{\mathsf{c}}^{\mathcal{S} ^2\left( \mathsf{a} \right) \otimes \mathcal{S} ^2\left( \mathsf{b} \right)}$ is isomorphic to  $		\oplus _{\mathsf{hi}}V_{\mathsf{c}}^{\mathsf{i}}\otimes V_{\mathsf{i}}^{\mathsf{h}}\otimes V_{\mathsf{h}}^{\mathsf{ab}}$. (c) Diagram for $\mathcal{S} \left( \mathcal{S} \left( \mathsf{a} \right) \otimes \mathcal{S} \left( \mathsf{b} \right) \right) $. Different $\mathsf{d}$, $\mathsf{e}$, $\mathsf{j}$, $\mu$, $\nu$, $\alpha$, and $\beta$ label different orthogonal vectors. The tensor product construction in this diagram indicates that the whole space is isomorphic to  $\oplus _{\mathsf{dej}}V_{\mathsf{c}}^{\mathsf{j}}\otimes V_{\mathsf{j}}^{\mathsf{de}}\otimes V_{\mathsf{e}}^{\mathsf{b}}\otimes V_{\mathsf{d}}^{\mathsf{a}}$.
	}
	\label{fig_hie_shrinking_fusion}
\end{figure}

Having defined the hierarchical shrinking diagram, we can further consider representing $\mathcal{S} ^2\left( \mathsf{a} \right) \otimes \mathcal{S} ^2\left( \mathsf{b} \right) $, $\mathcal{S} ^2\left( \mathsf{a}\otimes \mathsf{b} \right) $ and $\mathcal{S} \left( \mathcal{S} \left( \mathsf{a} \right) \otimes \mathcal{S} \left( \mathsf{b} \right) \right) $ diagrammatically. As shown in figure~\ref{fig_hie_shrinking_fusion}, we stack the shrinking diagrams and fusion diagrams to describe the processes above. These diagrams are still understood as vectors and they can be constructed by tensor product. Since in $5$D, hierarchical shrinking rules respect fusion rules: $\mathcal{S} ^2\left( \mathsf{a} \right) \otimes \mathcal{S} ^2\left( \mathsf{b} \right) =\mathcal{S} ^2\left( \mathsf{a}\otimes \mathsf{b} \right)$, we can expect that the diagrams for  $\mathcal{S} ^2\left( \mathsf{a} \right) \otimes \mathcal{S} ^2\left( \mathsf{b} \right) $ and $\mathcal{S} ^2\left( \mathsf{a}\otimes \mathsf{b} \right) $ describe the same physics. Also, as mentioned in section~\ref{s223}, for particles and loops, the relation $\mathcal{S} \left( \mathsf{c} \right) \otimes \mathcal{S} \left( \mathsf{d} \right) =\mathcal{S} \left( \mathsf{c}\otimes \mathsf{d} \right)$ still holds. In $5$D, for any excitation $\mathsf{c}\in \Phi _{1}^{5}$, we can always find an $\mathsf{a}\in \Phi _{0}^{5}$ such that $\mathcal{S}\left(\mathsf{a}\right)$ has shrinking channels to $\mathsf{c}$. Thus, we can replace $\mathsf{c}$ and $ \mathsf{d} $ with $\mathcal{S}\left(\mathsf{a}\right)$ and $\mathcal{S}\left(\mathsf{b}\right)$ respectively and obtain $\mathcal{S} ^2\left( \mathsf{a} \right) \otimes \mathcal{S} ^2\left( \mathsf{b} \right) =\mathcal{S} \left( \mathcal{S} \left( \mathsf{a} \right) \otimes \mathcal{S} \left( \mathsf{b} \right) \right) $. Similarly, we expect that the diagrams for $\mathcal{S} \left( \mathcal{S} \left( \mathsf{a} \right) \otimes \mathcal{S} \left( \mathsf{b} \right) \right)$ describe the same physics as $\mathcal{S} ^2\left( \mathsf{a} \right) \otimes \mathcal{S} ^2\left( \mathsf{b} \right)$. Now we can say that the three diagrams shown in figure~\ref{fig_hie_shrinking_fusion} correspond to three different bases of the space $V_{\mathsf{c}}^{\mathcal{S} ^2\left( \mathsf{a} \right) \otimes \mathcal{S} ^2\left( \mathsf{b} \right)}$. We can calculate the dimension of $V_{\mathsf{c}}^{\mathcal{S} ^2\left( \mathsf{a} \right) \otimes \mathcal{S} ^2\left( \mathsf{b} \right)}$ from these three diagrams as:
	\begin{align}
		\left(\text{a}\right)\,&\mathrm{dim}\left( V_{\mathsf{c}}^{\mathcal{S} ^2\left( \mathsf{a} \right) \otimes \mathcal{S} ^2\left( \mathsf{b} \right)} \right) =\mathrm{dim}\left( \oplus _{\mathsf{defg}}V_{\mathsf{c}}^{\mathsf{fg}}\otimes V_{\mathsf{g}}^{\mathsf{e}}\otimes V_{\mathsf{e}}^{\mathsf{b}}\otimes V_{\mathsf{f}}^{\mathsf{d}}\otimes V_{\mathsf{d}}^{\mathsf{a}} \right)=\sum_{\mathsf{defg}}{N_{\mathsf{c}}^{\mathsf{fg}}\mathrm{S}_{\mathsf{g}}^{\mathsf{e}}\mathrm{S}_{\mathsf{e}}^{\mathsf{b}}}\mathrm{S}_{\mathsf{f}}^{\mathsf{d}}\mathrm{S}_{\mathsf{d}}^{\mathsf{a}}\,,
		\\
		\left(\text{b}\right)\,&\mathrm{dim}\left( V_{\mathsf{c}}^{\mathcal{S} ^2\left( \mathsf{a} \right) \otimes \mathcal{S} ^2\left( \mathsf{b} \right)} \right)=\mathrm{dim}\left( \oplus _{\mathsf{hi}}V_{\mathsf{c}}^{\mathsf{i}}\otimes V_{\mathsf{i}}^{\mathsf{h}}\otimes V_{\mathsf{h}}^{\mathsf{ab}} \right)=\sum_{\mathsf{hi}}{N_{\mathsf{h}}^{\mathsf{ab}}\mathrm{S}_{\mathsf{c}}^{\mathsf{i}}\mathrm{S}_{\mathsf{i}}^{\mathsf{h}}}\,,
		\\
		\left(\text{c}\right)\,&\mathrm{dim}\left( V_{\mathsf{c}}^{\mathcal{S} ^2\left( \mathsf{a} \right) \otimes \mathcal{S} ^2\left( \mathsf{b} \right)} \right)=\mathrm{dim}\left( \oplus _{\mathsf{dej}}V_{\mathsf{c}}^{\mathsf{j}}\otimes V_{\mathsf{j}}^{\mathsf{de}}\otimes V_{\mathsf{e}}^{\mathsf{b}}\otimes V_{\mathsf{d}}^{\mathsf{a}} \right)=\sum_{\mathsf{dej}}{N_{\mathsf{j}}^{\mathsf{de}}\mathrm{S}_{\mathsf{c}}^{\mathsf{j}}\mathrm{S}_{\mathsf{e}}^{\mathsf{b}}\mathrm{S}_{\mathsf{d}}^{\mathsf{a}}}.
		\end{align}
Thus we have:
\begin{align}
	\sum_{\mathsf{defg}}{N_{\mathsf{c}}^{\mathsf{fg}}\mathrm{S}_{\mathsf{g}}^{\mathsf{e}}\mathrm{S}_{\mathsf{e}}^{\mathsf{b}}}\mathrm{S}_{\mathsf{f}}^{\mathsf{d}}\mathrm{S}_{\mathsf{d}}^{\mathsf{a}}=\sum_{\mathsf{hi}}{N_{\mathsf{h}}^{\mathsf{ab}}\mathrm{S}_{\mathsf{c}}^{\mathsf{i}}\mathrm{S}_{\mathsf{i}}^{\mathsf{h}}}=\sum_{\mathsf{dej}}{N_{\mathsf{j}}^{\mathsf{de}}\mathrm{S}_{\mathsf{c}}^{\mathsf{j}}\mathrm{S}_{\mathsf{e}}^{\mathsf{b}}\mathrm{S}_{\mathsf{d}}^{\mathsf{a}}}\,.
	\label{eq4}
\end{align}
Eq.~(\ref{eq4}) is the consistent condition for fusion and shrinking coefficients in $5$D topological orders~\cite{Huang2023}.

\begin{figure}
	\centering
	\includegraphics[scale=0.6,keepaspectratio]{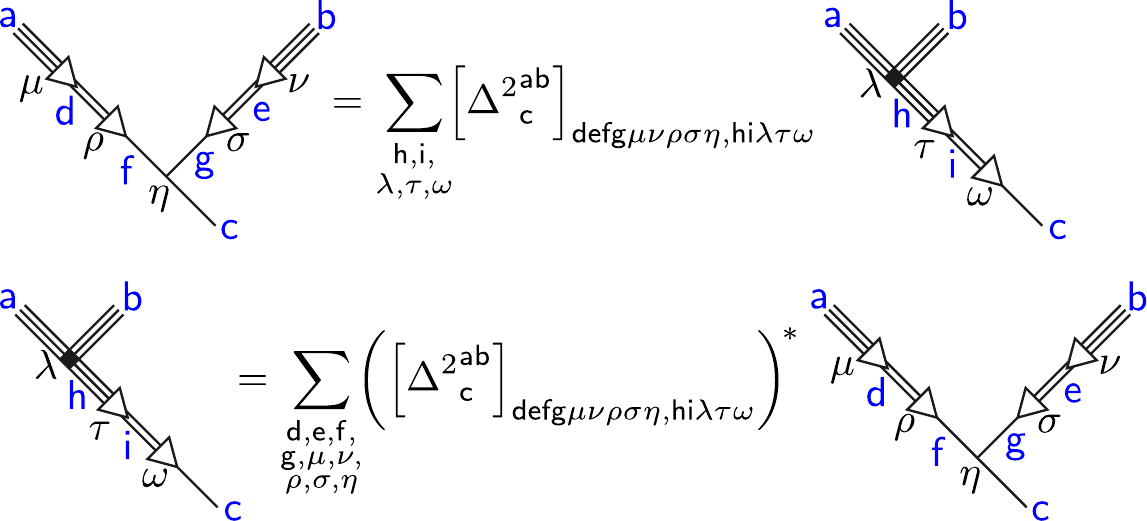}
	\caption{Definition of $\Delta^2$-symbols. The left and right diagrams describe the same physics in different bases. We can use a unitary $\Delta^2$-symbol to change bases.  ``$2$'' means that the diagrams involve hierarchical shrinking processes. 
	}
	\label{fig_delta2}
\end{figure}

Now we consider using unitary matrices to change bases. Since generally $\mathcal{S} \left( \mathsf{a} \right) \otimes \mathcal{S} \left( \mathsf{b} \right)=\mathcal{S} \left( \mathsf{a}\otimes \mathsf{b} \right) $ does not hold  for all excitations in $5$D, we need to introduce a new set of unitary matrices: the $\Delta^2$-symbols as shown in figure~\ref{fig_delta2}, to transform from the diagram for $\mathcal{S} ^2\left( \mathsf{a} \right) \otimes \mathcal{S} ^2\left( \mathsf{b} \right) $ to the diagram for $\mathcal{S} ^2\left( \mathsf{a}\otimes \mathsf{b} \right) $. Note that the $\Delta^2$-symbol is not the product of two $\Delta$-symbols.  The number  $2$ in $\Delta^2$  means that there are two successive shrinking processes. Generally, hierarchical shrinking rules and fusion rules in an $\mathcal{D}$-dimensional may satisfy $\mathcal{S} ^{\mathcal{D}-3}\left( \mathsf{a} \right) \otimes \mathcal{S} ^{\mathcal{D}-3}\left( \mathsf{b} \right) = \mathcal{S} ^{\mathcal{D}-3}\left( \mathsf{a}\otimes \mathsf{b} \right) $. In such a scenario, the symbol that transforms from the diagram for $\mathcal{S} ^{\mathcal{D}-3}\left( \mathsf{a} \right) \otimes \mathcal{S} ^{\mathcal{D}-3}\left( \mathsf{b} \right) $ to the diagram for $ \mathcal{S} ^{\mathcal{D}-3}\left( \mathsf{a}\otimes \mathsf{b} \right) $ is denoted as the $\Delta^{\mathcal{D}-3}$-symbol. In this convention, $\Delta^1$-symbol is nothing but $\Delta$-symbol. Unitarity of the $\Delta^2$-symbol demands:
	\begin{align}
		&\sum_{\substack{\mathsf{d},\mathsf{e},\mathsf{f}, \mathsf{g},\\ \mu ,\nu ,  \rho ,\sigma ,\eta}}\!\!{\left[ {\Delta ^2}_{\mathsf{c}}^{\mathsf{ab}} \right] _{\mathsf{defg}\mu \nu \rho \sigma \eta ,\mathsf{hi}\lambda \tau \omega}\left( \left[ {\Delta ^2}_{\mathsf{c}}^{\mathsf{ab}} \right] _{\mathsf{defg}\mu \nu \rho \sigma \eta ,\mathsf{h}^{\prime}\mathsf{i}^{\prime}\lambda^{\prime} \tau^{\prime} \omega^{\prime}} \right) ^{\!\!\ast}}=\delta _{\mathsf{hh}^{\prime}}\delta _{\mathsf{ii}^{\prime}}\delta _{\lambda \lambda^{\prime}}\delta _{\tau \tau^{\prime}}\delta _{\omega \omega^{\prime}}\,,
		\\
		&\sum_{\substack{\mathsf{h},\mathsf{i},\\ \lambda ,\tau ,\omega}}\!\!{\left[ {\Delta ^2}_{\mathsf{c}}^{\mathsf{ab}} \right] _{\mathsf{defg}\mu \nu \rho \sigma \eta ,\mathsf{hi}\lambda \tau \omega}\!\!\left( \left[ {\Delta ^2}_{\mathsf{c}}^{\mathsf{ab}} \right] _{\mathsf{d}^{\prime}\mathsf{e}^{\prime}\mathsf{f}^{\prime}\mathsf{g}^{\prime}\mu^{\prime} \nu^{\prime} \rho^{\prime} \sigma^{\prime} \eta^{\prime} ,\mathsf{hi}\lambda \tau \omega} \right) ^{\!\!\ast}}=\delta _{\mathsf{dd}^{\prime}}\delta _{\mathsf{ee}^{\prime}}\delta _{\mathsf{ff}^{\prime}}\delta _{\mathsf{gg}^{\prime}}\delta _{\mu \mu^{\prime}}\delta _{\nu \nu^{\prime}}\delta _{\rho \rho^{\prime}}\delta _{\sigma \sigma^{\prime}}\delta _{\eta \eta^{\prime}}\,.
	\end{align}

\begin{figure}
	\centering
	\includegraphics[scale=0.6,keepaspectratio]{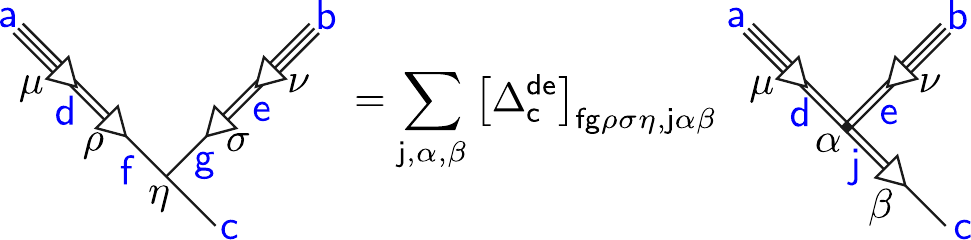}
	\caption{Basis transformation from $\mathcal{S} ^2\left( \mathsf{a} \right) \otimes \mathcal{S} ^2\left( \mathsf{b} \right) $ to $\mathcal{S} \left( \mathcal{S} \left( \mathsf{a} \right) \otimes \mathcal{S} \left( \mathsf{b} \right) \right) $. Since $\mathsf{d}$ and $\mathsf{e}$ can only be loops and particles, we are allowed to use the $\Delta$-symbol to change the basis. 
	}
	\label{fig_trans_1_3}
\end{figure}

\begin{figure}
	\centering
	\includegraphics[scale=0.6,keepaspectratio]{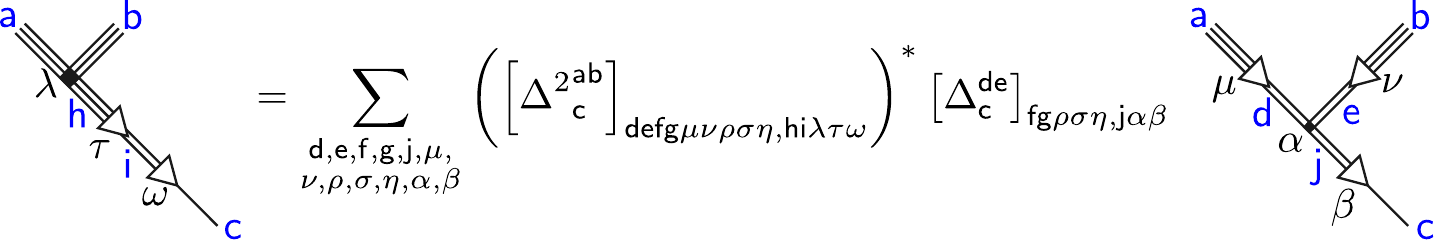}
	\caption{Basis transformation from $\mathcal{S} ^2\left( \mathsf{a}\otimes\mathsf{b} \right)  $ to $\mathcal{S} \left( \mathcal{S} \left( \mathsf{a} \right) \otimes \mathcal{S} \left( \mathsf{b} \right) \right) $. We can derive this transformation by using the lower diagram in figure~\ref{fig_delta2} and then use figure~\ref{fig_trans_1_3}.
	}
	\label{fig_trans_2_3}
\end{figure}

As for changing bases from $\mathcal{S} ^2\left( \mathsf{a} \right) \otimes \mathcal{S} ^2\left( \mathsf{b} \right) $ to $\mathcal{S} \left( \mathcal{S} \left( \mathsf{a} \right) \otimes \mathcal{S} \left( \mathsf{b} \right) \right) $, we  can still use the $\Delta$-symbols shown in figure~\ref{fig_delta_matrix} because even in $5$D, loops and particles still obey $\mathcal{S} \left( \mathsf{c} \right) \otimes \mathcal{S} \left( \mathsf{d} \right)=\mathcal{S} \left( \mathsf{c}\otimes \mathsf{d} \right) $. Thus, as shown in figure~\ref{fig_trans_1_3} and~\ref{fig_trans_2_3}, we establish the basis transformations between the three diagrams in figure~\ref{fig_hie_shrinking_fusion}.

\subsection{Diagrammatics of the hierarchical shrinking-fusion hexagon equation}
\begin{figure}
	\centering
	\includegraphics[scale=0.6,keepaspectratio]{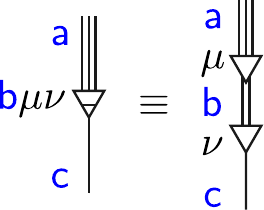}
	\caption{Hierarchical shrinking diagram in a more compact form. We introduce the diagram on the left hand side as the simplified hierarchical shrinking diagram. Physically, this simplified diagram retains all the essential information. The triangle with a line represents the complete hierarchical shrinking process.
	}
	\label{fig_simplify}
\end{figure}

Before investigating into diagrams where  three excitations together undergo shrinking and fusion processes, we present a more compact version of figure~\ref{fig_hierarchical_shrinking}, as shown in figure~\ref{fig_simplify}. We use a triangle decorated by a line inside the triangle, in order to represent the whole hierarchical shrinking process. If we only consider using the $\Delta^2$-symbols and $F$-symbols, it is convenient to use such a compact fashion. 
\begin{figure*}
	\centering
	\includegraphics[scale=0.55,keepaspectratio]{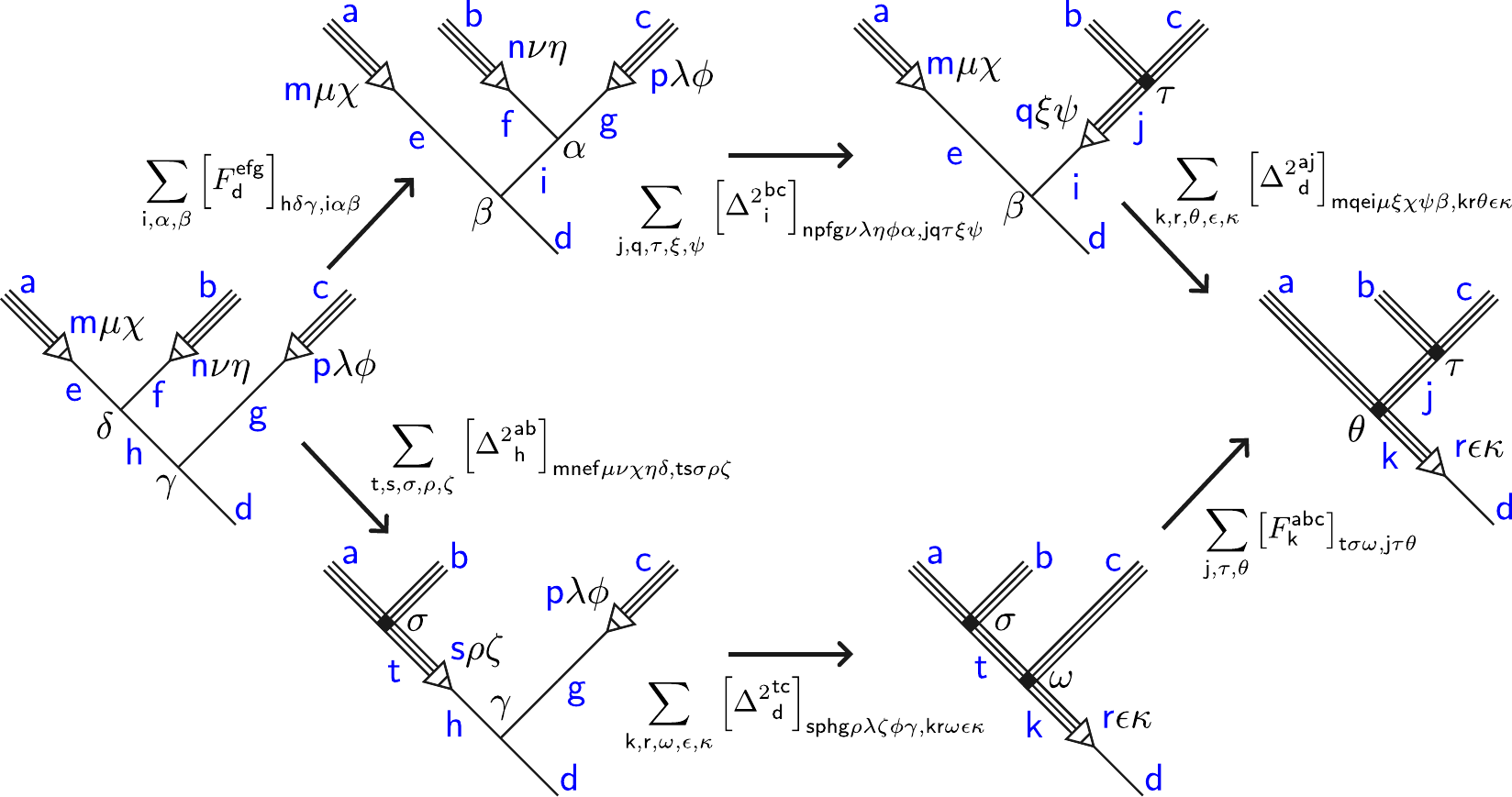}
	\caption{Diagrammatic representations of the hierarchical shrinking-fusion hexagon equation. Here we use compact hierarchical shrinking diagrams shown in figure~\ref{fig_simplify}. Similar to figure~\ref{fig_shrinking_fusion_hexagon}, we have two different paths to transform the far left diagram to the far right diagram. Comparing these two different paths, we obtained the hierarchical shrinking-fusion hexagon equation.
	}
	\label{fig_hexagon_5d}
\end{figure*}

Recall the figure~\ref{fig_shrinking_fusion_hexagon}, where three excitations undergo shrinking and fusion processes to finally get a particle. By using the $F$-symbols and $\Delta$-symbols, we transform diagrams from each other and finally derive the shrinking-fusion hexagon equation in $4$D. Now, we generalize it to $5$D by substituting the $\Delta$-symbols with the $\Delta^2$-symbols and employing the compact hierarchical shrinking diagrams shown in figure~\ref{fig_simplify}. The resulting diagram shown in figure~\ref{fig_hexagon_5d} gives the hierarchical shrinking-fusion hexagon equation in $5$D. The transformation from the far left diagram to the far right diagram differs along the upper and lower paths. By comparing the coefficients of the far right diagram, we derive the hierarchical shrinking-fusion hexagon equation:
	\begin{align}
		&\sum_{\mathsf{i},\mathsf{q}}{\sum_{\alpha =1}^{N_{\mathsf{i}}^{\mathsf{fg}}}{\sum_{\beta =1}^{N_{\mathsf{d}}^{\mathsf{ei}}}{\sum_{\xi =1}^{\mathrm{S}_{\mathsf{q}}^{\mathsf{j}}}{\sum_{\psi =1}^{\mathrm{S}_{\mathsf{i}}^{\mathsf{q}}}{}}}}}\left[ F_{\mathsf{d}}^{\mathsf{efg}} \right] _{\mathsf{h}\delta \gamma ,\mathsf{i}\alpha \beta}\!\left[ {\Delta ^2}_{\mathsf{i}}^{\mathsf{bc}} \right] _{\mathsf{npfg}\nu \lambda \eta \phi \alpha ,\mathsf{jq}\tau \xi \psi}\!\left[ {\Delta ^2}_{\mathsf{d}}^{\mathsf{aj}} \right] _{\mathsf{mqei}\mu \xi \chi \psi \beta ,\mathsf{kr}\theta \epsilon \kappa}\nonumber
		\\
		=\!&\sum_{\mathsf{t},\mathsf{s}}{\sum_{\sigma =1}^{N_{\mathsf{t}}^{\mathsf{ab}}}{\sum_{\omega =1}^{N_{\mathsf{k}}^{\mathsf{tc}}}{\sum_{\rho =1}^{\mathrm{S}_{\mathsf{s}}^{\mathsf{t}}}{\sum_{\zeta =1}^{\mathrm{S}_{\mathsf{h}}^{\mathsf{s}}}{}}}}}\left[ {\Delta ^2}_{\mathsf{h}}^{\mathsf{ab}} \right] _{\mathsf{mnef}\mu \nu \chi \eta \delta ,\mathsf{ts}\sigma \rho \zeta}\!\left[ {\Delta ^2}_{\mathsf{d}}^{\mathsf{tc}} \right] _{\mathsf{sphg}\rho \lambda \zeta \phi \gamma ,\mathsf{kr}\omega \epsilon \kappa}\!\left[ F_{\mathsf{k}}^{\mathsf{abc}} \right] _{\mathsf{t}\sigma \omega ,\mathsf{j}\tau \theta}.
		\label{eq_hie_shrinking_fusion_hexagon}
	\end{align}
Eq.~(\ref{eq_hie_shrinking_fusion_hexagon}) can be understood as the consistency relation between the $F$-symbols and $\Delta^2$-symbols.

If we do not utilize the compact diagram in figure~\ref{fig_simplify}, we can consider using the $\Delta$-symbols in figure~\ref{fig_hexagon_5d}. In this case, we will find that shrinking-fusion hexagon~(\ref{eq_shrinking_fusion_hexagon}) still holds. However, it is noteworthy that now $\mathsf{a}$, $\mathsf{b}$ and $\mathsf{c}$ in eq.~(\ref{eq_shrinking_fusion_hexagon}) can only belong to the subset $\Phi _{1}^{5}$. Such shrinking-fusion hexagon equation with a constraint on inputs is the consistency relation between the $F$-symbols and $\Delta$-symbols in $5$D.

Eq.~(\ref{eq_hie_shrinking_fusion_hexagon}) and eq.~(\ref{eq_shrinking_fusion_hexagon}) with constrained inputs are the key results in our $5$D diagrammatic representations. They establish the consistency relation between the $F$-symbols, $\Delta$-symbols, and $\Delta^2$-symbols. They also imply  constraints on fusion and shrinking coefficients, i.e., eq.~(\ref{eq4}).  {We conjecture that all anomaly-free $5$D topological orders should satisfy these two hexagon equations, ensuring consistent fusion, shrinking, and hierarchical shrinking rules.}

\section{Summary and outlook}\label{s5}
In this paper, with the help of our previous work on field-theoretical description of higher dimensional topological orders,  we have successfully constructed diagrammatic representations for higher-dimensional topological orders in 4D and 5D spacetime, extending the framework known for topologically ordered phases in 3D spacetime. By introducing and manipulating basic fusion and shrinking diagrams as vectors within corresponding spaces, we have demonstrated the construction of complex diagrams through their stacking. Using the $F$-, $\Delta$-, and $\Delta^2$-symbols, we have established unitary transformations between different bases in these vector spaces and discovered key consistency conditions, including the pentagon equations and (hierarchical) shrinking-fusion hexagon equations.
Our findings indicate that these consistency conditions are essential for ensuring the anomaly-free nature of higher-dimensional topological orders. Violations of these conditions suggest the presence of quantum anomalies, providing a diagnostic tool for identifying such anomalies in theoretical models.

Looking ahead, our work points towards several promising avenues for future research:
 \begin{enumerate}	 
	\item

In the literature (see references cited in section~\ref{s1}), braiding phases, fusion coefficients, shrinking coefficients, and quantum dimensions have been determined for a concrete topological order. It is also important to practically determine the unitary matrices (i.e., $F$-, $\Delta$-, and $\Delta^2$-symbols) introduced in this paper by solving pentagon equations and (hierarchical) shrinking-fusion hexagon equations. Solving these equations in the present work is expected to be very tedious (practical calculations in 3D can be found in references such as ref.~\cite{Wen2016}). While it is beyond the scope of the current paper, solving these equations undoubtedly represents a key step toward a more complete description of higher-dimensional topological orders. Recently, Quantinuum's H2 trapped-ion quantum processor~\cite{dreyer2024ionNature} successfully realized a non-Abelian topological order in 3D spacetime and demonstrated control of its anyons. This topological order can be formally regarded as the lower-dimensional counterpart of the Borromean rings topological order proposed in ref.~\cite{PhysRevLett.121.061601}. Efforts toward the experimental realization of 4D topological orders and control of loop excitations through the platform of trapped-ion quantum processors will provide experimental evidence for theoretical predictions of the topological data of loop-like topological excitations.

\item In this paper, we solely focus on diagrams pertaining to fusion and shrinking rules, although it is worth noting that braiding statistics  also hold significance in comprehending higher-dimensional topological orders. More concretely, we have defined $F$-, $\Delta$-, and  $\Delta^2$-symbols to perform unitary transformations associated with   fusion and shrinking processes; however, we have not   studied  yet the higher-dimensional spacetime counterpart of $R$-symbol in 3D. While diagrammatic representation of braiding processes is definitely intricate due to the involvement of spatially extended excitations, incorporating braiding processes into our diagrammatic representations presents an intriguing avenue for future exploration. Doing so may unveil new diagrammatic rules capable of encoding more complete algebraic structures relevant to braiding, fusion, and shrinking processes. Based on all data depicted by field theory and consistency conditions in diagrammatic representations, it is further interesting to study the connection between our field-theory-inspired diagrammatic representations and categorical approach in the future.

  \item $BF$ theory exhibits a close relationship with non-invertible symmetry and symmetry topological field theory (SymTFT). Various twisted terms, along with their fusion rules and braiding statistics, have been   involved in the context of SymTFT, including notable examples such as $BB$, $BBA$, and $AdAdA$ twisted terms, see, e.g., refs.~\cite{2023JHEP10053K,antinucci2024anomalies,brennan2024symtft,argurio2024symmetry}. These topological terms were previously a focus of investigation in the field-theoretical description of higher-dimensional topological orders~\cite{Zhang2023Continuum,zhang_topological_2022,Huang2023,PhysRevB.99.235137}. It would be intriguing to explore the incorporation of shrinking and hierarchical shrinking rules~\cite{Zhang2023fusion,Huang2023} into SymTFT and establish connections between SymTFT and our diagrammatic representations. Such investigations hold the potential to enrich our understanding of both SymTFT and the broader landscape of topological field theories.
    
  \item   In this paper, our focus is primarily on the diagrammatic representations of $4$D and $5$D topological orders. However, it is worth noting that the framework we have developed lends itself to potential generalization for arbitrarily higher-dimensional topological orders. Such generalization holds promise for providing deeper insights into the underlying structure and properties of higher-dimensional topological orders, warranting further exploration in future studies.
 \end{enumerate}

\acknowledgements
 P.Y. thanks Z.-C. Gu for the warm hospitality during the visit to the Chinese University of Hong Kong, where part of this work was conducted.
This work was partially supported by the National Natural Science Foundation of China (NSFC) under Grant Nos. 12474149 and 12074438. The calculations reported were performed on resources provided by the Guangdong Provincial Key Laboratory of Magnetoelectric Physics and Devices (No. 2022B1212010008).

\appendix
\section{Review of anyon diagrams}\label{ap1}
In this appendix, we will briefly review the basic concepts in the diagrammatic representations of anyons. The diagrammatic rules and algebra set up the structure of anyon theories and they are closely related to TQFT descriptions.

\subsection{Fusion rules and diagrammatic representations}\label{ss21}
Since shrinking rules are trivial in $3$D, we do not need to put excitations in different sets and treat them differently. Consequently, we can easily recover the diagrammatic representations for anyons by drawing all fusion diagrams in a single-line fashion.

\begin{figure}
	\centering
	\includegraphics[scale=0.6,keepaspectratio]{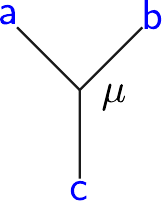}
	\caption{Fusion diagram of two anyons.  $\mathsf{a}$, $\mathsf{b}$ and $\mathsf{c}$ denote anyons, $\mu=\left\{1,2,\cdots,N_{\mathsf{c}}^{\mathsf{a}\mathsf{b}}\right\}$ labels different fusion channels to $\mathsf{c}$.}
	\label{fig_fusion}
\end{figure}

\begin{figure}
	\centering
	\includegraphics[scale=0.6,keepaspectratio]{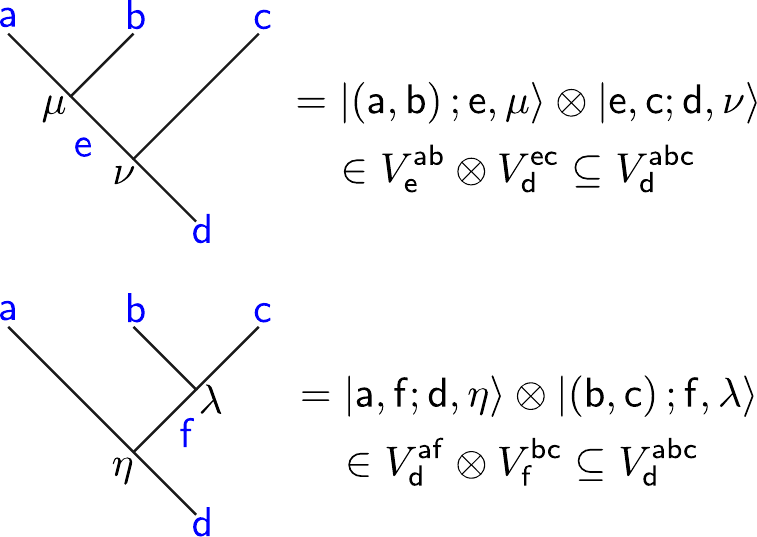}
	\caption{Fusion diagram of three anyons. $\mathsf{a}$, $\mathsf{b}$, and $\mathsf{c}$ are input anyons and they finally fuse to $\mathsf{d}$. The anyons in the bracket fuse together first. $\otimes$ mean tensor product here. In the upper diagram, the set $\left\{\ket{\left( \mathsf{a},\mathsf{b} \right) ;\mathsf{e},\mu} \otimes\ket{\mathsf{e},\mathsf{c};\mathsf{d},\nu}\right\} $ spans the whole  space $V_{\mathsf{d}}^{\mathsf{abc}}$, where different $\mu$, $\nu$ and $\mathsf{e}$ label different orthogonal vectors. Such space $V_{\mathsf{d}}^{\mathsf{abc}}$ is isomorphic to $\oplus _{\mathsf{e}}V_{\mathsf{e}}^{\mathsf{ab}}\otimes V_{\mathsf{d}}^{\mathsf{ec}}$ and  thus $\text{dim}(V_{\mathsf{d}}^{\mathsf{a}\mathsf{b}\mathsf{c}})=\sum_{\mathsf{e}}{N_{\mathsf{e}}^{\mathsf{ab}}}N_{\mathsf{d}}^{\mathsf{ec}}$. For the lower diagram, the case is similar and we have $V_{\mathsf{d}}^{\mathsf{abc}}\cong \oplus _{\mathsf{f}}V_{\mathsf{d}}^{\mathsf{af}}\otimes V_{\mathsf{f}}^{\mathsf{bc}}$, which indicates $\text{dim}(V_{\mathsf{d}}^{\mathsf{a}\mathsf{b}\mathsf{c}})=\sum_{\mathsf{f}}{N_{\mathsf{d}}^{\mathsf{af}}}N_{\mathsf{f}}^{\mathsf{bc}}=\sum_{\mathsf{e}}{N_{\mathsf{e}}^{\mathsf{ab}}}N_{\mathsf{d}}^{\mathsf{ec}}$.
	}
	\label{fig_3fusion_1}
\end{figure}

\begin{figure}
	\centering
	\includegraphics[scale=0.6,keepaspectratio]{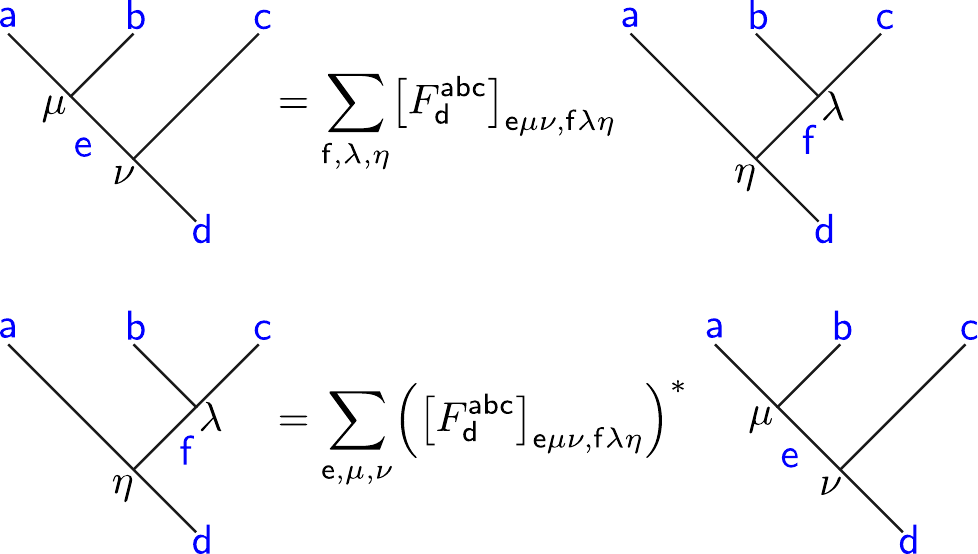}
	\caption{Definition of $F$-symbol. The left and right diagrams describe the same physics in different bases. We use the unitary $F$-symbol to change the basis.}
	\label{fig_Fmatrix}
\end{figure}

Suppose the anyonic fusion process $\mathsf{a}\otimes \mathsf{b}$ has fusion channels to $\mathsf{c}$, we can represent this fusion process diagrammatically in figure~\ref{fig_fusion}. $\mu=\left\{1,2,\cdots,N_{\mathsf{c}}^{\mathsf{a}\mathsf{b}}\right\}$ labels different fusion channels to $\mathsf{c}$. The diagram can be defined as a vector $\ket{\mathsf{a},\mathsf{b};\mathsf{c},\mu} $, where different $\mu$ represent orthogonal vectors in the fusion space $V_{\mathsf{c}}^{\mathsf{a}\mathsf{b}}$ with $\text{dim}(V_{\mathsf{c}}^{\mathsf{a}\mathsf{b}})=N_{\mathsf{c}}^{\mathsf{a}\mathsf{b}}$. For fusing three anyons, suppose $\left(\mathsf{a}\otimes \mathsf{b}\right)\otimes \mathsf{c}$ has fusion channels to $\mathsf{d}$. Diagrammatically, we depict the process as the upper diagram shown in figure~\ref{fig_3fusion_1}. Associativity indicates that the upper and the lower diagrams in figure~\ref{fig_3fusion_1} represent different bases and we can use the  unitary  $F$-symbol to change the basis. The definition of the $F$-symbol is given by figure~\ref{fig_Fmatrix}.

\begin{figure}
	\centering
	\includegraphics[scale=0.55,keepaspectratio]{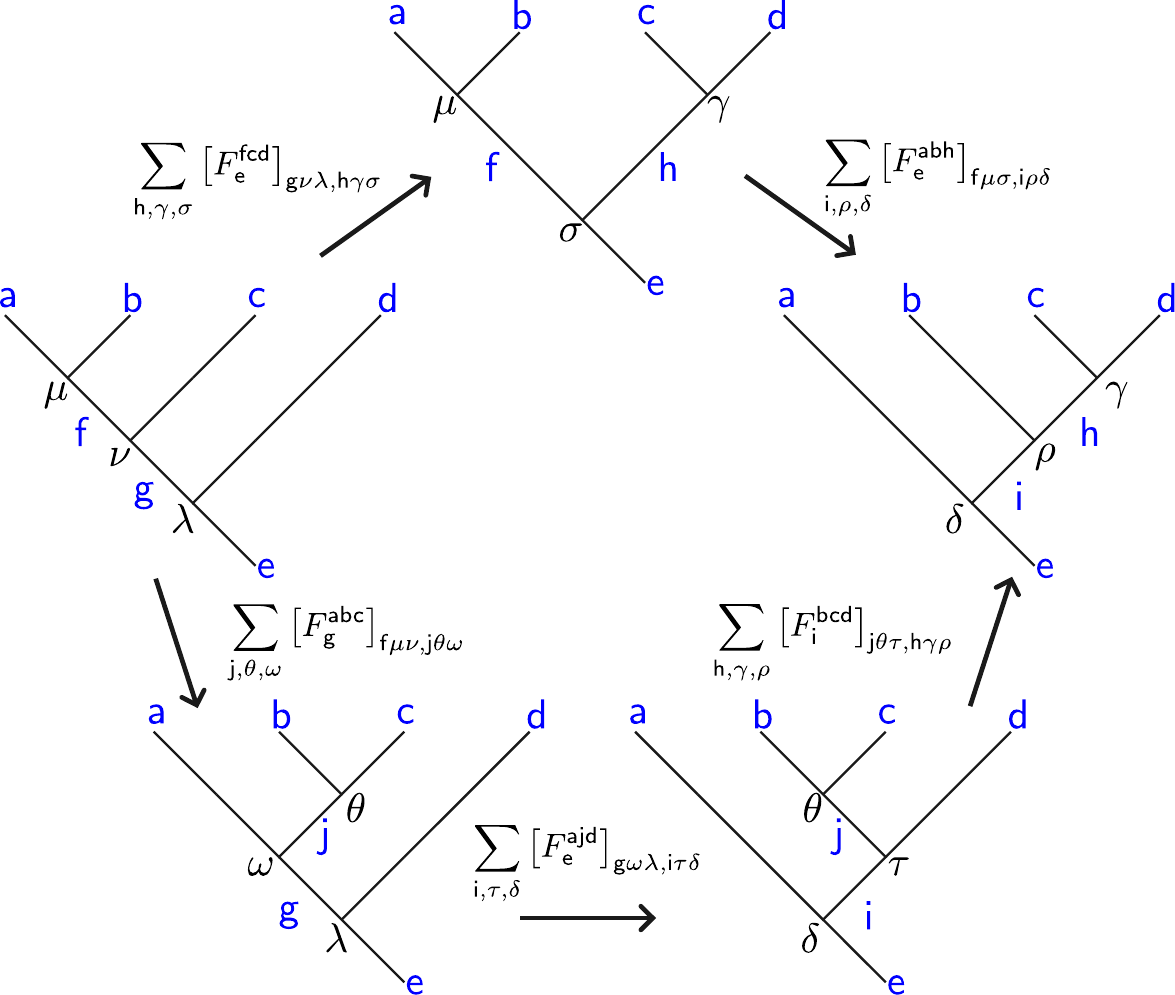}
	\caption{Pentagon equation. We highlight anyons in blue. Starting from the far left diagram, we can go to the far right diagram through either the upper path or the lower path. Comparing these two different paths, we can derive the pentagon equation.}
	\label{fig_pentagon}
\end{figure}

For the diagrams involving four anyons, as shown in figure~\ref{fig_pentagon}, we use the $F$-symbols to transform diagrams and derive the pentagon equation:
\begin{align}
	&\sum_{\sigma =1}^{N_{\mathsf{e}}^{\mathsf{fh}}}{\left[ F_{\mathsf{e}}^{\mathsf{fcd}} \right] _{\mathsf{g}\nu \lambda ,\mathsf{h} \gamma \sigma}}\left[ F_{\mathsf{e}}^{\mathsf{abh}} \right] _{\mathsf{f}\mu \sigma ,\mathsf{i} \rho \delta}=\sum_{\mathsf{j}}{\sum_{\omega =1}^{N_{\mathsf{g}}^{\mathsf{aj}}}{\sum_{\theta =1}^{N_{\mathsf{j}}^{\mathsf{bc}}}{\sum_{\tau =1}^{N_{\mathsf{i}}^{\mathsf{jd}}}{\!\left[ F_{\mathsf{g}}^{\mathsf{abc}} \right] _{\mathsf{f}\mu \nu ,\mathsf{j} \theta \omega}\left[ F_{\mathsf{e}}^{\mathsf{ajd}} \right] _{\mathsf{g}\omega \lambda ,\mathsf{i} \tau \delta}\left[ F_{\mathsf{i}}^{\mathsf{bcd}} \right] _{\mathsf{j}\theta \tau ,\mathsf{h} \gamma \rho}}}}}.
	\label{eq_pentagon}
\end{align}
No more identities beyond the pentagon equation can be derived by drawing more complicated fusion diagrams. A set of unitary $F$-symbols satisfying the pentagon equation form a unitary fusion category.

\begin{figure}
	\centering
	\includegraphics[scale=0.6,keepaspectratio]{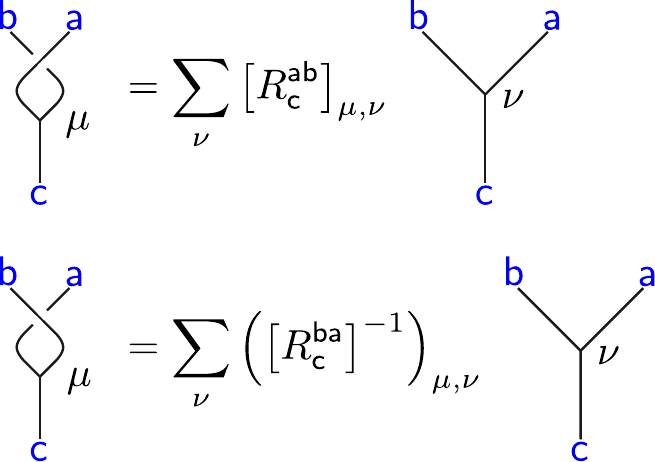}
	\caption{Definition of $R$-symbol. The upper and lower diagrams correspond to over-crossing and under-crossing respectively. $-1$ denotes the inverse matrix. Since the $R$-symbol is unitary, $\left( \left[ R_{\mathsf{c}}^{\mathsf{ba}} \right] ^{-1} \right) _{\mu ,\nu}$ can be written as $\left( \left[ R_{\mathsf{c}}^{\mathsf{ba}} \right] _{\nu ,\mu} \right) ^{\ast}$.}
	\label{fig_Rmatrix}
\end{figure}
\begin{figure}
	\centering
	\includegraphics[scale=0.55,keepaspectratio]{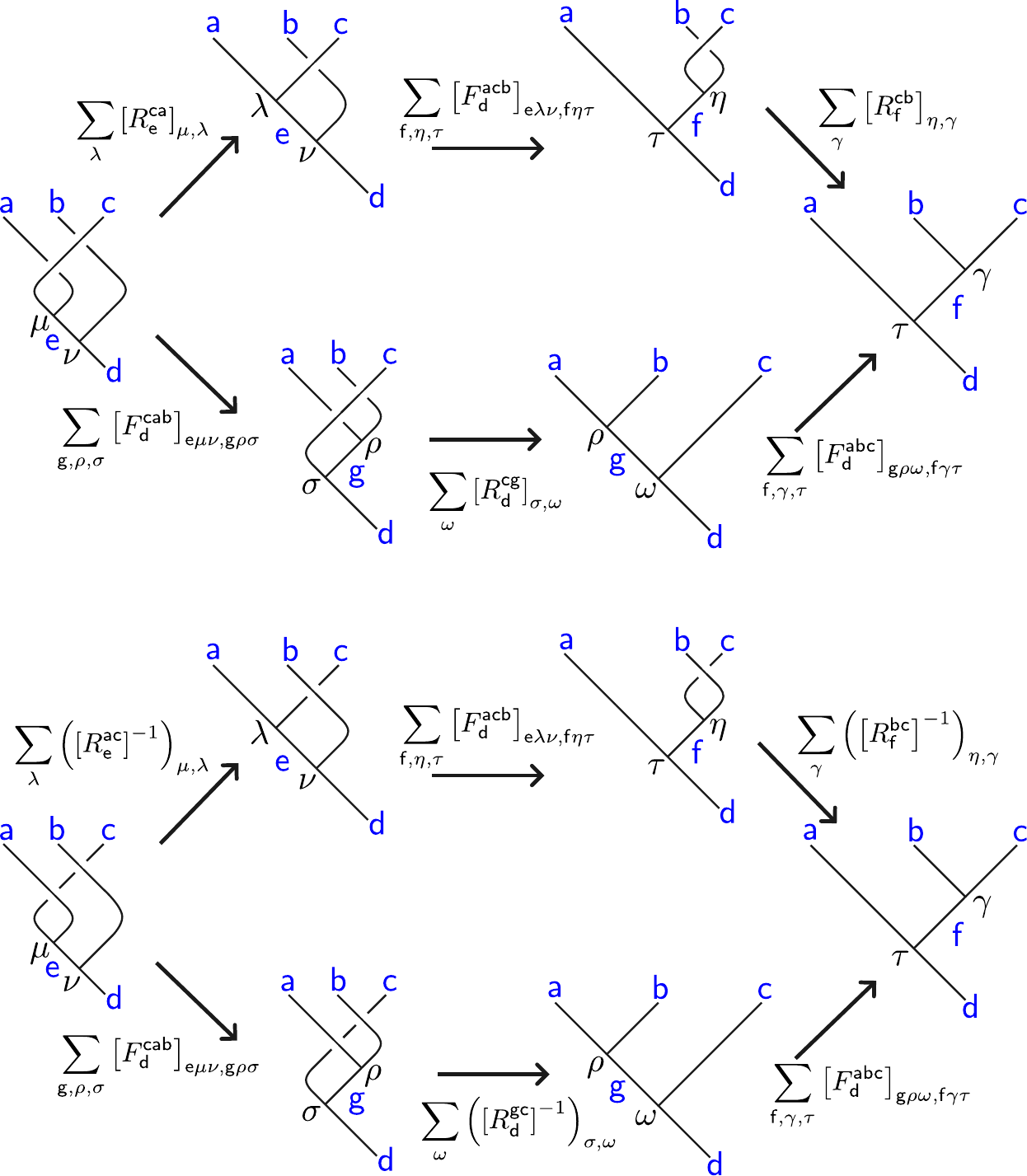}
	\caption{Hexagon equations. We highlight anyons in blue. Eq.~(\ref{eq_hexagon1}) and eq.~(\ref{eq_hexagon2}) correspond to the upper (over-crossing) and the lower (under-crossing) hexagon diagrams. Similar to figure~\ref{fig_pentagon}, for each hexagon diagram, we have two paths to transform the far left to the far right. Comparing the upper and the lower paths directly leads to eq.~(\ref{eq_hexagon1}) and eq.~(\ref{eq_hexagon2}).}
	\label{fig_hexagon}
\end{figure}

\subsection{Braiding statistics and diagrammatic representations\label{ss22}}
In section~\ref{ss21}, we only consider fusion processes, thus worldlines will not cross each other. In this section, we allow worldlines to cross over and under each other to form braiding processes. For Abelian anyons, a braiding process only accumulates a phase $e^{i\theta}$, which can be regarded as a one-dimensional representation of the braid group. However, for non-Abelian anyons, a braiding process is equivalent to multiplying a unitary matrix to the initial wavefunction of the system. This unitary matrix can be regarded as a higher-dimensional representation of the braid group.

Suppose $\mathsf{a}$ and $\mathsf{b}$ fuse to $\mathsf{c}$. This process can be viewed as a vector in the fusion space $V^{\mathsf{ab}}_{\mathsf{c}}$. Due to the principle of locality, if we (half) braid $\mathsf{a}$ and $\mathsf{b}$ first and then fuse them together, the fusion output remains $\mathsf{c}$. Consequently, this whole process can also be viewed as a vector in the space $V^{\mathsf{ab}}_{\mathsf{c}}$. These two vectors are related to each other through the $R$-symbol, as shown in figure~\ref{fig_Rmatrix}. The upper and lower diagrams in figure~\ref{fig_Rmatrix} correspond to over-crossing and under-crossing respectively. $\left[ R_{\mathsf{c}}^{\mathsf{ba}} \right] ^{-1}$ is the inverse of $\left[ R_{\mathsf{c}}^{\mathsf{ba}} \right] $. Similar to the $F$-symbol, the $R$-symbol is also unitary:
\begin{gather}
	\sum_{\nu}{\left[ R_{\mathsf{c}}^{\mathsf{ab}} \right] _{\mu ,\nu}}\left( \left[ R_{\mathsf{c}}^{\mathsf{ab}} \right] _{\mu^{\prime},\nu} \right) ^{\ast}=\delta _{\mu \mu^{\prime}}\,.
\end{gather}

Consider using the $F$-symbols and $R$-symbols in the diagrams involving three anyons, as shown in figure~\ref{fig_hexagon}, we can derive consistency relations for the $F$-symbols and $R$-symbols in $3$D topological order, known as the hexagon equations:
\begin{align}
	&\sum_{\lambda=1}^{N_{\mathsf{e}}^{\mathsf{ac}}}{\sum_{\eta=1}^{N_{\mathsf{f}}^{\mathsf{cb}}}{\left[ R_{\mathsf{e}}^{\mathsf{ca}} \right] _{\mu ,\lambda}\left[ F_{\mathsf{d}}^{\mathsf{acb}} \right] _{\mathsf{e}\lambda \nu ,\mathsf{f}\eta \tau}\left[ R_{\mathsf{f}}^{\mathsf{cb}} \right] _{\eta ,\gamma}}}\nonumber
	\\
	=\!&\sum_{\mathsf{g}}{\sum_{\rho=1}^{N_{\mathsf{g}}^{\mathsf{ab}}}{\sum_{\sigma=1}^{N_{\mathsf{d}}^{\mathsf{cg}}}{\sum_{\omega=1}^{N_{\mathsf{d}}^{\mathsf{gc}}}\!{\left[ F_{\mathsf{d}}^{\mathsf{cab}} \right] _{\mathsf{e}\mu \nu ,\mathsf{g}\rho \sigma}\left[ R_{\mathsf{d}}^{\mathsf{cg}} \right] _{\sigma ,\omega}\left[ F_{\mathsf{d}}^{\mathsf{abc}} \right] _{\mathsf{g}\rho \omega ,\mathsf{f}\gamma \tau}}}}}\,,\label{eq_hexagon1}
	\\
	&\sum_{\lambda=1}^{N_{\mathsf{e}}^{\mathsf{ac}}}{\sum_{\eta=1}^{N_{\mathsf{f}}^{\mathsf{cb}}}{\left( \left[ R_{\mathsf{e}}^{\mathsf{ac}} \right] ^{-1} \right) _{\mu ,\lambda}\left[ F_{\mathsf{d}}^{\mathsf{acb}} \right] _{\mathsf{e}\lambda \nu ,\mathsf{f}\eta \tau}\left( \left[ R_{\mathsf{f}}^{\mathsf{bc}} \right] ^{-1} \right) _{\eta ,\gamma}}}\nonumber
	\\
	=\!&\sum_{\mathsf{g}}{\sum_{\rho=1}^{N_{\mathsf{g}}^{\mathsf{ab}}}{\sum_{\sigma=1}^{N_{\mathsf{d}}^{\mathsf{cg}}}{\sum_{\omega=1}^{N_{\mathsf{d}}^{\mathsf{gc}}}\!{\left[ F_{\mathsf{d}}^{\mathsf{cab}} \right] _{\mathsf{e}\mu \nu ,\mathsf{g}\rho \sigma}\!\left( \left[ R_{\mathsf{d}}^{\mathsf{gc}} \right] ^{-1} \right) _{\sigma ,\omega}\!\left[ F_{\mathsf{d}}^{\mathsf{abc}} \right] _{\mathsf{g}\rho \omega ,\mathsf{f}\gamma \tau}}}}}.\label{eq_hexagon2}
\end{align}
Eq.~(\ref{eq_hexagon1}) and eq.~(\ref{eq_hexagon2}) correspond to the upper and the lower diagrams in figure~\ref{fig_hexagon} respectively. A set of unitary $F$-symbols and unitary $R$-symbols satisfying the pentagon and hexagon equations form a unitary braided tensor category. All $3$D anyon theories must be in this form. Given all fusion rules, there are only finite gauge inequivalent solutions of the pentagon and hexagon equations.
\begin{figure}
	\centering
	\includegraphics[scale=0.5,keepaspectratio]{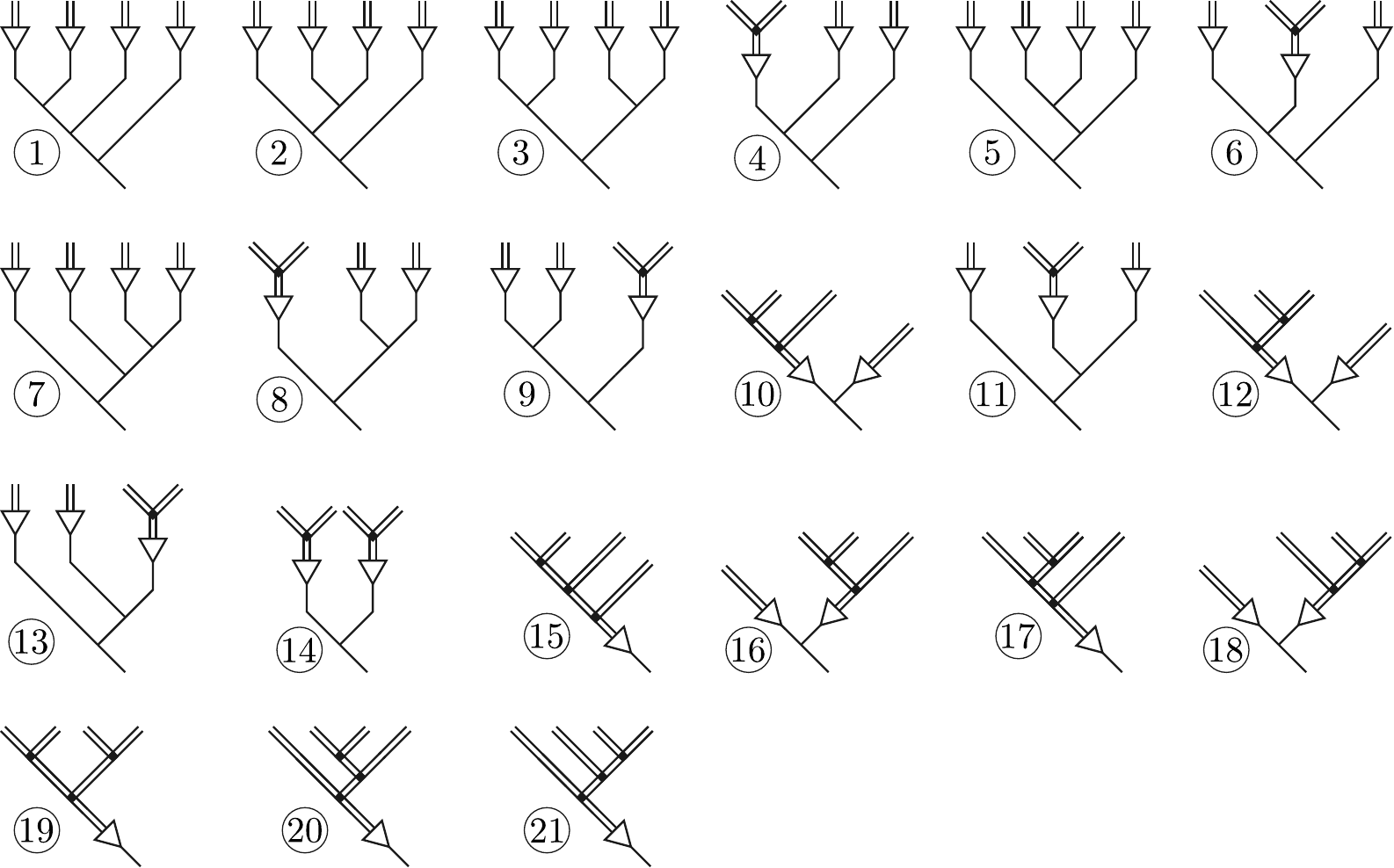}
	\caption{All $21$ diagrams that involve four topological excitations. We label them from $1$ to $21$ and their relations are given in figure~\ref{fig_connect}.}
	\label{fig_21diagram}
\end{figure}
\begin{figure}
	\centering
	\includegraphics[scale=0.5,keepaspectratio]{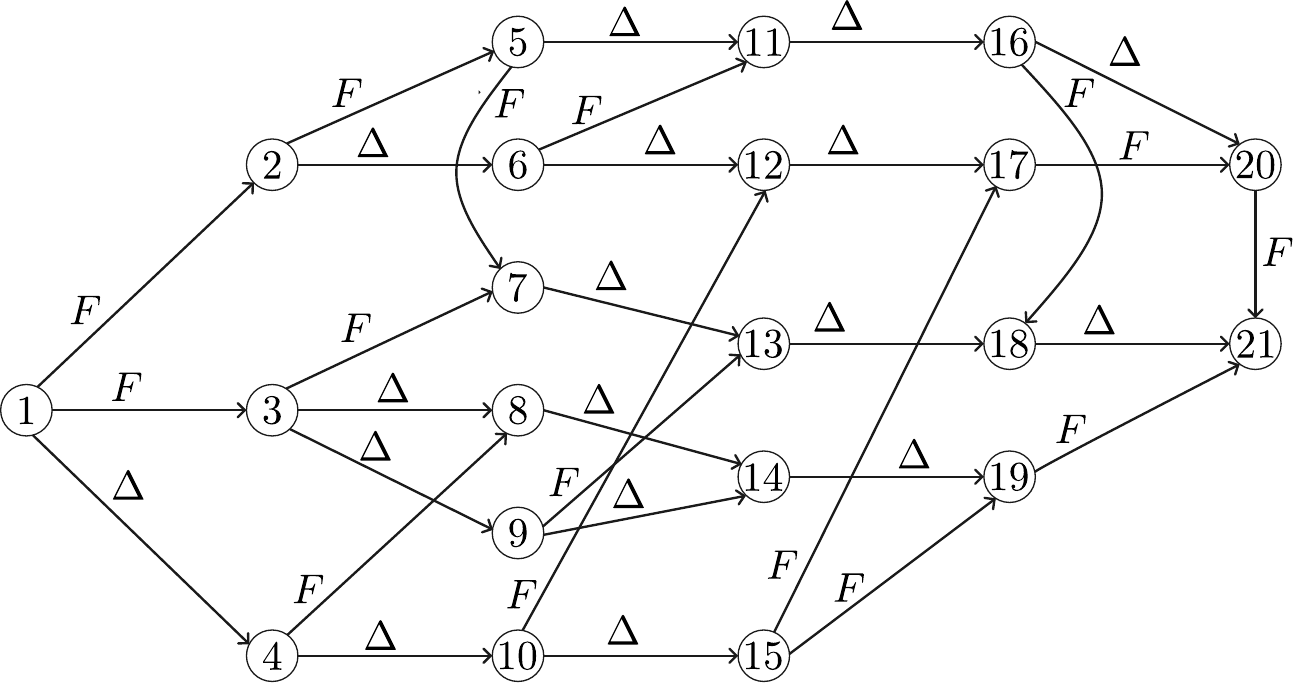}
	\caption{The relations between the $21$ diagrams. An arrow with $F$ or $\Delta$ means that we can use $F$- or $\Delta$-symbols to transform the diagram. For example, diagram $1$ can be transformed to diagram $3$ by using an $F$-symbol, diagram $3$ can be transformed to diagram $9$ by using a $\Delta$-symbol.}
	\label{fig_connect}
\end{figure}
\begin{figure}
	\centering
	\includegraphics[scale=0.5,keepaspectratio]{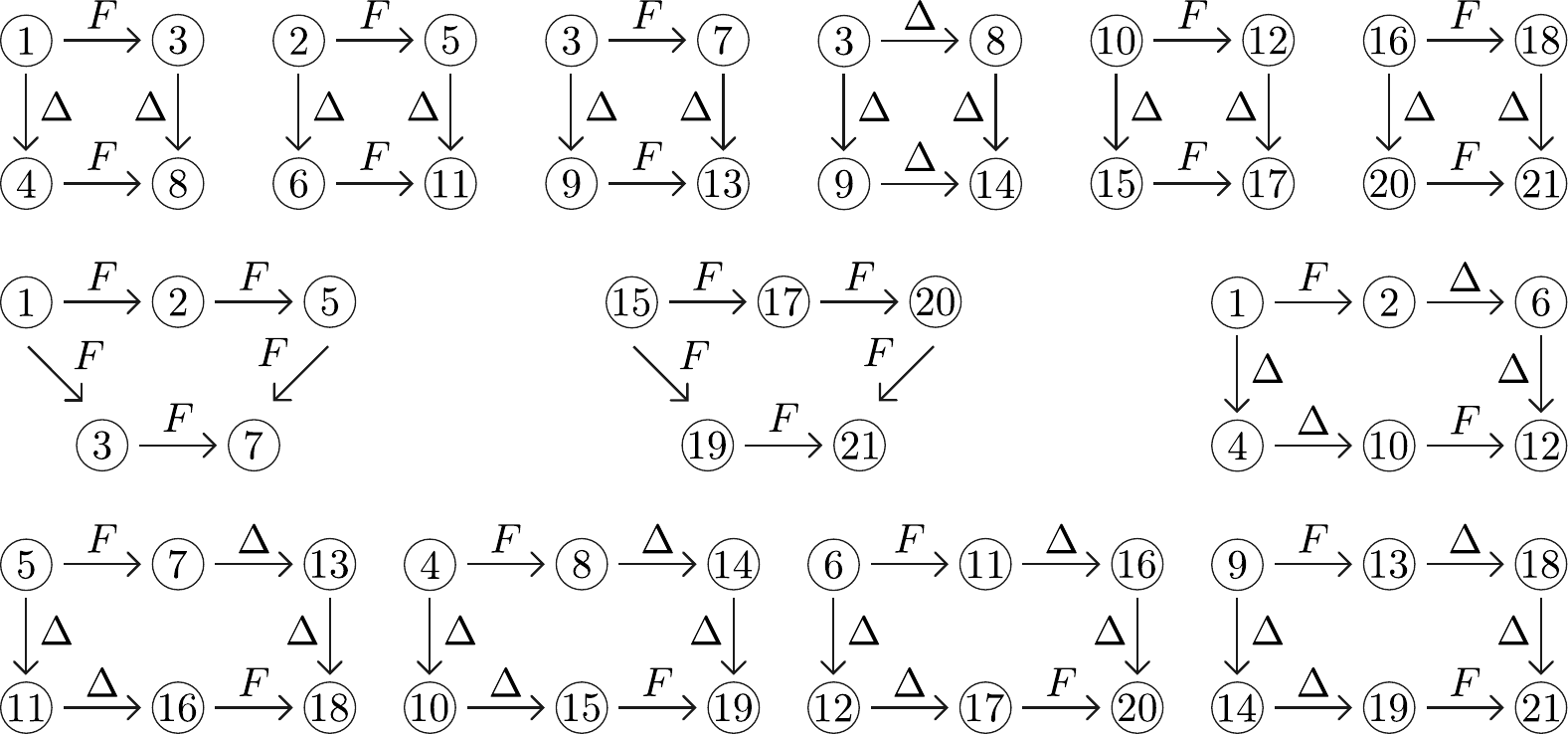}
	\caption{$13$ polygons from figure~\ref{fig_connect}. It turns out that all quadrilateral equations are trivial. Pentagons and hexagons essentially give the independent pentagon and shrinking-fusion hexagon equations.}
	\label{fig_candidate}
\end{figure}

\section{Diagrams involving four excitations}\label{ap2}
In this appendix, we consider diagrams that involve four excitations and show that  we cannot obtain new consistency relations besides the pentagon and shrinking-fusion hexagon equations from these diagrams. 

Starting with $\left[ \left[ \mathcal{S} \left( \mathsf{a} \right) \otimes \mathcal{S} \left( \mathsf{b} \right) \right] \otimes \mathcal{S} \left( \mathsf{c} \right) \right] \otimes \mathcal{S} \left( \mathsf{d} \right) $, we can finally generate a total of $21$ distinct diagrams by utilizing the $F$- and $\Delta$-symbols. We list all these diagrams in figure~\ref{fig_21diagram} and omit all excitation and channel labels for simplicity. The relations between these $21$ diagrams are shown in figure~\ref{fig_connect}, where we use arrows with $F$ or $\Delta$ to indicate transformations via the $F$- or $\Delta$-symbols. From figure~\ref{fig_connect} we can find $13$ candidates for independent polygon equations as shown in figure~\ref{fig_candidate}. However, upon recovering the excitation and channel labels and attempting to write down the consistent equations, we find that only the pentagon and shrinking-fusion hexagon equations are independent, while the other equations are automatically satisfied and thus trivial.

To be more specific, we consider the quadrilateral equation given by diagrams $1$, $3$, $4$, and $8$ as an example. The transformations $1\rightarrow3\rightarrow8$ and $1\rightarrow3\rightarrow4$ give coefficients
\begin{align}
	&1\rightarrow 3\rightarrow 8:\quad \sum_{\mathsf{m},\gamma ,\eta}{\sum_{\mathsf{n},\delta ,\tau}{\left[ F_{\mathsf{k}}^{\mathsf{igh}} \right] _{\mathsf{j}\sigma \lambda ,\mathsf{m}\gamma \eta}\left[ \Delta _{\mathsf{i}}^{\mathsf{ab}} \right] _{\mathsf{ef}\mu \nu \rho ,\mathsf{n}\delta \tau}}}\nonumber
	\\
	&1\rightarrow 4\rightarrow 8:\quad \sum_{\mathsf{n},\delta ,\tau}{\sum_{\mathsf{m},\gamma ,\eta}{\left[ \Delta _{\mathsf{i}}^{\mathsf{ab}} \right] _{\mathsf{ef}\mu \nu \rho ,\mathsf{n}\delta \tau}\left[ F_{\mathsf{k}}^{\mathsf{igh}} \right] _{\mathsf{j}\sigma \lambda ,\mathsf{m}\gamma \eta}}}\nonumber
\end{align}
respectively. We can see that they equal to each other automatically and thus the quadrilateral equation is trivial. For a similar reason, other quadrilateral equations are also trivial. The two pentagons in figure~\ref{fig_candidate}  respectively give the pentagon equations for $\left\{\mathsf{a},\mathsf{b},\mathsf{c},\mathsf{d}\right\}$ and $\left\{\mathcal{S}\left(\mathsf{a}\right),\mathcal{S}\left(\mathsf{b}\right),\mathcal{S}\left(\mathsf{c}\right),\mathcal{S}\left(\mathsf{d}\right)\right\}$, where $\mathsf{a},\mathsf{b},\mathsf{c},\mathsf{d}\in \Phi _{0}^{4}$. Since $\mathcal{S}\left(\mathsf{a}\right),\mathcal{S}\left(\mathsf{b}\right),\mathcal{S}\left(\mathsf{c}\right),\mathcal{S}\left(\mathsf{d}\right)\in \Phi _{1}^{4}\subseteq \Phi _{0}^{4}$, the later pentagon equation is automatically satisfied when we demand that the former pentagon equation holds. The five hexagons shown in figure~\ref{fig_candidate} give the shrinking-fusion hexagon equations for $\left\{\mathsf{a},\mathsf{b},\mathsf{c}\right\}$, $\left\{\mathsf{b},\mathsf{c},\mathsf{d}\right\}$, $\left\{\left(\mathsf{a}\otimes\mathsf{b}\right),\mathsf{c},\mathsf{d}\right\}$, 
$\left\{\mathsf{a},\left(\mathsf{b}\otimes\mathsf{c}\right),\mathsf{d}\right\}$ and  
$\left\{\mathsf{a},\mathsf{b},\left(\mathsf{c}\otimes\mathsf{d}\right)\right\}$ respectively and essentially they are the same  equation. Thus we conclude that there are only two independent equations in figure~\ref{fig_candidate}, which are the pentagon and shrinking-fusion hexagon equations.

 
%

\end{document}